\documentclass{article}

\usepackage[utf8]{inputenc}
\usepackage[T1]{fontenc}
\usepackage{lmodern}
\usepackage{microtype}

\usepackage{arxiv}

\usepackage{hyperref}       % hyperlinks
\usepackage{url}            % simple URL typesetting
\usepackage{booktabs}       % professional-quality tables
\usepackage{amsfonts}       % blackboard math symbols
\usepackage{nicefrac}       % compact symbols for 1/2, etc.
\usepackage{lipsum}		% Can be removed after putting your text content
\usepackage{graphicx}
\usepackage{natbib}
\usepackage{doi}
\usepackage{ulem}

\usepackage{algorithm}
\usepackage{algpseudocode}
\usepackage{amsmath,amssymb,amsfonts}
\usepackage{amsthm}
\usepackage{mathrsfs}
\usepackage{bm}
\allowdisplaybreaks
\usepackage{mathtools}

\usepackage{graphicx}
\usepackage{caption}
\usepackage{subcaption}
\usepackage{booktabs}
\usepackage{multirow}
\usepackage{threeparttable}
\usepackage{float}
\usepackage{pdflscape}
\usepackage{enumitem}

\usepackage[title]{appendix}

\usepackage{tikz}
\usetikzlibrary{arrows.meta, positioning}

\usepackage{natbib}
\bibpunct{(}{)}{;}{a}{,}{,}
\usepackage{url}
\usepackage{hyperref}

\theoremstyle{definition}

\newtheorem{proposition}{Proposition}
\newtheorem{remark}{Remark}
\newtheorem{theorem}{Theorem}
\newtheorem{lemma}[theorem]{Lemma}

\newtheorem{corollary}{Corollary}
\newtheorem{assumption}{Assumption}
\newtheorem{procedure}{Procedure}

\newcommand{\mutil}{\widetilde{\mu}}

\newcommand{\ntil}{\widetilde{n}}

\newcommand{\md}{\mathsf{d}}

\newcommand{\Niter}{N_{\mathrm{iter}}}

\newcommand{\qR}{q_{\mathrm{R}}}
\newcommand{\qL}{q_{\mathrm{L}}}
\newcommand{\qthin}{q_{\mathrm{thin}}}
\newcommand{\qfill}{q_{\mathrm{fill}}}
\newcommand{\Pois}{\mathrm{Pois}}

\newcommand{\hi}{\mathrm{hi}}
\newcommand{\lo}{\mathrm{lo}}

\title{Scalable Bayesian inference for time-discretised Markov jump processes via exchangeable particle Gibbs}

\author{
  Lanya Yang \\
  School of Mathematical Sciences \\
  Lancaster University \\
  \texttt{l.yang28@lancaster.ac.uk}
  \And
  Lloyd A. C. Chapman \\
  School of Mathematical Sciences \\
  Lancaster University \\
  \texttt{l.chapman4@lancaster.ac.uk}
  \And
  Chris Sherlock \\
  School of Mathematical Sciences \\
  Lancaster University \\
  \texttt{c.sherlock@lancaster.ac.uk}
}

\date{}

\renewcommand{\shorttitle}{}

\hypersetup{
pdftitle={Scalable Bayesian inference for time-discretised Markov jump processes via exchangeable particle Gibbs},
pdfsubject={Bayesian inference for time-discretised Markov jump processes},
pdfauthor={ Lanya Yang, Lloyd A. C. Chapman, Chris Sherlock},
pdfkeywords={particle MCMC, particle Gibbs, Markov jump process, particle filter, reaction networks},
}

\begin{document}

\pagebreak

\maketitle

\begin{abstract}
 Inference in stochastic reaction-network models, such as the susceptible-exposed-infectious-recovered (SEIR) epidemic model, is crucial for understanding the dynamics of interacting systems in epidemiology, ecology, and systems biology. These models are typically represented as Markov jump processes (MJPs), with discrete, noisy observations and intractable likelihoods. Particle Markov chain Monte Carlo (PMCMC) offers a natural mechanism for inference on these hidden Markov models, but, even in moderate dimensional state spaces, particle degeneracy  typically leads to poor mixing of the PMCMC algorithm.
This paper focuses on improving the efficiency of particle Gibbs for inference on reaction networks by addressing the degeneracy problem. Building on recent work on the exchangeable particle Gibbs (xPG) sampler for diffusions, this paper develops a novel version of xPG tailored to discrete-state-space reaction networks, where randomness is driven by Poisson processes rather than Brownian motion. We focus on the $\tau$-leap discretisation of the MJPs. The proposed method retains the exchangeability framework of xPG while adapting it to the structural and computational challenges specific to reaction networks. The new method is applied to several examples, including four coupled SEIR models, where it substantially improves mixing efficiency compared with standard particle Gibbs. In this example, xPG improves sample efficiency by more than an order of magnitude relative to standard particle Gibbs, making inference feasible in hours rather than weeks. As part of our tuning strategy, we provide a mechanism for tuning the number of particles that is applicable more generally, to any particle Gibbs algorithm.
\end{abstract}

\noindent%
\vfill

\newpage

\section{Introduction}
\label{sec.intro}

Reaction networks are commonly used to describe dynamic processes in chemistry, epidemiology, ecology and biology. Their evolution can be represented by Markov jump processes (MJPs), which are continuous-time discrete-state stochastic processes. Compared to ordinary differential equations or stochastic differential equations (SDEs), MJPs provide a more realistic description of the evolution of such systems when there is a small number of individuals (or molecules) of each species (or reactant).

Given a sequence of noisy observations of the number of individuals of  a subset of the species at a discrete set of times, we focus on performing Bayesian inference for the parameters which control the Markov jump process. In other words, we consider Hidden Markov models (HMMs) whose latent processes are MJPs. 

Particle Markov chain Monte Carlo (PMCMC) algorithms \citep{Andrieu2010} enable inference on HMMs. However, a major issue with these algorithms is particle degeneracy, where over successive resampling steps in the internal particle filter most of the ancestor trajectories get discarded leaving very few unique particles to represent the historical path of the system. For particle Gibbs algorithms, particle degeneracy leads to poor mixing of the states of the latent process at early time points. This occurs especially in moderate- or high-dimensional state spaces and/or when the time series of observations is long. We explain particle degeneracy in more detail in Section \ref{sec:particle_degeneracy}. 

One approach that has been introduced to tackle particle degeneracy is particle Gibbs \textit{with backward simulation} (PGBS) \citep{whiteley2010discussion, whiteley2010efficient, Lindsten2013}. This strategy significantly improves the sample efficiency of particle Gibbs, and was extended to particle Gibbs \textit{with ancestor sampling} (PGAS) by \cite{Lindsten2014}. However, PGBS and PGAS are only usable when the transition densities of the MJP are tractable. In addition, they only address the particle degeneracy issue caused by long time series; the issue persists when the dimensionality of the state space is high \citep{Finke2023}. 

Exchangeable particle Gibbs (xPG) \citep{malory2021bayesian}, iterated random-walk conditional sequential Monte Carlo (i-RW-CSMC) \citep{Finke2023} and Particle-MALA and Particle-mGRAD \citep{corenflos2024} have been introduced to improve the sample efficiency of particle Gibbs in moderate and high-dimensional state spaces. These approaches modify the \textit{conditional} particle filter, also known as \textit{conditional} SMC (CSMC), within particle Gibbs to mitigate the particle degeneracy issue. However, i-RW-CSMC, particle-MALA, and particle-mGRAD all require an analytical form of the transition density in the HMM, whereas xPG does not. This represents a significant advantage when the transition density is intractable. For instance, when the latent process evolves as a diffusion in continuous time, the density of the state at a given observation time (conditional on the previous state) typically lacks a closed-form expression. \cite{malory2021bayesian} applies xPG to such diffusions. By correlating the proposed particles with those in the reference path, xPG generates particles with higher weights. As a result, it is less likely to resample the same particle excessively, thereby mitigating particle degeneracy.

In this paper we develop xPG for HMMs whose latent processes are Markov jump processes (MJPs). We focus on the tau-leap method \citep{Gillespie1977}, a time discretisation of the continuous-time MJP. The MJP transition density is difficult to evaluate for two main reasons:
\begin{itemize}
    \item For some reaction networks, such as the susceptible-infectious-susceptible (SIS) model, there is no one-to-one map between a pair of states and the number of each type of reaction required to move from one state to the other.
    \item Even if the count of each type of reaction is uniquely determined when a pair of states is given, evaluating the transition mass function at consecutive observation times requires considering all possible intermediate values between the two states, which can make the computation prohibitively expensive.
\end{itemize}

We apply xPG to simulated data for different HMMs and compare its performance against PG and, where tractable transitions permit, PGBS and xPG with backward simulation (xPGBS). The mathematical formulation of the models, the particle Gibbs algorithm for MJPs, and the issue of particle degeneracy are introduced in Section~\ref{sec: background}. Section~\ref{sec: exPG for MJPs} presents the construction of xPG for MJPs, together with a tuning strategy and a backward sampling step for our algorithm. We prove that our algorithm satisfies detailed balance with respect to the intended posterior, and describe the properties relevant to tuning. The algorithm has two key parameters: $\delta$, which controls the correlation between the proposed path and the reference path within the CSMC, and $M$, the number of proposed paths in the CSMC. We propose a strategy for tuning $\delta$ and $M$ to maximise sample efficiency, defined as the minimum effective sample size (ESS) across all state dimensions and observation time points divided by computational cost. Section \ref{sec: efficiency_metric} explains why we choose this to quantify sample efficiency. In Section~\ref{sec: experiments}, we compare the sample efficiency of xPG with PG on data simulated from SIS, susceptible-infectious-recovered (SIR), and susceptible-exposed-infectious-recovered (SEIR) models with univariate and multivariate state spaces, and data simulated from the gene autoregulatory model from \cite{GolightlyWilkinson2005}. In the simulation studies of Section \ref{subsec: SIR} and \ref{subsec: SEIR}, xPG produces much larger improvements in efficiency compared with PG than backwards sampling does, but xPGBS is the most efficient of the four algorithms. Section \ref{sec: discussion} concludes the paper and points out some further research directions. 

\subsection{Notation}
We employ the standard notation: $a:b :=\{a, a+1, \ldots, b\}$ if $b \geq a$; otherwise, $a:b:=\emptyset$; $x_{a:b}$ is the set $\{x_c : c \in a:b\}$; $\mathrm{i.i.d.}$ means independent and  identically distributed; $\mathrm{Pois}(x; \mu)$ and $\mathrm{Bin}(x;n, p)$ are, respectively, the probability mass functions (pmfs) of a Poisson random variable with expectation $\mu$ and binomial random variable with $n$ trials and success probability $p$, evaluated at $x$. Specific symbols are reserved as follows:
\begin{description}[leftmargin=1.5cm, style=nextline]
   \setlength{\itemsep}{0pt} % Lanya I added these to try to reduce the spacing
  \setlength{\parskip}{0pt}
  \setlength{\parsep}{0pt}
  \item[$n$, $\tau$] Number of time steps in the  observation window and length of each step.
    \item[$T$] Total length of the full observation window, $T = n \times \tau$.
    \item[$L$, $t_{0:L}$] Number of observations, start time and observation times; $t_0=0$, $t_L=T$.
    \item[$M$] Number of proposed paths within each run of the conditional particle filter.
    \item[$\delta$] $\in [0, 1]$. Tuning parameter. Increasing $\delta$ decreases the correlation between the reference and the proposals (xPG with $\delta = 1$ is particle Gibbs).
    \item[$N$] Population size.
    \item[$\Niter$] Number of iterations.
    \item[$B^m_\ell$] Index of the weighted state at observation time $t_\ell$ in the path traced back from the $m$-th particle at the terminal time when running the conditional particle filter, so we have $B^m_{L}=m$.
    \item[$A^m_\ell$]  Index of the weighted, time-$t_{\ell}$ ancestor of the unweighted particle $\Tilde{x}_{t_\ell}^m$ at time point $t_\ell$; equivalently, index of the weighted time-$t_{\ell}$ ancestor of the weighted particle $x_{t_{\ell+1}}^m$, when running the conditional particle filter, so we have $a_{\ell}^{\,b_{\ell+1}} = b_{\ell}$, $\ell=1, \dots, L-1$. 
    \item[$ X_{t, \, j}^{(m),\,i}$] The $m$-th particle of the $j$-th species at time $t$ in the $i$-th MCMC iteration when running the conditional particle filter.
    \item[$-m$] The index set $\{0,\dots,M\}\setminus\{m\}$.
    \item[$a_{1:\ell-1}^{-b_{2:\ell}}$] Shorthand for the collection
$a_1^{-b_2}, a_2^{-b_3}, \dots, a_{\ell-1}^{-b_\ell}$.
    \item[$U$] The joint set of the reference path and all the random variables generated in a single run of the conditional particle filter.
    \item[$U_-$] Random variables $U$ except for the reference path.
\end{description}

\newpage
\section{Background}\label{sec: background}
\subsection{Hidden Markov model}
The Hidden Markov model (HMM) is a mathematical framework modelling cases where there are noisy and/or partial observations of an underlying Markov process. The underlying Markov process is referred to as the \textit{state process} or \textit{latent process}, while the observations are referred to as the \textit{observation process}.  

 If the state process evolves in continuous time, the observations can be made at any time $t\in [0, T]$. We assume there are $L$ observations, denoted by $\{y_1, \dots, y_L\}$, which occur at time $t_1, \dots, t_L$ where $t_1>0$ and $t_L=T$. The mathematical formulation of a time-homogeneous HMM in continuous time is given by $X_0\sim p_0(\cdot\mid\theta)$, 
\begin{align*}
X_{(t_{\ell-1}, t_{\ell}]} \mid (x_{[0, t_{\ell-1}]}, y_{1:\ell-1},\theta) &\sim p( \cdot \mid x_{t_{\ell-1}}, \theta), \quad \\
Y_{\ell} \mid (x_{[0, t_{\ell}]}, y_{1:\ell-1}, \theta) &\sim g(\cdot \mid x_{t_\ell}, \theta).
\end{align*}
where $\ell=1, \dots, L$. The initial state of the state process follows a prior distribution $p_0$ parameterised by some vector, $\theta$. The latent state at time $(t_{\ell-1}, t_\ell]$ follows a conditional distribution $p$, conditional on the latent state at the previous observation time $t_{\ell-1}$ and parameterised by $\theta$. Conditional on the state value $x_{t_{\ell}}$, each observation has a distribution, $g$, again parameterised by $\theta$. The observations are, thus, conditionally independent given the hidden states at the corresponding time points. 

 % We discretise the full observation interval into small time steps so that each observation time aligns with the end of one time step and the start of another. In this paper, we divide the time horizon into time steps of length $\tau$, such that each of $\{t_0, t_1, \dots, t_L\}$ is an integer multiple of $\tau$.

\subsection{Reaction networks}
Reaction networks have applications across a variety of settings. For generality, they will be explained in terms of computational biology, using the terms ``\textit{molecule}'' and ``\textit{species}''. For an epidemic model, these could become ``\textit{individual}'' and ``\textit{infection status}''.

Suppose there are $u$ reactant species and $v$ types of reactions in a dynamical system. The general form of reaction $\mathcal{R}_i$ is given by:
$$
\sum_{j=1}^u a_{ij} \mathcal{X}_j \xrightarrow{h_i} \sum_{j=1}^u b_{ij} \mathcal{X}_j, \quad\quad i = 1, \dots, v.
$$
Here, $\mathcal{X}_j$ denotes a molecule of species $j$, and $a_{ij}$ and $b_{ij}$ are non-negative integers known as \textit{stoichiometric coefficients}, representing the number of molecules of $\mathcal{X}_j$ consumed and produced, respectively, in reaction $\mathcal{R}_i$. The term $h_i$ refers to the \textit{hazard rate}, which governs the frequency at which individual events of reaction $\mathcal{R}_i$ occur. If an event of reaction $\mathcal{R}_i$ occurs at time $t$,
the change of the number of species $\mathcal{X}_j$ is $b_{ij}-a_{ij}$, $j=1, \dots, u$. The state updates according to $X_t = X_{t^-} + \Gamma^i$,
%$$
%X_t = X_{t^-} + \Gamma^i,
%$$
where $X_{t^-}$ denotes the state immediately before $t$, and $\Gamma^i = (b_{i1}-a_{i1}, \dots, b_{iu}-a_{iu})^T$. The $u \times v$ matrix $\Gamma$, whose $i$-th column is $\Gamma^i$, is the \textit{stoichiometry matrix} of the reaction network. We now describe the specific reaction networks that will be used in this article; the corresponding stoichiometry matrices are provided in Appendix~\ref{appendix:reactionNetworks}.

\subsubsection*{Susceptible-Infectious-Susceptible (SIS) model} 
% are two important examples of reaction networks in epidemiology. 
The SIS model assumes two categories of individuals in a population: susceptible ($\mathsf{S}$) and infectious ($\mathsf{I}$). There are two reaction constants, $\beta>0$ and $\gamma>0$ and two reactions:
\begin{align*}
\mathcal{R}_1: \mathsf{S}+\mathsf{I}\xrightarrow{\beta SI/N}\mathsf{2I} \quad(\text{Infection}), \qquad      &\mathcal{R}_2: \mathsf{I}\xrightarrow{\gamma I}\mathsf{S}\quad (\text{Recovery}), 
\end{align*}
where $S$ and $I$ are, respectively, the current $\mathsf{S}$ and $\mathsf{I}$ frequencies. 

\subsubsection*{Susceptible-Infectious-Removed (SIR) model}
The SIR model incorporates a third species: recovered ($\mathsf{R}$), and has two reaction types:
\begin{align*}
    &\mathcal{R}_1: \mathsf{S}+\mathsf{I}\xrightarrow{\beta SI/N}2\mathsf{I} \quad(\text{Infection}), \qquad    &\mathcal{R}_2: \mathsf{I}\xrightarrow{\gamma I}\mathsf{R}\quad (\text{Recovery}).
\end{align*}

\subsubsection*{Susceptible-Exposed-Infectious-Recovered (SEIR) model}
The SEIR model includes an additional species, exposed ($\mathsf{E}$) individuals, who are infected but not yet infectious: 
\begin{align*}
    &\mathcal{R}_1: \mathsf{S}+\mathsf{I}\xrightarrow{\beta SI/N}\mathsf{E}+ \mathsf{I}
    \quad(\text{Infection}),\qquad    & \mathcal{R}_2: \mathsf{E}\xrightarrow{\eta E}\mathsf{I}\quad (\text{Progression to infectious}),\\
    &\mathcal{R}_3: \mathsf{I}\xrightarrow{\gamma I}\mathsf{R}\quad (\text{Recovery}).
\end{align*}

\subsubsection*{Coupled epidemic models}
Suppose a pandemic is spreading across $d$ regions, with the epidemiological dynamics in each region described by a separate epidemic model. Infectious individuals from region $j$ may infect susceptible individuals from region $i$ if either travels to the other's region. Therefore, we need to couple the epidemic models via a $d\times d$ connectivity matrix, $C$, where $C_{ij}$ is the proportion of individuals in region $j$ travelling to region $i$ $(i, j=1, \dots, d)$. So the hazard rate of the infection reaction in region $i$ becomes
\[
h_i(S,I)
=
\beta S_i \frac{\sum_{j=1}^d C_{ij} I_j}{N_i},
\qquad i=1,\dots,d.
\]
where \(S_j\) and \(I_j\) represent the numbers of \(\mathsf{S}\) and \(\mathsf{I}\) individuals in region \(j\), respectively, and \(N_j\) is the population size of region \(j\), for \(j=1,\dots,d\).

\subsubsection*{Autoregulatory model} \label{subsec: autoregulatory model intro}\cite{GolightlyWilkinson2005} models a gene regulation system in which DNA makes RNA, and RNA makes a protein $\mathsf{P}$. Two protein $\mathsf{P}$ molecules may form a protein pair, $\mathsf{P_2}$, which can also break apart into two $\mathsf{P}$ molecules. Both $\mathsf{RNA}$ and $\mathsf{P}$ can degrade. The protein $\mathsf{P_2}$ can bind to the $\mathsf{DNA}$, thereby reducing the production of more $\mathsf{RNA}$. Details of the eight reactions and their effects on the five species are given in Appendix~\ref{appendix:reactionNetworks}

\subsection{Markov jump processes}\label{subsec: MJPs}
A Markov jump process (MJP) is a continuous-time, discrete-state-space process. It models the stochastic evolution of a reaction network.

An MJP may be simulated exactly using \textit{Gillespie's direct method} \citep{Gillespie1977}. We focus on a popular approximation of the evolution of an MJP: the \textit{tau-leap discretisation} \citep{gillespie2001approximate}, given in Algorithm \ref{alg: tau-leap}. After dividing the continuous time window into small time steps of length $\tau$ such that each observation time $t_j$ is an integer multiple of $\tau$, we set the current time to 0 and calculate the hazard rate of each reaction at the current time. We assume the number of occurrences of reaction $\mathcal{R}_i$ within the next time step follows a Poisson distribution with a mean equal to the product of the hazard rate, $h_i$, of $\mathcal{R}_i$ at the current time and the length of the time step, $\tau$. By sampling these Poisson counts for all reactions, we update the system state at the end of the step. This procedure is repeated until the terminal time is reached.

 \begin{algorithm}[H]
    \caption{Tau-leap method}\label{alg: tau-leap}
    \begin{algorithmic}[1]
    \State Divide the full observation interval into time steps of length $\tau$.
    \State Set $t=0$. Start from the initial value $x_{0}=(x_{0, 1}, \dots, x_{0, u})^\prime$, 
    \State Calculate the hazard rate of each reaction, $h_i(x_t),\, i=1,\dots, v$. \label{step: calc_hazard}
    \State Simulate the number of each reaction $\mathcal{R}_i$ within the next time step $[t, t+\tau]$,
    $N^{\mathcal{R}_i}\sim \text{Poisson}(h_i(x_t)\tau)$.
    \State Set $x_{t+\tau} :=x_t+ \sum_{i=1}^v N^{\mathcal{R}_i}S^i$
    \State Set $t := t+\tau$. Save $x_t$, $N^{\mathcal{R}_1},\dots, N^{\mathcal{R}_v}$. 
    \State If $t<T$, return to step \ref{step: calc_hazard}. 
    \end{algorithmic}
  \end{algorithm}
  
  The step size $\tau$ could vary from one step through the loop to the next, and the number of steps between observations could also vary from one inter-observation interval to another.  For simplicity of exposition, we will be using a fixed $\tau$ and a fixed number of steps of size $\tau$ between observation times, $K$. That is $t_{\ell}=\ell K \tau$ and $T=LK\tau$.
%\[
%t_\ell = \ell K \tau
%~~~\mbox{and}~~~
%T=LK\tau.
% \]

\subsection{Particle Gibbs for MJPs}\label{sec: PG for MJP}
 In this context, a \emph{path} refers to the collection of latent states $\{x_{k\tau}\}_{k=0}^n$, where $n$ is the total number of time steps in the full observation window, i.e. $n=KL$. A reference path refers to the existing path on which the \textit{conditional particle filter} is conditioned, and from which proposed paths are generated through the resampling steps. 

Particle Gibbs alternates sampling the latent process and the model parameter from their conditional posterior distributions, $p(x_{(0:n)\tau}\mid y_{1:L}, \theta)$ and $p( \theta \mid y_{1:L}, x_{(0:n)\tau})$, respectively. In practice, the first update is performed using a conditional particle filter, while the second update is often straightforward because the prior for $\theta$ can usually be chosen so that the posterior distribution is conditionally conjugate given the latent path. After an initial burn-in period, the algorithm is run for many further iterations to obtain samples from the joint posterior distribution of the latent path and the parameters.

In each iteration, it first generates $M$ new paths based on the current reference path $\{x_{k\tau}^{(0)}\}_{k=0}^n$ using the conditional particle filter. These $M$ paths are the proposals of the latent process. Next, it updates the reference path by sampling a new path from these $M+1$ paths including the current reference path. Finally, it updates the model parameters according to their conditional distribution given the selected reference path. 

The conditional particle filter samples $M$ particles, $x_0^{(1:M)}$, independently from the proposal distribution $q_0$. Each particle, $x_0^{(m)}$, is then propagated forward using Algorithm \ref{alg: tau-leap}, %advancing first to time $\tau$, then to $2\tau$ and so on, up to
until it reaches the first observation time $t_1=K\tau$. The transition kernel corresponding to a single $\tau$-leap, conditional on the state at the start of the leap is denoted by $q_{\tau}$. 

At each observation time $t_\ell$, for $\ell=1, \dots, L$, the particles are assigned weights
\begin{equation}
    w_\ell^{(m)}
    =
    \left[
        \frac{p_0(x_0^{(m)}\mid\theta)}
             {q_0(x_0^{(m)}\mid\theta)}
    \right]^{\mathbb{I}\{\ell=1\}}
    \frac{
        p\left(
            x_{t_{\ell-1}+\tau(1:K)}^{(m)}
            \mid \tilde{x}_{t_{\ell-1}}^{(m)},\theta
        \right)
        g\left(y_\ell\mid x_{t_{\ell-1}+\tau (1:K)}^{(m)},\theta\right)
    }{
        q\left(
            x_{t_{\ell-1}+\tau(1:K)}^{(m)}
            \mid \tilde{x}_{t_{\ell-1}}^{(m)},\theta
        \right)
    },
    \quad m=0,\dots,M,
    \label{eq: weightT}
\end{equation}
where $t_0=0$, $\tilde{x}_0^{(m)}=x_0^{(m)}$, and
$\mathbb{I}\{\ell=1\}$ equals one when $\ell=1$ and zero otherwise.
Here,
\[
    q\left(
        x_{t_{\ell-1}+\tau(1:K)}
        \mid x_{t_{\ell-1}},\theta
    \right)
    =
    \prod_{k=1}^K
    q_\tau\left(
        x_{t_{\ell-1}+k\tau}
        \mid x_{t_{\ell-1}+(k-1)\tau},\theta
    \right)
\]
is the product of the proposal probability mass functions for the
steps of size $\tau$, and $q_0$ is the proposal probability mass
function for the initial state.

Resampling according to the normalised weights yields a sample of
unweighted particles
$\{\tilde{x}_{t_\ell}^{(m)}\}_{m=1}^M$.
Each resampled particle is then propagated forward to the next
observation time. We repeat the propagation, weighting and
resampling steps until the terminal observation time $t_L$.

In addition, at resampling time $t_\ell$, we define the index of the ancestor of the unweighted particle $\Tilde{x}_{t_\ell}^{(m)}$ to be $a_\ell^m$, so we have $\Tilde{x}_{t_\ell}^{(m)}=x_{t_\ell}^{(a_\ell^m)}$, $m=1, \dots, M$ and $\ell=1, \dots, L$. We also set $a_\ell^0=0$, $\ell=1, \dots, L$, indicating that the reference path does not change. Each run of the conditional particle filter produces $M+1$ trajectories sampled from an approximation to the conditional distribution of the latent process given all the observations and the model parameters. They are obtained by tracing back from the terminal states. The path traced back from the terminal state $x_{t_L}^{(m)}$ is denoted by
\begin{align*}
    x_{(0:n)\tau}^{(b_{1:L}^m)}
= \big(
x_{t_0}^{(b_1^m)},\,
x_{t_0+\tau}^{(b_1^m)}, \dots, x_{t_1}^{(b_1^m)},\dots,\,
x_{t_{\ell-1}+\tau}^{(b_\ell^m)}, \dots, x_{t_\ell}^{(b_\ell^m)}, \dots,\,
x_{t_{L-1}+\tau}^{(b_L^m)}, \dots, x_{t_L}^{(b_L^m)}
\big),
\end{align*}
where $b_\ell^m$ denotes the index of the weighted state at observation time $t_\ell$ in the path traced back from the $m$-th particle at the terminal time when running the conditional particle filter; $b_L^m=m$. The conditional particle filter is a special case of the exchangeable conditional particle filter, which is given in Algorithm~\ref{alg: EXconditionalPF}.

Within the conditional particle filter, if we initialise from the prior and propagate forward by directly sampling from the transition distribution $p$, the weights in \eqref{eq: weightT} reduces to those of the bootstrap particle filter, the likelihood of the corresponding observation. Consequently, the closed form of the transition density $p(x_{t_\ell}|x_{t_{\ell-1}})$ is not required. This strategy is used in xPG, which is introduced in Section \ref{sec: exPG for MJPs}. The particle Gibbs algorithm is shown in Algorithm \ref{alg: PG}.

\begin{algorithm*}
\caption{Particle Gibbs}
\label{alg: PG}
\begin{algorithmic}[1]
\Require the prior distribution of $\theta$, $\pi_0$; the conditional posterior distribution for $\theta$, $p(\theta|y_{t_{1:L}}, x_{(0:n)\tau})$; the proposal distribution $q_*$ of $U_-$ for the conditional particle filter; the total number of iterations, $N_{\text{iter}}$
\State Initialise $x_{(0:n)\tau}^{(0), 0}$
\For{$i = 1$ to $N_{\text{iter}}$}
    \State Sample $\theta\sim p(\cdot|y_{t_{1:L}}, x^{(0), i-1}_{(0:n)\tau})$
    \State Sample $u_-\sim q_*(\cdot|\theta, x^{(1), i-1}_{(0:n)\tau})$ to run a conditional particle filter targeting $p(x_{(0:n)\tau}|\theta, y_{t_{1:L}})$, obtaining $x_{(0:n)\tau}^{(1:M)}$, $a_{1:(L-1)}^{1:M}$ and $w_L^{0:M}$ \label{step: conPF}
    \State Set $u=\{u_-, x^{(0), i-1}_{(0:n)\tau}\}$
    \State Sample an index $m$ from $\{0, 1, \dots, M\}$ with probability proportional to $w_L^m$.\label{step: PG_sample_m}
    % Given $m$, use $u$ to get the path back from $x_T^{(m)}$
    \State Set $\theta^{i}=\theta$ and $x_{(0:n)\tau}^{(0), i}=x_{(0:n)\tau}^{(b_{1:L}^m)}$ \label{step: PG_trace_back}
\EndFor\\
\Return $\{\theta^{i}\}_{i=1}^{N_{\text{iter}}}$ and $\{x_{(0:n)\tau}^{(0), i} \}_{i=0}^{N_{\text{iter}}}$ 
\end{algorithmic}
\end{algorithm*}

\subsection{Particle degeneracy}\label{sec:particle_degeneracy}
 The multiple proposed paths generated by the conditional particle filter share fewer and fewer distinct ancestor trajectories when traced backwards from the terminal time $T$. This issue is known as \textbf{particle degeneracy}. It is illustrated in Figure~\ref{fig:particle_degeneracy} in Appendix~\ref{appendix:particleDegeneracy}, and becomes particularly problematic in high-dimensional state spaces and/or when the time horizon is long. In the plot, we generate $100$ paths based on the reference path in a single run of the conditional particle filter. However, due to the resampling steps, when tracing these $100$ paths backwards from the particles at the terminal time, most share the same ancestors as the reference path during the early part of the time interval. We describe this phenomenon as the paths coalescing backwards in time with the reference path.

\subsection{Particle Gibbs with Backward Sampling}
Backward sampling is an extension to particle Gibbs which seeks to counter particle degeneracy by allowing a point at time $t$ to choose a different time $t-1$ ancestor. Within each iteration of particle Gibbs, given the sampled index $b^m_L=m$ for the particle at the terminal time, instead of tracing back directly as in step~\ref{step: PG_trace_back} in Algorithm~\ref{alg: PG}, a new reference path is obtained by sampling the index backward as follows. For $\ell = L-1,\dots, 1$, draw index $b_\ell^m$ with $\mathbb{P}(B_\ell^m=i)\propto \tilde{w}^i_\ell$, where 
\begin{equation}
\tilde{w}^i_\ell = w_\ell^i \cdot p\left(x_{t_{\ell+1}}^{(b_{\ell+1}^m)} \mid x_{t_\ell}^{(i)}, \theta\right), \quad i=0, \dots,M.
    \label{eq: BS_weight1}
\end{equation}
Then we obtain a new reference path, $x_{(0:n)\tau}^{(b_{1:L}^m)}$. 

By doing this, the particle degeneracy issue is largely alleviated. However, 
this requires the transition probabilities from one observation time to the next to be tractable. They are generally intractable when the corresponding reaction counts are not uniquely determined by the latent state at the two times, or when the interval between observations contains multiple intermediate time steps.

\subsection{Exchangeable particle Gibbs for diffusions}\label{sec: xPGforDiffusions}
To address the particle degeneracy issue, \cite{malory2021bayesian} introduced a method called \textbf{exchangeable particle Gibbs} (xPG). This method is intended for discrete-time models with continuous state spaces.

Firstly, the latent processes were intractable, multidimensional diffusions $\{X_t\}_{t\in[0, T]}$, where \(X_t\) denotes the latent state at time \(t\)
and takes values in the state space \(\mathcal{X} \subseteq \mathbb{R}^d\). These processes satisfy the SDE,
$dX_t=\mu(X_t) dt +\sqrt{\Sigma(X_t)} dW_t$, and we, we use the Euler-Maruyama scheme to discretise them. Given $X_t=x$, we have 
 \begin{equation}
     X_{t+\tau} = x + \mu(x) \tau + \sqrt{\Sigma(x)} \Delta W_t
     \label{eq: E-M}
 \end{equation}
where $\Delta W_t = \sqrt{\tau} Z_t$ is the increment in Brownian motion and $Z_t \sim N(0,1)$.

The key idea of xPG applied to SDEs is to correlate both the initial state $X_0$ and the Brownian motions of the proposed paths with the equivalent quantities in the reference path, rather than simulating them independently. In addition, if the conditional SMC algorithm targets the correct distribution, then the reference path is a sample from its conditional distribution given the data and the current value of the parameter, $\pi(x_{[0,T]}|y_{1:T},\theta)$.
In exchangeable particle Gibbs, the $M$ non-reference particles are proposed jointly in such a way that if, instead, the reference path had arisen from its prior,
$\pi(x_{[0, T]}|\theta)$, then each of the proposals would, marginally, have this same distribution and, furthermore, the joint distribution of the $M+1$ paths would be exchangeable, \emph{i.e.}, the joint distribution would be invariant to a permutation of the $M+1$ particles. To simplify notation, we omit the dependence on the parameter $\theta$ from all densities. 
% \CS{LANYA: you had $\theta$ in the prior for $X_0$ but not in the transitions.}

Given the initial state of the reference path $x_0^{(0)}$, the proposal distribution of the initial states in the proposed paths is defined as 
\begin{equation}
    \Tilde{q}_0(x_0^{(1:M)}|x_0^{(0)})=\frac{q_0^{(M)}(x_0^{(0:M)})}{q_0(x_0^{(0)})},
    \label{eq: q_tilde_0}
\end{equation}
where $q^{(M)}_0$ has identical marginals of $q_0\equiv p_0$ and is an exchangeable density function: for any permutation, $\sigma$, on $0:M$,
\begin{equation} 
\begin{aligned}
    &q_0^{(M)}\left(x_0^{(0:M)}\right) = q_0^{(M)}\left(x_0^{(\sigma(0:M))}\right).
\end{aligned}
\label{eq: exPrior} 
\end{equation} 

Moreover, for the simulation from time $t-\tau$ to $t$ ($t=\tau, 2\tau, \dots, K\tau$), $M$ particles $x_t^{(1)}, \dots, x_t^{(M)}$ are sampled from a joint proposal density $\Tilde{q}$ conditional on $x^{(0:M)}_{t-\tau}$ and $x^{(0)}_t$. The joint proposal density is defined as 

\begin{equation}
\label{eq: q_tilde}
    \Tilde{q}(x^{(1:M)}_t|x^{(0)}_t,x^{(0:M)}_{t-\tau})=\frac{q^{(M)}(x^{(0:M)}_t|x^{(0:M)}_{t-\tau})}{q(x^{(0)}_t|x^{(0)}_{t-\tau})}
\end{equation}
where $q^{(M)}(\cdot|x^{(0:M)}_{t-\tau})$ is a joint conditional density being exchangeable for any permutation of the $M+1$ pairs $\{(x_t^{(0)},x_{t-\tau}^{(0)}), \dots, (x_t^{(M)},x_{t-\tau}^{(M)})\}$  and $q(x_t|x_{t-\tau})\equiv p_\tau(x_t|x_{t-\tau})$.

Sampling from both distributions requires a mechanism for obtaining $\mathbf{Z}=(Z^{(0)}, \dots, Z^{(M)})^\top$ with distribution $N(\mu_{M+1}, \Sigma_{M+1})$ where $\mu_{M+1}=\mathbf{0}_{(M+1)d}$ and $\Sigma\in \mathbb{R}^{(M+1)d\times(M+1)d}$ is the Kronecker product of $I_d$ and the $(M+1)\times (M+1)$ matrix with $1$s on the diagonal and with all off-diagonal terms equal to $\rho$, the correlation parameter.
%\[
%\mu_{M+1} = \mathbf{0}_{(M+1)d}, 
%\qquad
%\Sigma_{M+1}
%=
%\begin{bmatrix}
%I_d & \rho I_d & \cdots & \rho I_d \\
%\rho I_d & I_d & \cdots & \rho I_d \\
%\vdots & \vdots & \ddots & \vdots \\
%\rho I_d & \rho I_d & \cdots & I_d
%\end{bmatrix}
%\in \mathbb{R}^{(M+1)d \times (M+1)d}.
%\]
%where $\rho$ is the correlation parameter.

At the centre of the mechanism is the methodology of \cite{Tjelmeland2004UsingAllMHProposals} for generating correlated standard Gaussian random variables. Imagine $Z^{(0)}\sim N(0, 1)$, we define a common core $Z^*=\sqrt{1-\delta^2}Z^{(0)}+Z_\delta^{(0)}$ where $Z_\delta^{(0)}$ is independently sampled from $N(0, \delta^2)$. Then, for $m=1, \dots, M$, we generate $Z^{(m)}=Z^*+Z_\delta^{(m)}$ where $Z_\delta^{(1:M)}\overset{\mathrm{i.i.d.}}{\sim} N(0, \delta^2)$. More details on generating $x_t^{(1:M)}$ given $x_t^{(0)}$ and $x_{t-\tau}^{(0:M)}$ are presented in Appendix~\ref{appendix: xpg for diffusions}. 

In \cite{malory2021bayesian}, this methodology was applied to the setting in which the latent process is a diffusion, driven by Gaussian increments. Our contribution is the far-from trivial extension of this method to the case when the latent process is a discretised Markov jump process, driven by Poisson increments. 

\section{xPGibbs for MJPs}\label{sec: exPG for MJPs}

\subsection{Main ideas}
%We are interested in applying the exchangeable particle Gibbs methodology to Markov jump processes and, in particular, to those simulated through the tau-leap mechanism described in Section \ref{subsec: MJPs}.

In \cite{malory2021bayesian}, the reference and proposed initial values and the Gaussian Brownian increments were made exchangeable with respect to their prior distribution by correlating the unit Gaussians on which they were based. This meant that the paths were exchangeable with marginals given by their prior distribution. Here, we describe methodology for making the initial values and the Poisson increments over each time step exchangeable  between the reference path and the proposals. As in \cite{malory2021bayesian} the marginals are given by the priors; that is, in \eqref{eq: q_tilde_0} and \eqref{eq: q_tilde}, $q_0(x_0)=p_0(x_0)$ and $q(x_t|x_{t-\tau})=p_\tau(x_t|x_{t-\tau})$.%, which again means that all of the paths are exchangeable with respect to their prior distribution.  

The key steps in this process are jointly simulating:
\begin{enumerate}
\item Initial values for the $M$ particles given the reference initial value, such that all are exchangeable with marginals $q_0(x_0)=p_0(x_0)$. 

\item The number of events of a particular reaction over a single interval of length $\tau$ for each of the $M$ particles, given the number of such events in the reference path and the differing Poisson rates for the reference path and each of the $M$ proposals.
\end{enumerate}

The second task is the more difficult, and so we tackle that first. 

% \CS{LANYA: I found the structure unintuitive: the next part was part of the same section as the above. Then the simulation of the priors was in the same subsection as how to do the whole xPGibbs. I've re-structured. You may need to sort out the referencing, since the "how to do the whole thing" is now unlabelled.}

\subsubsection{Simulating event numbers}
\label{sec:simExEvents}
We divide the continuous time interval from $0$ to $T$ into time steps of fixed length $\tau$. The number of occurrences of reaction $\mathcal{R}_r$ in the the $m$th  path during the time interval $\left[(k-1)\tau,\, k\tau\right]$ is denoted by $N^{(m), \mathcal{R}_r}_k$. Within the tau-leap mechanism, $N^{(m), \mathcal{R}_r}_k$ is independent of $\{N^{(m), \mathcal{R}_s}_k; s\ne r\}$ so we can deal with each reaction type separately and, for simplicity, we, therefore omit the superscript $\mathcal{R}_r$ in the notation. Define $\mu_{k}^{(m)}$ to be the corresponding expected number of $\mathcal{R}_r$ reactions in the $k$th tau-leap time step.

At time point $(k-1)\tau$, we aim to simulate $N_k^{(1)}, \dots, N_k^{(M)}|N_k^{(0)}, \mu_k^{(0)}, \dots, \mu_k^{(M)}$; \textit{i.e.}, $\Tilde{q}(x^{(1:M)}_{k\tau}|x^{(0)}_{k\tau},x^{(0:M)}_{(k-1)\tau})$ from \eqref{eq: q_tilde}.

%, the number of reactions occurring in the proposed paths during the next time step of length $\tau$, denoted by $N_k^{(1)}, \dots, N_k^{(M)}$, conditioned on the number of reaction $\mathcal{R}_r$ in the reference path, $N_k^{(0)}$, and the resulting Poisson expectations for each path, $\mu_k^{(0)}, \dots, \mu_k^{(M)}$ for the next time step.

We first thin $N_k^{(0)}$ by sampling a core $\Tilde{N}_k^{(0)}$ from the binomial distribution as follows
\begin{equation}
    \Tilde{N}_k^{(0)}\mid N_k^{(0)}  \sim \text{Bin}(N_k^{(0)}, 1-\delta)
    \label{eq: firstBinthin}
\end{equation}

If $N_k^{(0)}\sim \text{Pois}(\mu_k^{(0)})$, then $\Tilde{N}_k^{(0)}\sim\text{Pois}(\{1-\delta\}\mu^{(0)}_k)$. Starting from $\Tilde{N}_k^{(0)}$, we generate the cores for the proposed paths sequentially so that if $\Tilde{N}_k^{(0)}\sim\text{Pois}(\mu_k^{(0)})$, then the joint distribution of $\Tilde{N}_k^{(0:M)}$ is that of a Poisson process with unit rate $1-\delta$ evaluated at times $\mu_k^{(0:M)}$. To achieve this, we order all the Poisson expectations, obtaining a ladder, $\mu_k^{(
s_{-\mathrm{lo}})} \le\dots\le\mu_k^{(s_{-1})}\le\mu_k^{(s_0)}\le\mu_k^{(s_1)}\le \dots\le\mu_k^{(s_{\mathrm{hi}})} $ where $s_0=0$,  corresponds to the reference path.  We generate the core to the right, via a Poisson update, 
\begin{equation}
\begin{aligned}
   \Tilde{N}_k^{(s_{i+1})}\mid \Tilde{N}_k^{(s_{i})} =\Tilde{N}_k^{(s_{i})}+\text{Pois}\left((1-\delta) \left(\mu_k^{(s_{i+1})}-\mu_k^{(s_{i})}\right)\right), 
\end{aligned}
    \label{eq: PoisUpdate}
\end{equation}
$i=0,1, \dots, \mathrm{hi}-1$. We generate the core to the left via binomial thinning,
\begin{equation}
    \Tilde{N}_k^{(s_{i-1})}\mid \Tilde{N}_k^{(s_{i})} \sim \text{Bin}\left(\Tilde{N}_k^{(s_{i})}, \frac{\mu_k^{(s_{i-1})}}{\mu_k^{(s_{i})}}\right),
    \label{eq: BinThinning}
\end{equation}
$i = 0, -1,\dots, -\mathrm{lo}+1$. The algorithm then upgrades all the cores of the non-reference paths to full counts of reaction numbers:
\begin{equation}
    N_k^{(j)}\mid \Tilde{N}_k^{(j)} = \Tilde{N}_k^{(j)}+\text{Pois}\left(\delta \mu_k^{(j)}\right) \quad j=1, \dots, M.
    \label{eq: finalPoisUpdate}
\end{equation}
Therefore, marginally $N_k^{(j)}\sim \text{Pois}(\mu_k^{(j)})$. Moreover, the mass function of  $\Tilde{N}_k^{(0:M)}|\mu_k^{(0:M)}$ is the same as that of $\Tilde{N}_k^{(\sigma(0:M))}|\mu_k^{(\sigma(0:M))}$, since each targets the time points $\mu_k^{(0:M)}$ on a homogeneous Poisson process with rate $1-\delta$. Adding appropriate independent Poisson noise at each time point and marginalising over $\Tilde{N}_k^{(0:M)}$ we see that the mass function of  $N_k^{(0:M)}|\mu_k^{(0:M)}$ is the same as that of $N_k^{(\sigma(0:M))}|\mu_k^{(\sigma(0:M))}$. Since $\mu_k^{(m)}$ depends only on $x_{(k-1)\tau}^{(m)}$ and, given $x_{(k-1)\tau}^{(m)}$, $x_{k\tau}^{(m)}$ only depends on $N_k^{(m)}$ (for each reaction channel). This gives the exchangeability required of $q^{(M)}$ just below \eqref{eq: q_tilde}, and is a key stepping stone in the proof of Theorem \ref{thm1}.

The correlation between $N_k^{(s_{i})}$ and $N_k^{(s_{j})}$, $i<j$, increases as $\delta$ decreases. When $\delta=1$, $N_k^{(0)}, \dots, N_k^{(M)}$ are independent. When $\delta=0$, or, equivalently, the thinning step is not performed, all proposed paths coincide with the reference path whenever they share its initial state. This generation mechanism is given as Algorithm~\ref{alg: generateProposedN} in Appendix~\ref{appendix:exPG}

% \CS{I think the algorithm is sufficiently well explained in the text that we can put this in the Appendix and just refer to Section 3.1.1 rather than Algorithm 4: This generation mechanism is given as Algorithm REF in Appendix REF} 

\subsubsection{Simulating initial values}
\label{subsec: prior}
We now turn to simulating exchangeably from the prior at time 0, conditional on the value in the reference path; \textit{i.e.}, $\Tilde{q}_0$ from \eqref{eq: q_tilde_0} This depends on the prior distribution, $p_0$. We consider three cases: (a) $X_0=x_0$, (b) $X_0 \sim \text{Pois}(a_0)$, and (c) $X_0 \sim \text{Bin}(N,p_0)$.

%\begin{enumerate}[label=\alph*)]
%  \item $X_0=x_0$;
%  \item $X_0 \sim \text{Pois}(a_0)$ \label{case: pois};
 % \item $X_0 \sim \text{Bin}(N,p_0)$ \label{case: bin}.
%\end{enumerate}

Case (a) is trivial. For case (b) we first thin $X_0^{(0)}$ to draw a core $\Tilde{X}_0\sim \text{Bin}(X_0, 1-\delta)$. Secondly we obtain the proposed initial states $X_0^{(1:M)}$ via 
\begin{equation}
    X_0^{(m)} = \Tilde{X}_0 + \text{Pois}(\delta a_0)\quad m = 1, \dots, M.
\end{equation}
Clearly $X_0^{(0:M)}$ are exchangeable with marginals of $\text{Pois}(a_0)$. 

For case (c), the core is generated by sampling from a hypergeometric distribution 
\begin{equation}
    \Tilde{X}_0\sim \text{Hypergeometric}(N, X_0, N_0)
    \label{eq: X_0_case_b}
\end{equation}
where $N_0=\left \lfloor (1-\delta)N\right \rfloor $. Drawing $N_0$ times without replacement from a population of size is $N$, containing $X_0$ successes, yields $\Tilde{X}_0$ successes in the $N_0$ draws. Thus $\Tilde{X}_0\sim\text{Bin}(N_0, p_0)$. We then set 
\[
X_0^{(m)} = \Tilde{X}_0+\text{Bin}(N-N_0, p_0), \quad m=1, \dots, M.
\]
Therefore, $X_0^{(0:M)}$ are exchangeable with marginals of $\text{Bin}(N, p_0)$. 

In cases (b) and (c) the pairwise correlation is $0$ when $\delta=1$ and $1$ when $\delta=0$.

\subsection{Exchangeable conditional particle filter for the tau-leap}

Given a sequence of observations $y_{t_1}, \dots, y_{t_L}$ where $t_1>0$ and $t_L=T$ and the choices $q_0(x_0)=p_0(x_0)$ and $q(x_t|x_{t-\tau})=p(x_t|x_{t-\tau})$, the conditional exchangeable particle filter for MJPs, first samples the initial states $x^{(1)}_0, \dots, x_0^{(M)}$ from a proposal distribution $\tilde{q}_0(\cdot|x_0^{(0)})$ using the strategy described in Section \ref{subsec: prior}, ensuring that $x_0^{(0)},x_0^{(1)},\dots,x_0^{(M)}$ are exchangeable with joint mass function $q_0^{(M)}$; see \eqref{eq: exPrior}. 

 Starting from the initial states, we update the number of each reaction $\mathcal{R}_i$, $i=1,\dots,v$, in the first time step $[0, \tau]$ for each proposed path using the mechanism in Section \ref{sec:simExEvents}. We then update the states for all the proposed paths at the starting point of the next time step $[\tau, 2\tau]$ and repeat the steps above until the observation time $t_1$. Next, we weight the particles according to their likelihood at time $t_1$, $g(y_{t_1}|x_{t_1}, \theta)$. We then resample $M$ times from the set $\{x_{t_1}^{(m)}\}_{m=0}^M$ according to their weights. Based on the resampled states, we repeat the steps above until the terminal time $T$. 

To obtain exchangeable particle Gibbs, we replace the conditional particle filter in Algorithm~\ref{alg: PG} with the exchangeable conditional particle filter presented in Algorithm~\ref{alg: EXconditionalPF}.

% The detailed steps of xPG for MJPs is very similar to xPG for diffusions described in Sec~\ref{sec: xPGforDiffusions}, with two key differences: the initialisation of the states at time zero, and the replacement of Algorithm~\ref{alg: conMultiGeneratorXPG_Xt} in step \ref{step: proposeJointly_X1} and \ref{step: proposeJointly} of Algorithm~\ref{alg: conditionalPF_xPG_Malory} by the mechanism in Section \ref{sec:simExEvents}, which requires the number of each reaction occurring at each time step of the existing reference path, $\{n_k^{(0), \mathcal{R}_i}\}_{k=1}^n$, $i=1, \dots, v$.

\begin{algorithm*}[h]
\caption{Exchangeable conditional particle filter}
\label{alg: EXconditionalPF}
\begin{algorithmic}[1]
\Require the parameter, $\theta$; an existing path, $x_{(0:n)\tau}^{(0)}$; the observations $y_{t_{1:L}}$; the proposal distribution $q_0(\cdot)$ for $X_0$; the proposal distribution $q(\cdot|x_{t-1})$ for $X_t$ given $X_{t-1}=x_{t-1}$
\State Sample $x^{(1:M)}_{0}$ jointly from $\Tilde{q}_0(\cdot|x_0^{(0)}, \theta)$ 
\State For $k=1$ to $K$, sample $x_{k\tau}^{(1:M)} $jointly from $ \tilde{q}(\cdot|x_{k\tau}^{(0)}, x_{(k-1)\tau}^{(1:M)}, \theta)$\label{step: proposeJointly_X1}
\State Compute the weights of $x^{(0)}_{t_1}, \dots, x^{(M)}_{t_1}$ 
\For{$\ell = 1$ to $L$}
    \State Set $\Tilde{x}_{t_{\ell-1}}^{(0)}=x_{t_{\ell-1}}^{(0)}$
    \State Sample $M$ times with replacement from $\{x_{t_{\ell-1}}^{(m)}\}_{m=0}^M$ with probability of obtaining $x_{t_{\ell-1}}^{(m)}$ being proportional to $w_{t_{\ell-1}}^{(m)}$ to obtain unweighted samples $\{\Tilde{x}_{t_{\ell-1}}^{(m)}\}_{m=1}^{M}$
    \For{$k=1$ to $K$}
    \State Sample $x_{t_{\ell-1}+k\tau}^{(1:M)}$ jointly from $ \tilde{q}(\cdot|x_{t_{\ell-1}+k\tau}^{(0)}, \Tilde{x}_{t_{\ell-1}+(k-1)\tau}^{(1:M)}, \theta)$ \label{step: proposeJointly}
    \State If $k<K$, set $\Tilde{x}_{t_{\ell-1}+k\tau}^{(m)}=x_{t_{\ell-1}+k\tau}^{(m)}$, $m=1, \dots, M$
    \EndFor
    \State Compute the weights of $x_{t_{\ell}}^{(0)},\dots, x_{t_{\ell}}^{(M)}$
\EndFor\\
\Return $\{(x_{(0:n)\tau}^{(m)}, a_{1:(L-1)}^{m}, w^{(m)}_{L})\}_{m=0}^M$
\end{algorithmic}
\end{algorithm*}

% \CS{LANYA: the paragraph just gone currently does not make much sense given where the Algorithms are. I suggest that you move Algorithm 2 to the appendix but create an Algorithm here for Exchangeable Conditional Particle Filter. Earlier, where in Section 2.4 you say "The conditional particle filter is presented in Algorithm 2, you could instead say "The conditional Particle filter is a special case of the Exchangeable conditional particle filter, which is given in Algorithm REF." Then in Algorithm 3, just remove the "Using Algorithm 2" part. You can point out that to turn the current Algorithm 3 into exchangeable particle Gibbs you use the exchangeable conditional particle filter instead of conditional particle filter.}

Theorem \ref{thm1}, below, is proved in Appendix \ref{sec: proof_thm1} and shows that the exchangeable particle Gibbs step targets the correct conditional distribution. For simplicity in the statement and proof, we take observations to be at integer times, $t_\ell=\ell$, $\ell=1,\dots,L$.

\begin{theorem}
   For any $i = 1,\dots,M$, the xPG kernel satisfies detailed balance with respect to the target posterior distribution of the latent states, $p(x_{(0:n)\tau}|y_{1:L}, \theta)$:
    \begin{align*}
&p\!\left(x_{(0:n)\tau}^{(0)} \mid y_{1:L}, \theta\right)
 \tilde{q}\!\left(
   u \setminus \left\{x_{(0:n)\tau}^{(0)}\right\}
   \mid x_{(0:n)\tau}^{(0)}, \theta
 \right)\times q_*(i \mid u, \theta)
\\
= \;&
p\!\left(x_{(0:n)\tau}^{(b_{0:L}^i)} \mid y_{1:L}, \theta\right)
  \tilde{q}\!\left(
   u \setminus \left\{x_{(0:n)\tau}^{(b_{0:L}^i)}\right\}
   \mid x_{(0:n)\tau}^{(b_{0:L}^i)}, \theta
 \right) \times q_*(0 \mid u, \theta).
\end{align*}
\label{thm1}
\end{theorem}

\subsection{Backward Sampling for exchangeable particle Gibbs}\label{subsec: xpgbs}
We now describe an approach for incorporating backward sampling into both particle Gibbs and exchangeable particle Gibbs for MJPs, yielding particle Gibbs with backward sampling (PGBS) and exchangeable particle Gibbs with backward sampling (xPGBS), respectively. This method is only practicable for reaction networks in which the reaction counts are uniquely determined by a given pair of states, such as the SIR and SEIR models. 

% Rather than sampling the particles backward only at the observation times, we also simulate the particles backward at the intermediate time points between consecutive observations. Therefore, it is necessary to store the indices of all time points. 

For the backward sampling, we take the state at observation time $t_1$ to be $\mathbf{x}_{1}=(x_0, x_{\tau}, \dots, x_{t_\ell})^\top$ and the state at observation time $t_\ell$ to be $\mathbf{x}_{\ell}=(x_{t_{\ell-1}+\tau}, x_{t_{\ell-1}+2\tau}, \dots, x_{t_\ell})^\top$, $\ell=2, \dots, L$. Let $\tilde{B}_{\ell}$ denote the index of the selected particle $\mathbf{x}_\ell$. Within each iteration of the particle Gibbs algorithm, the backward sampling step is performed immediately after step~\ref{step: PG_sample_m}. Starting from the index at the terminal time $\tilde{B}_L=m$, for $\ell=L-1, \dots, 1$, we draw index $\tilde{b}_\ell^m$ with $\mathbb{P}(\tilde{B}_{\ell}^m=i)\propto \tilde{w}_\ell^i$, where
\[
\tilde{w}_\ell^i =  w_{\ell}^i  \cdot p\left(\mathbf{x}_{\ell+1}^{(\tilde{b}_{\ell+1})} \mid \mathbf{x}^{(i)}_{\ell}, \theta\right)=w_{\ell}^i \cdot p\left(x_{t_{\ell}+\tau}^{(\tilde{b}_{\ell+1})} \mid x^{(i)}_{t_\ell}, \theta\right),\qquad i=0, \dots, M.
\]

The validity of this approach is proved in Appendix~\ref{Appendix: Backward sampling}. Under this construction, we only need to compute the transition probability between states across one time step. 

% We will show later in Section \ref{sec: experiments} that this extra step slightly improves the mixing of both the latent states and the parameter $\theta$. \CS{is it substantial? I know it was before we had to change it. What about: In the simulation studies of Section \ref{sec: experiments}, xPGibbs produces much larger improvements in efficiency than backwards sampling does, when both are compared with PGibbs; however, the combination of both xPGibbs and backwards sampling is the most efficient of the four algorithms. (Though this might be better text for the introduction)}

\subsection{Properties}
\subsubsection{Efficiency metric} \label{sec: efficiency_metric}
% We collect the initial state of the selected path in each iteration, $x_0^1, x_0^2, \dots, x_0^{\Niter}$. They are targeting the same marginal distribution, $\pi(x_0|y_{t_{0:L}}, \theta)$. 

For some test function $h: \mathcal{X}\to \mathbb{R}$, the \textit{effective sample size} (ESS) \citep[e.g.][]{Fearnhead_etal_2025_ch2} is the number $n$ of i.i.d. samples, $Z_1, \dots, Z_n\sim \pi$, such that 
\[
\mathrm{Var}\left[\frac{1}{n}\sum_{i=1}^n h(Z_i)\right]=\mathrm{Var}\left[\hat{I}_{\Niter}(h)\right],\qquad \mathrm{where }\quad \hat{I}_{\Niter}(h)=\frac{1}{\Niter}\sum_{i=1}^{\Niter} h(X_i),
\]
when $X_0\sim \pi$. Mathematically, it is given by $\text{ESS}=\Niter/\mathsf{IACT}_h$, where $\mathsf{IACT}_h=1+2\sum_{k=1}^\infty\rho_k$ is the \textit{integrated autocorrelation time} and $\rho_k:=\text{Corr}[h(X_0), h(X_k)]=\text{Corr}[h(X_j), h(X_{j+k})]$ is the lag-$k$ autocorrelation. 

Define $\sigma^2_h=\mathrm{Var}(Z)$, where $Z\sim\pi$.  We also define the asymptotic variance of the estimator $\hat{I}_n(h)$ as $\lim_{n\to\infty} n \, \operatorname{Var}\!\left[\hat{I}_n(h)\right]$.
%\[
%\lim_{n\to\infty} n \, %\operatorname{Var}\!\left[\hat{I}_n(h)\right]
%\]
For a stationary Markov chain satisfying $\sum_{k=1}^{\infty} |\rho_k| < \infty$, 
the asymptotic variance of $\hat{I}_n(h)$ is equal to $\sigma_h^2 \, \mathsf{IACT}_h$ \citep[e.g.][]{Fearnhead_etal_2025_ch2}. 

Another commonly used measure of Markov chain mixing is the \textit{expected squared jump distance} (ESJD) \cite[e.g.][]{Sherlock2009}. For a test function $h$, it is 
$$\text{ESJD}=\mathbb{E}_{X_j\sim \pi}\left[\{h(X_{j+1})-h(X_{j})\}^2\right].$$

Maximising ESJD minimises the lag-1 autocorrelation  \cite[e.g.,][]{SherlockFearnheadRoberts2010}. In the high-dimensional limit, for certain RWM and MALA algorithms \citep[e.g.][]{Gareth2001}, ESJD is the speed of the limiting diffusion for the process, so maximising ESJD maximises the speed of the diffusion, which is equivalent to minimising the integrated auto-correlation time and, hence, maximising the (theoretical) ESS.

Consider two successive reference paths generated by xPG without backward sampling, $x_{[0,T]}$ and $x'_{[0,T]}$. At each iteration, the reference path may or may not be resampled. If the entire reference path is resampled, then $x'_{[0,T]} = x_{[0,T]}$. Otherwise, the new reference path may coincide with the previous one up to a certain observation time $t^* \in \{t_1, \dots, t_L\}$, in which case $x'_{[0,t^*]} = x_{[0,t^*]}$; alternatively, it may not coalesce with the previous path at any observation time.

If $x'_t \ne x_t$ then the new reference path from time $t$ onwards, $x'_{[t, T]}$, is not the old reference path from time $t$ onwards, $x_{[t,T]}$. However, both $x_t$ and $x'_t$ could share a common ancestor at time $s$, $s<t$. which would mean that $x'_{[0,s]}=x_{[0,s]}$. 

\begin{remark}
    For an MJP, the state space is discrete; therefore, even though a new path might not be the reference path---in that the $x_T$ that is resampled is not $x^{(0)}_T$---it is possible for the two paths to be identical at some or even all time points, including time points when they have not yet coalesced.
    \label{remark1}
\end{remark}

 %Remark \ref{remark1} indicates that even if at time $t$ the new path and the reference path have not coalesced, it is still possible for $x'_t=x_t$ since the state space is discrete.

% See a comment early in Section 4.4 for how you could add more meat here and then make your life easier in Section 4.4.

% Due to the structure of xPG and particle Gibbs, the hidden states at time zero would have the lowest ESS among all time points. For this reason, we focus on the ESS of $X_0$ as a representative diagnostic. In Section \ref{subsec: change_point}, we use an example to illustrate the bias in the estimated posterior for the parameter $\theta$ that results when the early states of the latent process barely move. 

Poor mixing of certain components in the reaction network may lead to incomplete exploration of the posterior of $\theta$, and affect the overall performance. Thus, we focus on the minimum ESS of $X_t$ across all state dimensions and observation times. 

In addition, for any algorithm, we can increase the ESS by a factor of $k$ by just running the algorithm for $k$ times the number of iterations, but this would increase the CPU time by a factor of $k$, too. Therefore, a natural measure of the efficiency of a chain is the quotient of the ESS and the CPU time. Our two tuning parameters are $M$, the number of particles, and $\delta$, the correlation tuning parameter ($\delta$=1 indicates particle Gibbs, no correlation between proposals). For xPG, changing $\delta$ has little impact on the CPU time per iteration, whereas increasing the number of particles by a factor of $k$ increases the amount of simulation by a factor of $k$ and hence increases the CPU time by approximately a factor of $k$. Thus, for standard xPG and PG, we use M as a proxy for CPU time, which leads to the following metric to assess an algorithm's efficiency when tackling the same inference problem and comparing across tuning parameters :
\begin{equation}
    \mathrm{Eff} = \frac{\min_{\ell, d}\mathrm{ESS}(X_{t_\ell}^d)}{M}.
    \label{eq: Eff}
\end{equation}
where $X_{t_\ell}^d$ denotes the $d$-th component of the hidden state at observation time $t_\ell$. 

% For xPGBS and PGBS, the backward-simulation step is the most computationally intensive part, and involves sampling from \(M+1\) particles at each time point. Accordingly, we compute their sample efficiency by replacing \(M\) with \(M+1\) in Equation~\ref{eq: Eff}. \CS{You still do this?}

\subsubsection{Tradeoff between smaller jumps and higher acceptance rate}
% Recall that the Metropolis-Hastings Independence sampler (MHIS) and Random walk Metropolis (RWM) sampler are two special cases of the Metropolis-Hastings algorithm. The proposal distribution of MHIS is independent of the current state, i.e. $q(x'|x^{0})=q(x')$ where $x^0$ is the current state and $x'$ is the proposal. The proposal distribution $q$ is required to be an approximation of the target distribution $\pi$. The higher the dimensionality of the space, the more accurate the approximation needs to be \cite[e.g.][]{Fearnhead_etal_2025_ch2}. Therefore, MHIS is not preferable in high dimensions.

% The proposal distribution of the RWM is a Gaussian distribution centred at the current state $x^0$, i.e. $q(x^1|x^0)=N(x^0, \lambda^2 I_d)$. As the dimension $d$ increases, to stabilise the acceptance rate of RWM, the standard deviation of the proposal distribution $\lambda$ is suggested \citep{Roberts1996, Sherlock2009} to be chosen as 
% $$\lambda = \frac{l}{\sqrt{d}}.$$ 
% Setting $l=2.38$ further helps the RWM to achieve the optimal acceptance rate $0.234$, which maximises the ESS of the Markov chain generated by the RWM. Although this choice is derived in the high-dimensional limit, it performs well even in dimensions as low as 5.
 For Metropolis--Hastings algorithms such as the random-walk Metropolis (RWM), the trade off between smaller jumps and higher acceptance rate is well known \cite[e.g.][]{Fearnhead_etal_2025_scaling}. For particle Gibbs and exchangeable particle Gibbs, at any given time, acceptance is equivalent to the new reference path at that time not being the old reference path. To be clear, in light of Remark \ref{remark1}, acceptance can include a new reference path which has the same state as the old reference path at that time point, as long as it is not the reference path itself. With this definition we find a similar trade-off: the smaller $\delta$ is, the closer proposed paths are to the reference path (smaller jumps) but the larger the chance that they will fit with the data and, hence, be sampled as the new reference path.
 
As in, \cite{Roberts1997}, for example, in higher dimensions, a smaller jump is required. Contour plots in Section \ref{subsec: SIS} show that, for fixed $M$, increasing the dimensionality leads to a smaller optimal $\delta$, indicating a correspondingly smaller optimal jump size.

\subsection{Tuning particle Gibbs and exchangeable particle Gibbs} \label{sec: optimal_scaling}
\subsubsection{Optimal acceptance rate}\label{subsec: optimal acceptance rate}
 We define the acceptance rate for $X_t$ at time $t$, $\alpha(t)$, as the long-run proportion of iterations where at time $t$, the newly sampled reference path is not the current reference path, $x_t^{(0)}$. We will consider the case of xPG without backwards sampling and will show that for a fixed correlation parameter, $\delta$, to balance computational cost against mixing efficiency, the number of non-reference particles, $M$, is approximately optimal when $\alpha(0)\approx e^{-1}\approx 0.368$.

Due to the structure of xPG and particle Gibbs, the empirical acceptance rates are non-decreasing over time; that is, $\alpha(s) \leq \alpha(t)$ for $s < t$, where $s,t \in \{0, t_1,\dots, t_L\}$; since the first observation is at time $t_1>0$, there is no resampling at time $0$, so $\alpha(0)=\alpha(t_1)$. Thus, the hidden states at the first observation time have the lowest acceptance rate among all observation time points. For this reason, we focus on the acceptance rate of $X_0$, $\alpha(0)$, as a representative diagnostic in this section. We shorten $\alpha(0)$ to $\alpha$ in the proofs.

Within a single run of the conditional exchangeable particle filter, recall that at observation time $t_\ell$, particles $x_{t_\ell}^{(0:M)}$ are resampled $M$ times, with probabilities proportional to their weights, $w_\ell^{(0:M)}$, giving $\tilde{x}_{t_{\ell}}^{(1:M)}$. Furthermore, the reference path is preserved: $\tilde{x}_{t_\ell}^{(0)}=x_{t_\ell}^{(0)}$. 

Let \(W^{(0)}_{\ell}\) denote the weight of the reference path at observation time \(t_\ell\), and let \(\overline{W}_\ell\) denote the average weight of the $M$ proposed particles at time $t_\ell$. Define the ratio
\[
R_\ell = \frac{W^{(0)}_{\ell}}{\overline{W}_\ell}.
\]
Further, for \(\ell=1,\dots,L\), let $M^G_\ell$ denote the number of `good' particles; \emph{i.e.}, particles which, after resampling at time $t_\ell$, are neither the reference particle nor have the reference particle in their ancestry at time $t_j$, $j<\ell$). The theoretical results behind our tuning advice require several simplifying assumptions which we now detail and justify.

\begin{assumption}
For each observation time \(t_\ell\), \(\ell=1,\dots,L\), the average weight of particles whose time-zero ancestor is the reference path is equal to the average weight of particles whose time-zero ancestor is a non-reference path.
\label{assumption: average weight}
\end{assumption}

\begin{assumption}\label{assumption: independence}
For $\ell=2, \dots, L$, $W^{(0)}_{\ell}$ and $\overline{W}_\ell$ are independent of $M_{\ell-1}^G$, and their distributions do not depend on $M$.
\end{assumption}

\begin{assumption}
\(\{R_\ell\}_{\ell\in\mathbb{N}}\) is a stochastic process with:
\begin{enumerate}
    \item a stationary distribution \(\pi_R\) such that $\mathbb{E}_{\pi_R}\left[R^2\right]<\infty$;
    \item a finite asymptotic variance, denoted by \(v_R\).
\end{enumerate}
\label{assumption: R_k Markov chain}
\end{assumption}

We do not expect Assumptions~\ref{assumption: average weight}, \ref{assumption: independence}, and \ref{assumption: R_k Markov chain} to hold exactly, but to be reasonable approximations; see Appendix~\ref{appndix_subsec:assumptions} for a rationale and discussion.

Define $p_\ell(G)$ as the probability that a random particle, after resampling at time $t_\ell$, is not the the reference path and does not have the reference path in its ancestry, that is, at any time point in $[0,t_\ell]$. We focus on the expectation $\mathbb{E}[p_L(G)]$, with the expectation taken over $M_1^G, \dots, M_{L-1}^G$ (and, implicitly, conditioning on $X_{t_{0:L}}^{(0:M)}$) because this is the probability that the $X_0^{(0)}$ component of the MCMC Markov chain moves from the current reference path to a new path. It is therefore directly related to the $\mathrm{ESJD}$ (and, hence, $\mathrm{ESS}$) of $X_0$. The following is proved in Appendix \ref{app.proveExpectPoverMG}.

\begin{lemma}
\label{lemma.expectPoverMG}
For $\ell= 1, \dots, L,$ let
   \[
    P_\ell^{(M)} := \frac{M\overline{W}_\ell}{M\overline{W}_\ell + W_{\ell}^{(0)}}
\equiv \frac{1}{1 + R_\ell/M}.
   \]
If Assumptions~\ref{assumption: average weight} 
and~\ref{assumption: independence} hold then, for each 
\(\ell=2,\dots,L\), $    \mathbb{E}_{M_{1:\ell-1}^G}\left[p_\ell(G)|X_{t_{0:\ell}}^{(0:M)}\right] = \prod_{k=1}^\ell P_k^{(M)}$.
%    \[
%    \mathbb{E}_{M_{1:\ell-1}^G}\left[p_\ell(G)|X_{t_{0:\ell}}^{(0:M)}\right] = \prod_{k=1}^\ell P_k^{(M)}.
 %   \]
\end{lemma}

We are, therefore, interested in 
\begin{equation}
\label{eqn.PGood}
    P^{(L, M)} := \prod_{\ell=1}^L P_\ell^{(M)}.
\end{equation}
Defining $E_\ell := \mathbb{E}[R_\ell]$, we will show that 
\begin{equation}
\label{eqn.EGood}
    E^{(L, M)}:=\exp\!\left(-\frac{1}{M}\sum_{\ell=1}^L E_\ell\right)
\end{equation}
is a good approximation to $P^{(M)}$ as $M\to \infty$. The following, proved in Appendix \ref{Appendix: Tuning suggestion}, provides the key stepping stone:
\begin{lemma}
\label{lemma.bound}
\[
\mathbb{E}\!\left[
\left|
P^{(L, M)}-E^{(L,M)}
\right|
\right]
\le
\frac{1}{M^2}\,
\mathbb{E}\!\left[\sum_{\ell=1}^L R_\ell^2\right]
+
\frac{1}{M}\,
\operatorname{Var}\!\left[\sum_{\ell=1}^L R_\ell\right]^{\frac{1}{2}}.
\]
\end{lemma}
\begin{corollary}
Subject to Assumption \ref{assumption: R_k Markov chain}, if $M=\lceil cL\rceil$ then 
\[
L^{1/2}\mathbb{E}\!\left[
\left|
P^{(L, M)}-E^{(L, M)}
\right|
\right]
\to \frac{\sqrt{v_R}}{c}
\]
as $L\to \infty$.
\end{corollary}
\begin{proof}
    Parts 1 and 2 of Assumption \ref{assumption: R_k Markov chain}, respectively, imply that 
\[\frac{1}{L}\mathbb{E}\left[\sum_{\ell=1}^LR_\ell^2\right]\to\mathbb{E}_{\pi_R}[R^2]
~~~\mbox{and}~~~
\frac{1}{L}\text{Var}\left[\sum_{\ell=1}^LR_\ell\right]\to v_R,
\]
as $L\to \infty$. %Part 2 of Assumption \ref{assumption: R_k Markov chain}  shows that as $L\to \infty$,
%\[
%\frac{1}%{L}\text{Var}\left[\sum_{\ell=1}^LR_\ell\right]\to v_R.
%\]
 Under Assumption~\ref{assumption: R_k Markov chain}, Lemma \ref{lemma.bound} gives
\[
\mathbb{E}\left[\left|P^{(L, M)}-E^{(L, M)}\right|\right]\le \frac{L}{M^2}\mathbb{E}_{\pi_R}[R^2]+\frac{\sqrt{L}}{M}\frac{1}{\sqrt{L}}\text{Var}\!\left[\sum_{\ell=1}^L R_\ell\right]^{\frac{1}{2}}.
\]
So 
\[
\sqrt{L}\mathbb{E}\left[\left|P^{(L, M)}-E^{(L, M)}\right|\right]\le  \frac{L^{\frac{3}{2}}}{M^2}\mathbb{E}_{\pi_R}[R^2]+\frac{L}{M}\frac{1}{\sqrt{L}}\text{Var}\!\left[\sum_{\ell=0}^L R_\ell\right]^{\frac{1}{2}}.
\]
If we set \(M=\lceil cL\rceil\), then \(L/M \to 1/c\) as \(L\to\infty\). Hence, the right hand side of the inequality above converges to $\sqrt{v_R}/c$ as $L\to \infty$. 
\end{proof}
Thus, with $M\propto L$, the error in approximating $P^{(L, M)}$ by $E^{(L, M)}$ is  $\mathcal{O}(L^{-1/2})=\mathcal{O}(M^{-1/2})$. We, therefore, deduce tuning heuristics from $E^{(L,M)}$. Let $E:=\sum_{\ell=1}^LE_\ell$, rewrite the approximation of the acceptance rate as $\alpha\approx\exp(-E/M)$. 

Let $J(\delta,M)$ be the ESJD of $X_0$ conditional on the newly sampled $X_0$ not being the old reference path $X_0^{(0)}$. We expect this to depend strongly on $\delta$, since $\delta$ governs the size of the attempted jump, and only weakly on $M$ since $M$ mainly affects the acceptance probability itself not the movement conditional on acceptance.
Since $X_0$ can only move in the MCMC if the newly sampled $X_0$ is not the old reference-path at time $0$, we write the the ESJD as the product of  $J(\delta,M)$ and the expected acceptance rate at time $0$. Then the overall ESJD is given by
$P^{(M)}J(\delta,M)\approx \exp(-E/M)J(\delta,M)$.
Taking computational effort into account, as in  Section \ref{sec: efficiency_metric}, the sample efficiency is approximately equal to 
\[
\frac{1}{M} \exp\left(-\frac{E}{M}\right)J(\delta,M).
\]
Assumption~\ref{assumption: independence} indicates that $E$ does not depend on $M$. Therefore, with $\delta$ fixed, and ignoring any weak dependence of $J(\delta,M)$ on $M$, the optimal value of $M$, which maximizes the sample efficiency is $\hat{M}=E$. The corresponding optimal acceptance rate is $\hat{\alpha}:=\exp(-1)\approx 0.368$. 

Although the minimum acceptance rate is attained by the first component of the latent process, the minimum ESS may, in some cases, occur at an observation time other than the first. Therefore, minimizing $\text{ESS}(X_0)/M$ is not equivalent to minimizing the worst-case sample efficiency defined in Section \ref{sec: efficiency_metric}. Nevertheless, the two quantities typically behave similarly, so the advice still provides useful guidance for achieving relatively high sample efficiency. We will delve further into this in Section~\ref{sec: experiments}. 

% \begin{figure*}[t]
%     \centering
%     \includegraphics[width=0.7\linewidth]{pic/Eff_SIR_high2.png}
%     \caption{$\text{ESS}(X_0)/M$ against $\alpha_{\text{ref}}$; the vertical red dashed line represents $\alpha_{\text{ref}} = 0.368$; the true latent process is the product of two independent SIR models with shared model parameters $\beta = 0.6$ and $\gamma = 0.2$.}
%     \label{fig:Eff_SIR_high2}
% \end{figure*}

\subsubsection{Tuning strategy}
 From approximation of $\alpha$ derived in Section \ref{subsec: optimal acceptance rate}, the optimal $M$ for a fixed value of $\delta$ can be expressed by 
\begin{equation}
    \hat{M}_{\delta}=-M^*\log\left(\alpha_\delta^{M^*}\right).
    \label{eq: tune_strategy}
\end{equation}
where $\alpha_\delta^{M^*}$ is the actual acceptance rate obtained by running xPG with $\delta$ and $M=M^*$. Equation \eqref{eq: tune_strategy} provides an approximation of the optimal value of $M$ that maximises $\alpha/M$, given a fixed value of $\delta$. However, it's not sufficient to just look at $\alpha/M$. It is also necessary to account for the effect of different correlation parameters - and empirically we find that simply multiplying $\alpha/M$ by $\delta^2$ does not yield the efficiency measure Eff defined in Section \ref{sec: efficiency_metric}, since $\delta$ serves as a correlation parameter rather than a true scaling parameter.

In practice, to maximise efficiency over $(\delta,M)$ we restrict attention to the curve $\alpha_\delta^M\approx \hat{\alpha}$ using the following procedure. First set $\delta=0.5$, choose a large $M=M^*$ such as $100$ or $1000$, and run the algorithm for a moderate number of iterations, noting the actual acceptance rate $\alpha_{0.5}^{M^*}$. If $\alpha_{0.5}^{M^*}<\hat{\alpha}$ derived in Section~\ref{subsec: optimal acceptance rate}, we then reduce the value of $\delta$ and run xPG until $\alpha_{\delta}^{M^*}\simeq \hat \alpha$. Conversely, if $\alpha_{0.5}^{M^*}>\hat{\alpha}$, we increase $\delta$ until $\alpha_{\delta}^{M^*}\simeq \hat \alpha$. Then, starting from this $\delta$ and $M=M^*$, we reduce $\delta$ and $M$ together, while keeping the acceptance rate close to $\hat{\alpha}$, and continue this process until the highest efficiency is achieved. 

\section{Numerical Experiments}\label{sec: experiments}
In this section, we apply xPG and particle Gibbs to synthetic data. We simulate the true latent process and generate observations based on it. We then run xPG and particle Gibbs given the observations to infer the hidden state and the model parameters. The code and data used in these experiments are available at \url{https://github.com/yanglanya980929/xPGTauLeap}. 

We conduct the experiments on the cases where the latent process is governed by the following classes of models: (i) The product of multiple independent SIS models, with noisy observations of the number of infectious people in each region. (ii) Multiple coupled SIR models, with noisy observations of the number of infectious people in each region. (iii) Multiple coupled SEIR models, with noisy observations of the number of infectious people in each region. (iv) A single autoregulatory model, with exact observations of the number of molecules of $\mathsf{RNA}$, $\mathsf{P}$ and $\mathsf{P_2}$.

In the first three cases, each region has population $N=1000$ and is governed by an individual epidemic model; in the second and third cases, infectious individuals in one region may infect susceptible individuals in another. %is permitted to spread from one region to another can be regarded as representing the epidemiology of a region. For example, in the first case, the evolution of the outbreak in each region follows an SIS model, and there is no interaction between regions. This means that the disease spreads independently within each region. On the other hand, in the second and third cases, we assume that people can travel to other cities and infect the residents there, so the epidemic models are coupled. 

When reporting results we use the following abbreviations: PG for particle Gibbs, xPG for exchangeable particle Gibbs, PGBS for particle Gibbs with backwards sampling and xPGBS for exchangeable particle Gibbs with backwards sampling.

\subsection{SIS}\label{subsec: SIS}
We run xPG in multi-dimensional space, where the state in each dimension follows an independent SIS model. All the SIS models share the same true values of $\beta=1.2$ and $\gamma=0.6$, and the time-step length is set to $\tau=0.2$. We have $21$ observations of the number of infectious people in each model at time time $t=0, 1, \dots, 20$. We set $d=1, 3, 5$ and $7$. For any $t \in \{0, 1, \dots,20\}$, given $I_t^i$, the number of infectious people in region $i$ ($i\in\{1,\dots,d\})$, the observation $Y_t^i$ is generated by sampling from a truncated discretised Gaussian distribution with the following probability mass function $g(Y_t^i|X_t^i)$. For $Y_t^i\in \mathbb{Z}$, 
\[
g(Y_t^i|I_t^i) = \begin{cases}
\displaystyle \frac{\phi\left(|Y_t^i - I_t^i|\right)}{\sum_{k=-2\sigma}^{2\sigma}\phi(k)}, & \text{if } |Y_t^i - I_t^i| \le 2\sigma \\[6pt]
0, & \text{otherwise}
\end{cases},
\]
where $\phi$ is the density function of a Gaussian distribution with mean $0$ and standard deviation $\sigma$, i.e., $N(0, \sigma^2)$, and $\sigma$ is a positive integer. Therefore, the observation $Y_t=(Y_t^1, \dots, Y_t^d)$ is multi-dimensional when $d>1$. In addition, the prior distribution of the initial infectious state at time zero in each region is 
\[
I_0^i\sim \operatorname{Bin}(N,0.05),
\qquad i=1,\ldots,d.
\]
For each region $i$, we propose the initial infectious states exchangeably with respect to the prior using the method described in Sec~\ref{subsec: prior}, Case (c). 

In both data simulation and inference, we set $\sigma$ to be $5$. The parameter priors are  
\begin{equation}
\beta \sim \text{Gamma}(1,1)\quad \text{and}\quad \gamma \sim \text{Gamma}(1,1).
\label{eq: paramPriorSIS}
\end{equation}
Their initial values are $\beta^{0}=1$ and $\gamma^0=1$.

 We ran xPG and particle Gibbs (equivalent to xPG with $\delta = 1$) for $M \in$\\ $\{1,5,10,30,50,70,100,120,150,180\}
    \cup \{200+50k: k=0, 1, \dots, 9\}$, and $\delta\in$\\
    $\left\{10^{-2.5+0.2k}: k=0,1,\dots,12 \right\}\cup \{1\} $. %The total population size of each region is fixed at $n=1000$. 
    At each time point, each model in the product has $S+I=N$, so the ESSs of the $S$ and corresponding $I$ states are equal.
    %contains only two compartments, it is sufficient to consider the ESS of the latent states corresponding to either the susceptible or the infectious population, as the ESS of the two compartments are equal. %This then allows the sample efficiency, as defined in Section~\ref{sec: efficiency_metric}, to be computed in a natural way.
    
\begin{figure}[h]
\captionsetup{font=footnotesize}
\captionsetup[subfigure]{labelformat=empty,font=footnotesize}
\centering
% \begin{figure}[H]
% \captionsetup{font=footnotesize}
% \captionsetup[subfigure]{labelformat=empty,font=footnotesize}
% \centering
\begin{subfigure}{0.49\linewidth}
  \centering
    \captionsetup{
    justification=centering,
    singlelinecheck=false,
    margin={0pt,12mm}
  }
  \caption{1D}
  \includegraphics[width=\linewidth,height=0.31\textheight,keepaspectratio]{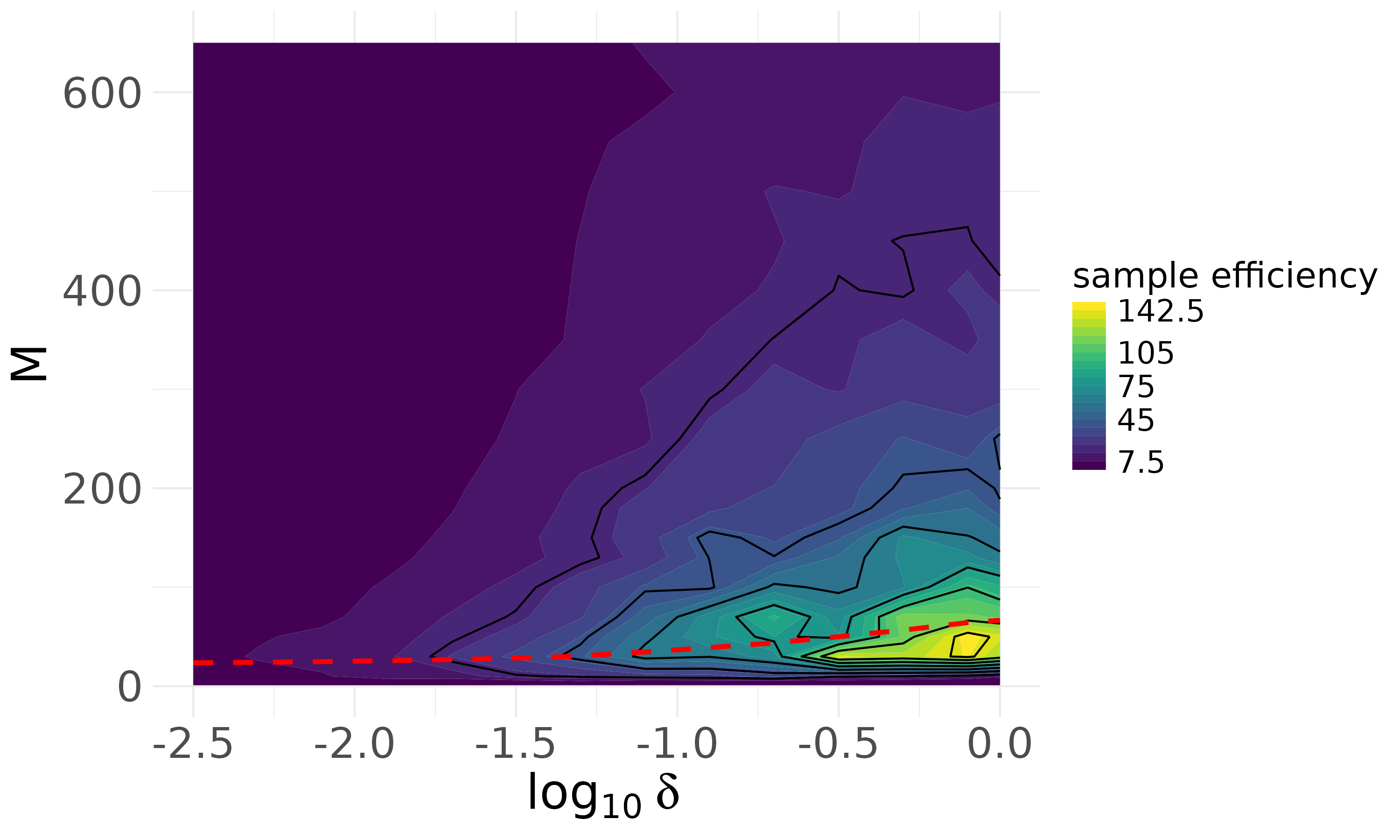}
  \label{fig:SIS_high1}
\end{subfigure}
\hfill
\begin{subfigure}{0.49\linewidth}
  \centering
      \captionsetup{
    justification=centering,
    singlelinecheck=false,
    margin={0pt,12mm}
  }
  \caption{3D}
  \includegraphics[width=\linewidth,height=0.31\textheight,keepaspectratio]{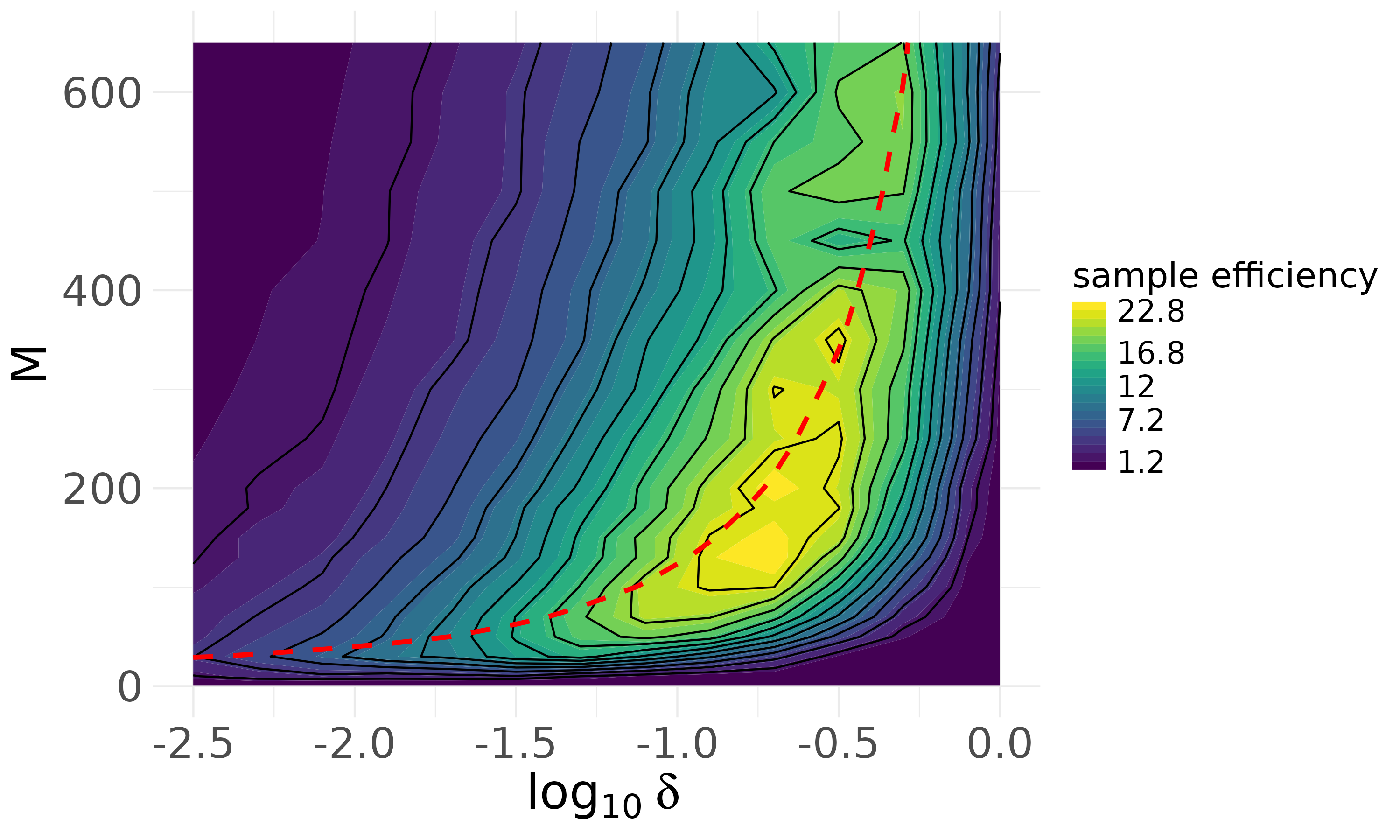}
  \label{fig:SIS_high3}
\end{subfigure}

\vspace{2mm}

\begin{subfigure}{0.49\linewidth}
  \centering
      \captionsetup{
    justification=centering,
    singlelinecheck=false,
    margin={0pt,12mm}
  }
  \caption{5D}
  \includegraphics[width=\linewidth,height=0.31\textheight,keepaspectratio]{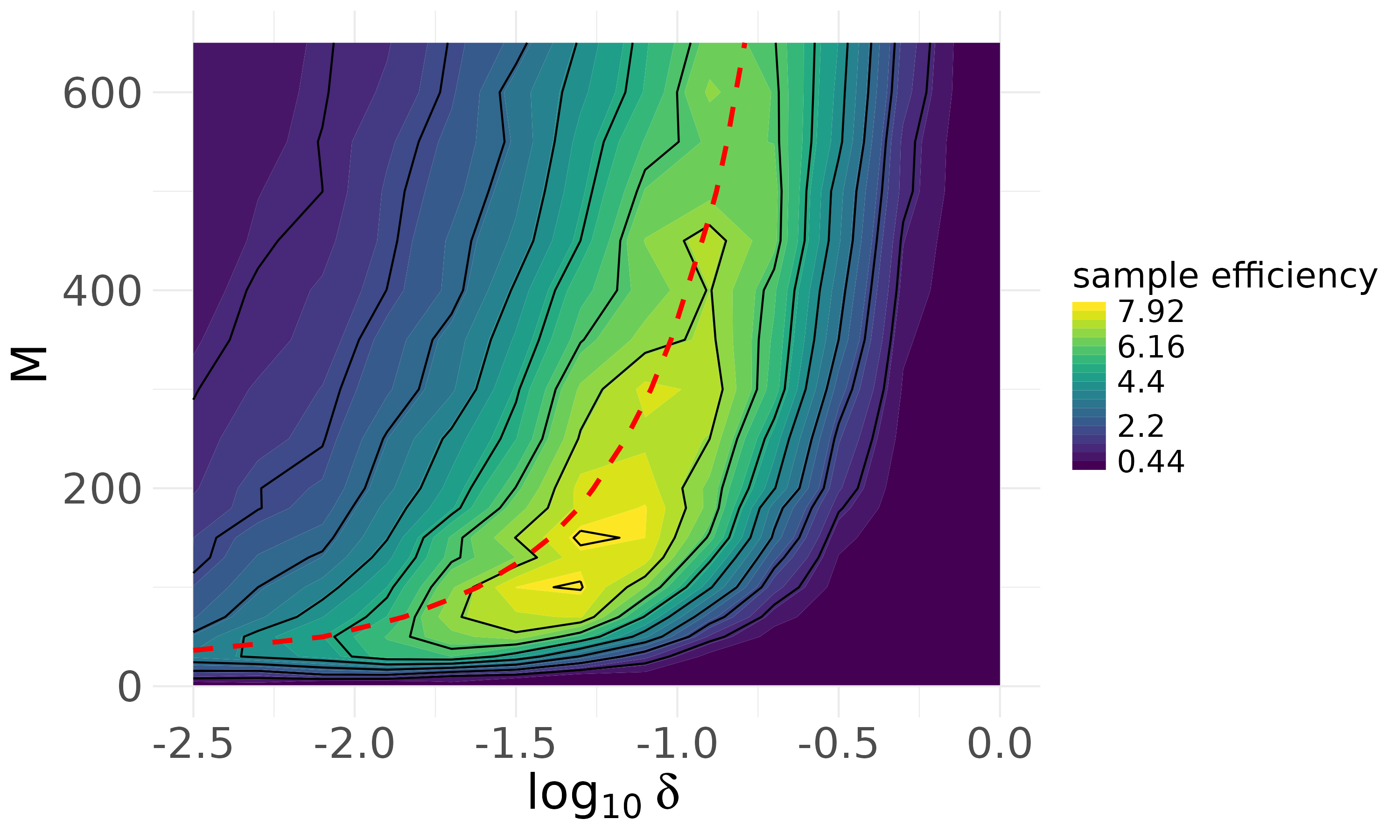}
  \label{fig:SIS_high5}
\end{subfigure}
\hfill
\begin{subfigure}{0.49\linewidth}
  \centering
      \captionsetup{
    justification=centering,
    singlelinecheck=false,
    margin={0pt,12mm}
  }
  \caption{7D}
  \includegraphics[width=\linewidth,height=0.31\textheight,keepaspectratio]{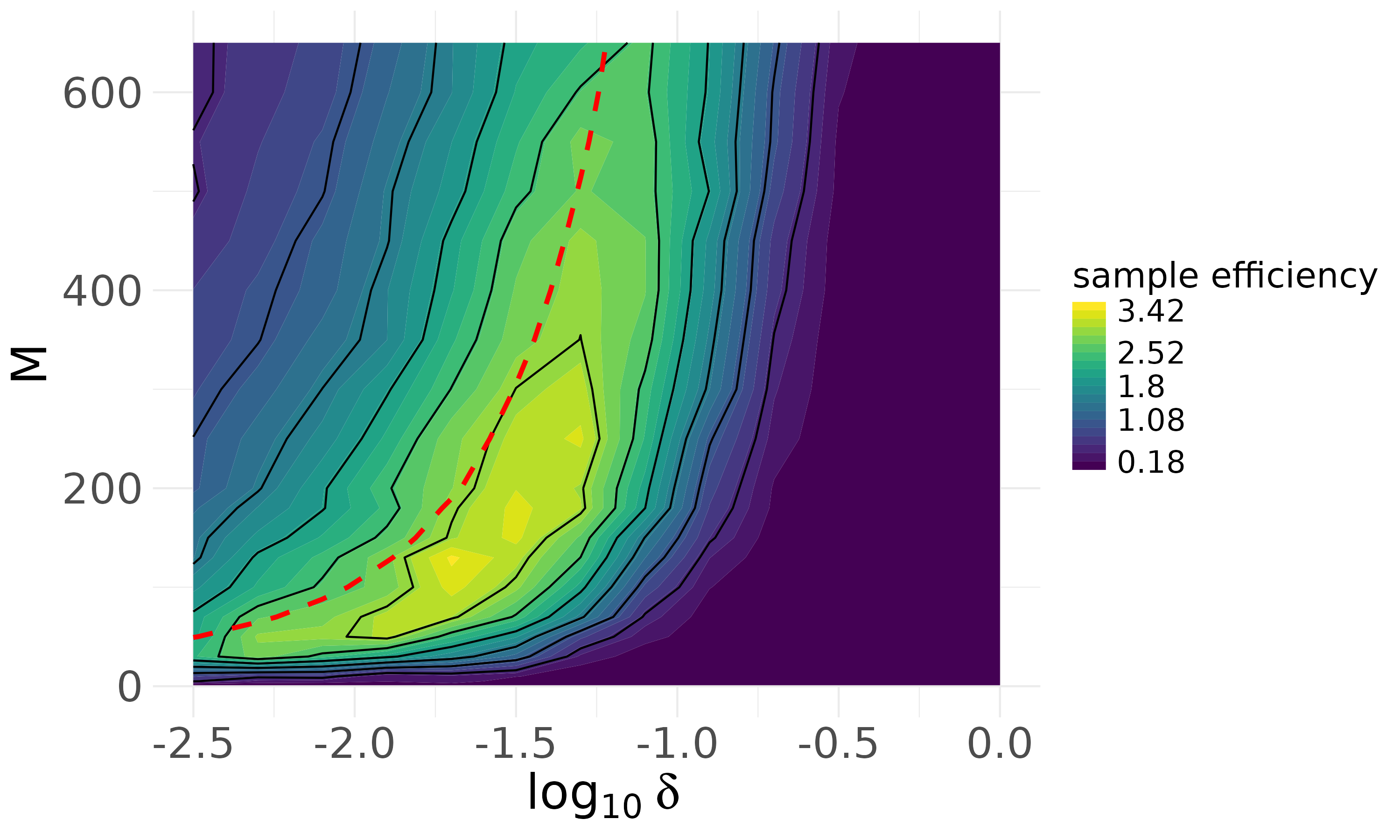}
  \label{fig:SIS_high7}
\end{subfigure}
\caption{
\small Contour plots of sample efficiency after running xPG for $10^5$ iterations in 1D, 3D, 5D, and 7D; latent states are products of independent SIS models with shared parameters $\beta=1.2$ and $\gamma=0.6$. The red dashed line indicates the points at which the acceptance rate at the first observation time $t_1$ is \(\exp(-1)\).
}
\label{fig: SIS_four}
\end{figure}

Figure \ref{fig: SIS_four} shows the contour plots of the sample efficiency of xPG after running $10^5$ iterations in the one-, three-, five- and seven-dimensional cases. The maximum sample efficiency of xPG is achieved at progressively smaller values of $\delta$ as the dimension increases. In dimensions $1$, $3$, $5$ and $7$, the maximum is achieved at  $(\delta, M)=(0.79,50)$, \((0.20,130)\), \((0.05,150)\) and  \((0.02,130)\), respectively.

In addition, the plots suggest the presence of a tipping point in the relative efficiency of xPG vs particle Gibbs at around three dimensions. Beyond this point, xPG shifts from being only slightly more efficient than particle Gibbs to being substantially better, and further increasing $M$ no longer improves performance (i.e., particle Gibbs does not reach the same efficiency as xPG even for large $M$). In the three-dimensional case in particular, particle Gibbs achieves its maximum efficiency at $M = 650$, with a value of $4$. However, the maximum efficiency of xPG is $24$, approximately $6$ times higher than that of xPG. 

Moreover, as shown in Figure~\ref{fig: SIS_four}, the red dashed line, which corresponds to the optimal acceptance rate derived in Section~\ref{sec: optimal_scaling}, generally passes through the circle indicating the highest sample efficiency for each fixed value of $\delta$. This therefore provides a useful guideline for choosing the optimal value of $M$ given $\delta$, although it slightly overestimates the number of particles in the $7$-dimensional case.

\subsection{SIR}\label{subsec: SIR}
We apply xPG to $d=4$ coupled SIR models, all of which share the same model parameters, $\beta=0.6$ and $\gamma=0.2$. Appendix \ref{appendix:reactionNetworks} gives the $d\times d$ connectivity matrix between regions.

The length of a time step, $\tau$, is $0.2$, and observations $Y_t^i$ $(t=1,\dots,20)$ are still of the number of infectious people in each region $i$ ($i\in\{1,2,3,4\}$, but they follow a binomial distribution given the corresponding actual number of infected individuals $I_t^i$
\begin{equation}
\label{eq.BinObs}
Y_t^i \sim \text{Bin}\left(I_t^i, 0.9\right).
\end{equation}
For each region, the initial state at time zero is assumed to be fixed and known. The priors and initial values of $\beta$ and $\gamma$ are the same as those for the SIS model (Equation \eqref{eq: paramPriorSIS}).

\begin{figure}[h]
\captionsetup{font=footnotesize}
\captionsetup[subfigure]{labelformat=empty,font=footnotesize}
\centering

\begin{subfigure}{0.49\linewidth}
  \centering
      \captionsetup{
    justification=centering,
    singlelinecheck=false,
    margin={0pt,12mm}
  }
  \caption{Susceptible state, xPG}
  \includegraphics[width=\linewidth,height=0.31\textheight,keepaspectratio]{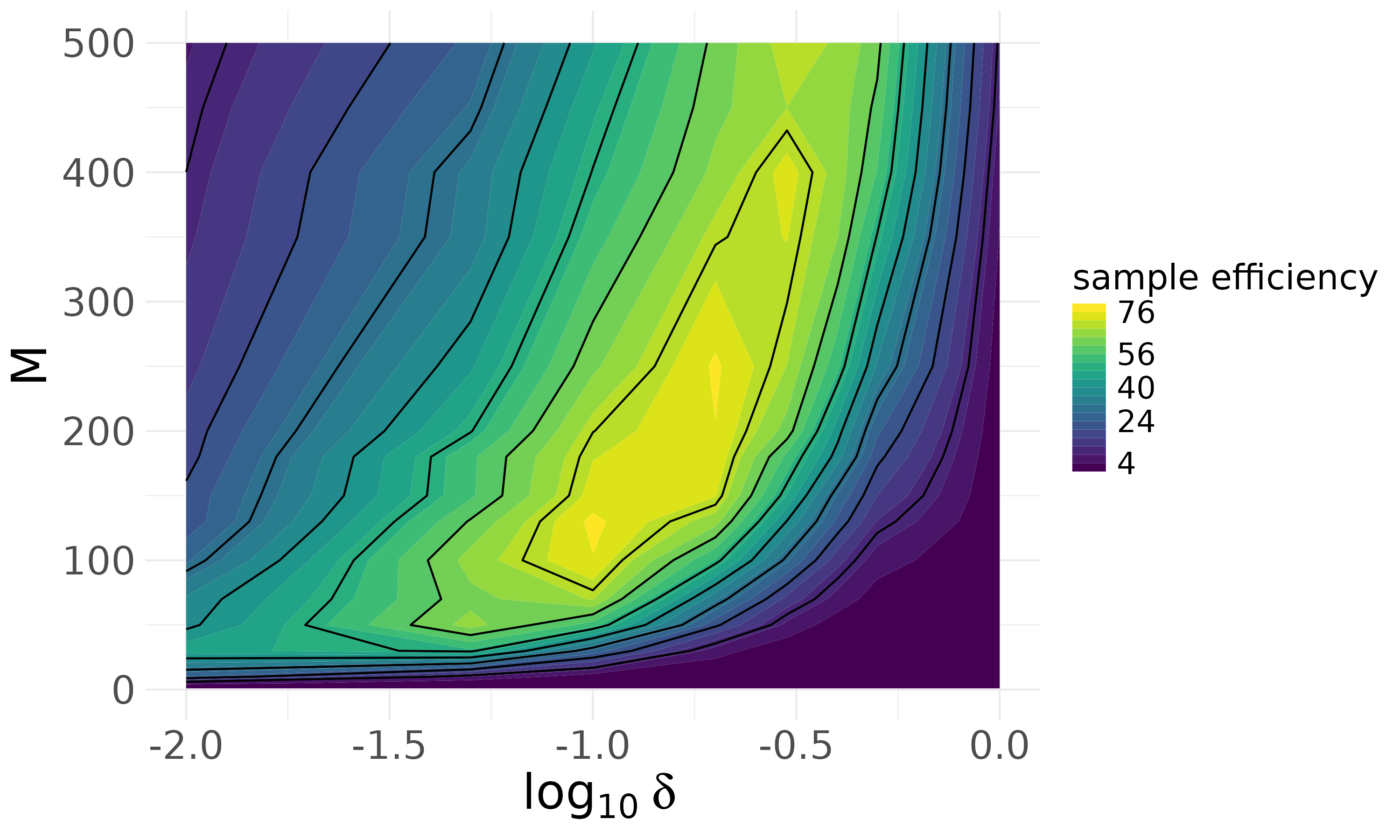}
  \label{fig:SIR_high4_S}
\end{subfigure}
\hfill
\begin{subfigure}{0.49\linewidth}
  \centering
        \captionsetup{
    justification=centering,
    singlelinecheck=false,
    margin={0pt,12mm}
  }
  \caption{Susceptible state, xPGBS}
  \includegraphics[width=\linewidth,height=0.31\textheight,keepaspectratio]{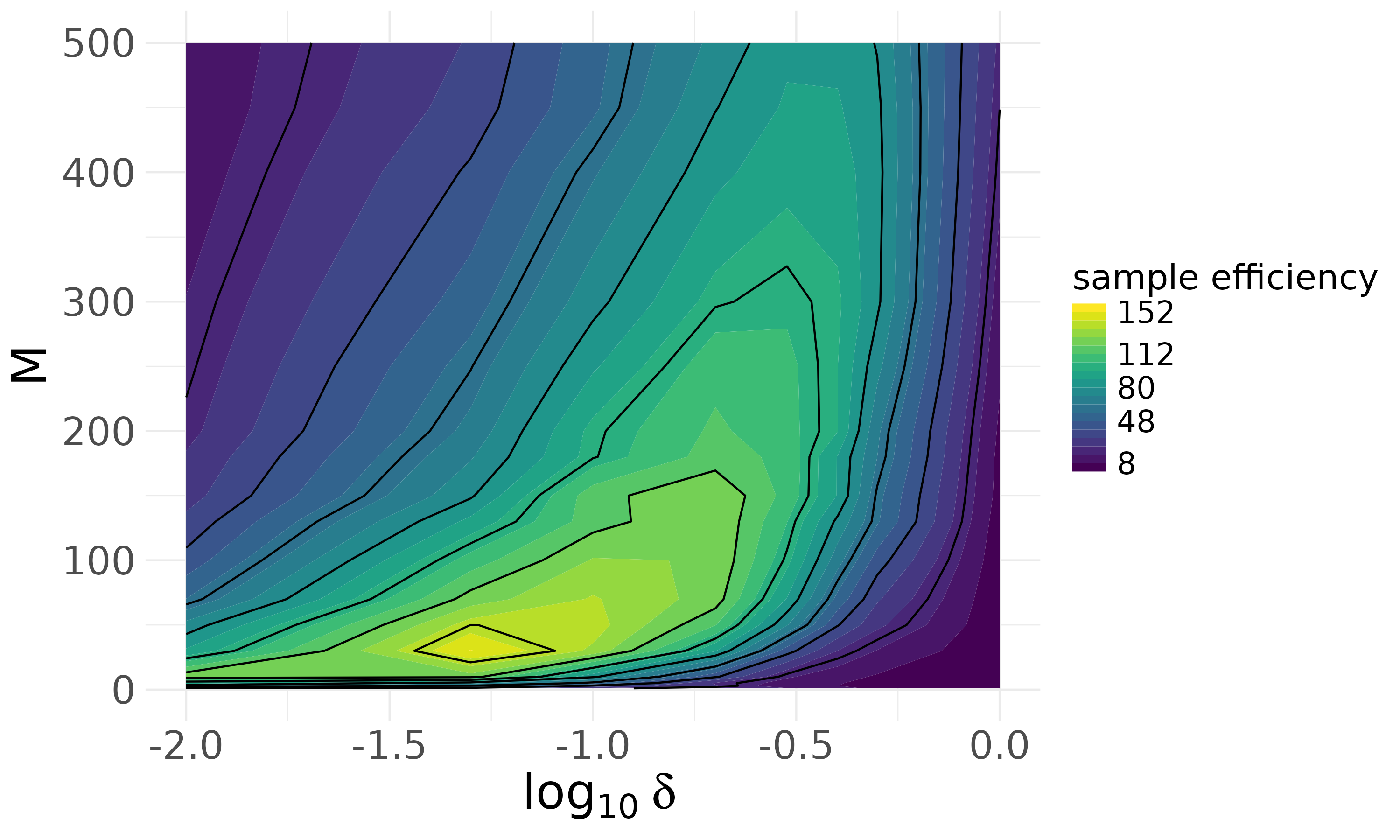}
  \label{fig:SIR_BS_high4_S}
\end{subfigure}

\vspace{2mm}

\begin{subfigure}{0.49\linewidth}
  \centering
        \captionsetup{
    justification=centering,
    singlelinecheck=false,
    margin={0pt,12mm}
  }
  \caption{Infectious state, xPG}
  \includegraphics[width=\linewidth,height=0.31\textheight,keepaspectratio]{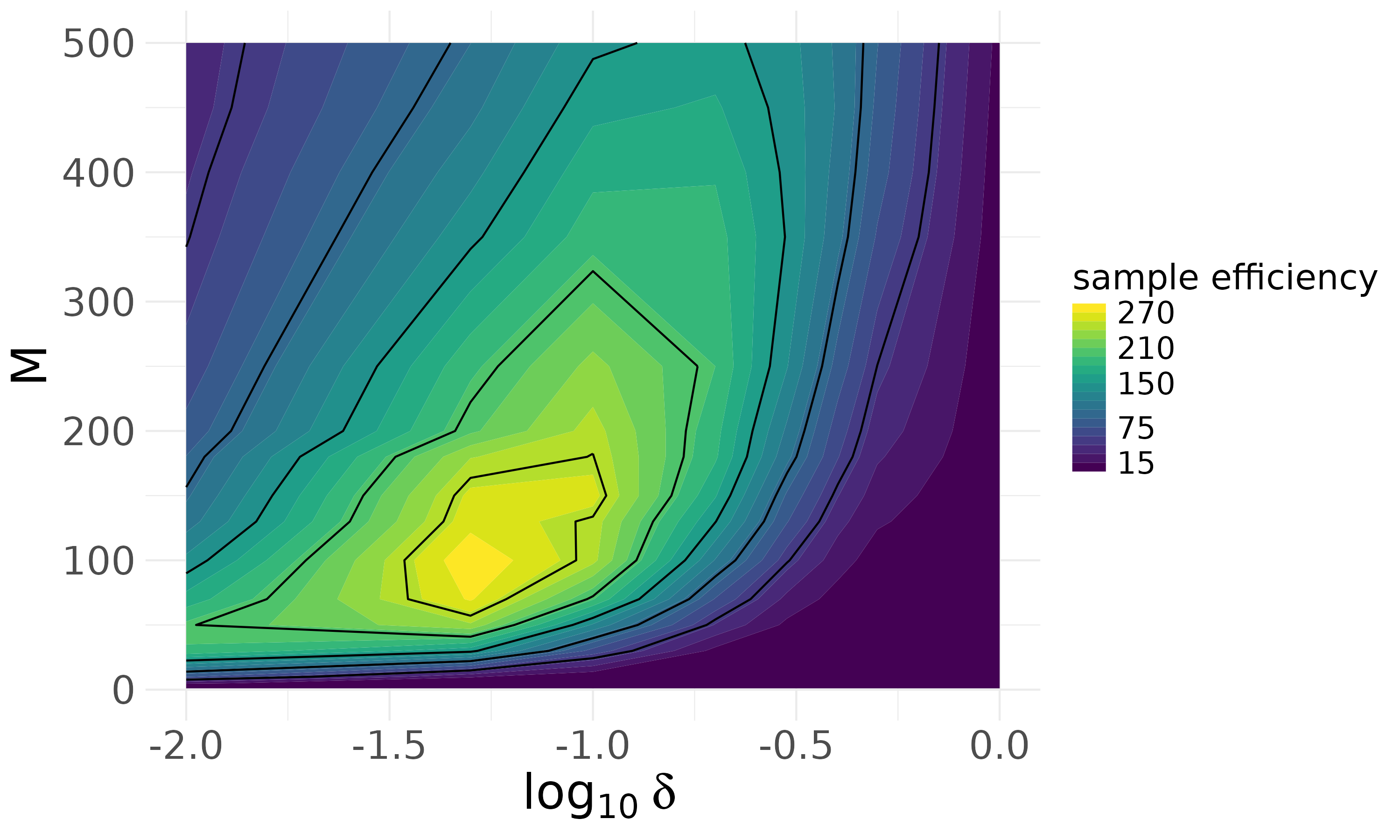}
  \label{fig:SIR_high4_I}
\end{subfigure}
\hfill
\begin{subfigure}{0.49\linewidth}
  \centering
        \captionsetup{
    justification=centering,
    singlelinecheck=false,
    margin={0pt,12mm}
  }
  \caption{Infectious state, xPGBS}
  \includegraphics[width=\linewidth,height=0.31\textheight,keepaspectratio]{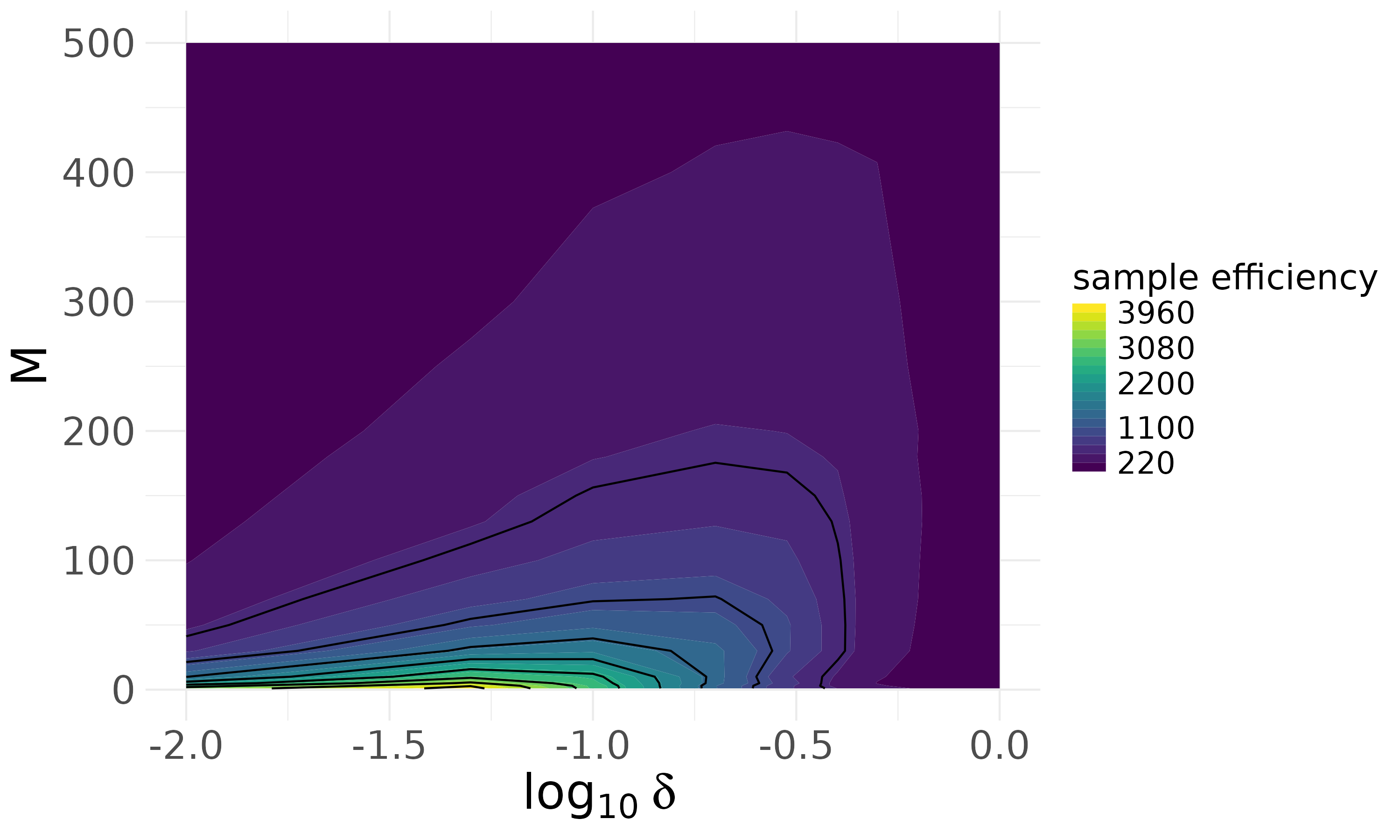}
  \label{fig:SIR_BS_high4_I}
\end{subfigure}

\caption{
Contour plots of sample efficiency in terms of the number of susceptible and infectious individuals after running xPG and xPGBS for $10^6$ iterations in 4D. The latent states are four coupled SIR models with shared parameters $\beta=0.6$ and $\gamma=0.2$.
}
\label{fig:SIR_xPG_xPGBS}
\end{figure}

We run PG, xPG, PGBS (equivalent to xPGBS with $\delta=1$) and xPGBS for $M\in\{1,3,5,10,  30, 50, 70, 100, 130, 150, 180, (4:10)\times 50 \}$, and $\delta \in \{0.01, 0.05, (1:5)\times 0.1, 1\}$. There are three species in the SIR model, so we focus on the ESS values for the susceptible and infectious individuals. Figure~\ref{fig:SIR_xPG_xPGBS} shows the contour plots of the sample efficiency in terms of susceptible and infectious populations of xPG and xPGBS after running $10^6$ iterations, respectively. As shown in both figures, xPG and xPGBS achieve substantially lower sample efficiency for the susceptible state than for the infectious state. It is because only the numbers of infectious people are observed, so the posterior distribution of the infectious state is more concentrated than that of the susceptible state, making it easier for xPG to explore the infectious state space efficiently. As a consequence, the optimal values of the tuning parameters, $(\hat{\delta}, \hat{M})$, are defined as those that give the maximum sample efficiency for the susceptible state. 

\begin{table}[h]
\centering
\caption{{Efficiency of different particle Gibbs algorithms run on simulated data from four coupled SIR models. All runs were for $10^6$ iterations with thinning interval of $10$. The algorithm attains its maximum sample efficiency at $M=\hat{M}$ and $\delta=\hat{\delta}$.}}
\renewcommand{\arraystretch}{1.0} % increase row height
\begin{tabular}{lcccc}
\hline
Alg & PG & xPG  & PGBS  & xPGBS  \\
\hline
$\hat{M}$ & $500^{*}$ & $130$ & $500^{\dagger}$ & $30$ \\
$\hat{\delta}$ & $1$ & $0.1$ & $1$ & $0.05$ \\
$\min\limits_{t,d} \text{ESS}(S_t^d)/\hat{M}$ & $9.06$ & $77.25$ & $22.37$ & $149.14$ \\
$\min\limits_{t,d} \text{ESS}(I_t^d)/\hat{M}$ & $ 10.08$ & $247.61$ & $52.96$ & $1481.00$ \\
$\text{ESS}(\beta)/\hat{M}$ & $14.93$ & $66.74$ & $23.76$ & $142.14$ \\
$\text{ESS}(\gamma)/\hat{M}$ & $163.97 $ & $679.74$ & $185.99$ & $2552.51$ \\
\hline
\end{tabular}
\label{tab:alg_comparison}
\vspace{1mm}
\begin{minipage}{0.9\linewidth}

\footnotesize
$^{*} (^{\dagger})$ $\hat{M}$ may be $>500$ because PG and xPG (PGBS and xPGBS) were only run up to $M=500$ as it took $6$ days $4$ hours (7 days 2 hours) to run PG (PGBS) with this $M$.
%
%$^{\dagger}$ $\hat{M}$ may be $>400$ because PGBS and xPGBS were only run up to $M=400$ as it took $8$ days 14 hours to run xPGBS with $M=400$.
\end{minipage}
\end{table}
Table~\ref{tab:alg_comparison} presents the optimal parameter values of the four algorithms and the corresponding sample efficiency for the susceptible state, infectious state, and model parameters $\beta$ and $\gamma$. xPGBS achieves the highest sample efficiency, whereas PG performs the worst. The xPG algorithm substantially improves over the efficiency of PG---the sample efficiency for the worst mixing component of the state for PG being over eight times higher for xPG---and xPGBS similarly provides a marked improvement over PGBS---the sample efficiency for the worst mixing component for PGBS being approximately seven times higher for xPGBS. Relative to PG, xPG improves sample efficiency more than PGBS.

\subsection{SEIR}\label{subsec: SEIR}
In this section, we consider four coupled SEIR models representing four different regions. We set $\tau=0.1$ and, as in Section~\ref{subsec: SIR}, the data are distributed according to \eqref{eq.BinObs} at times $t=1,\ldots,20$. The connectivity matrix is given in Appendix \ref{appendix:reactionNetworks}.

 The underlying true values of the model parameters are $\beta=0.6$, $\eta=0.6$ and $\gamma=0.2$. Their independent prior distributions are $
\beta, \eta \sim \text{Gamma}(1,1)$ and $\gamma \sim \text{Gamma}(20, 100)$,
and the initial states at time zero are assumed to be fixed and known.

\begin{figure}[h]
\captionsetup{font=footnotesize}
\captionsetup[subfigure]{labelformat=empty,font=footnotesize}
\centering

\begin{subfigure}{0.49\linewidth}
  \centering
        \captionsetup{
    justification=centering,
    singlelinecheck=false,
    margin={0pt,12mm}
  }
  \caption{ Minimum across all states, xPG}
  \includegraphics[width=\linewidth,height=0.31\textheight,keepaspectratio]{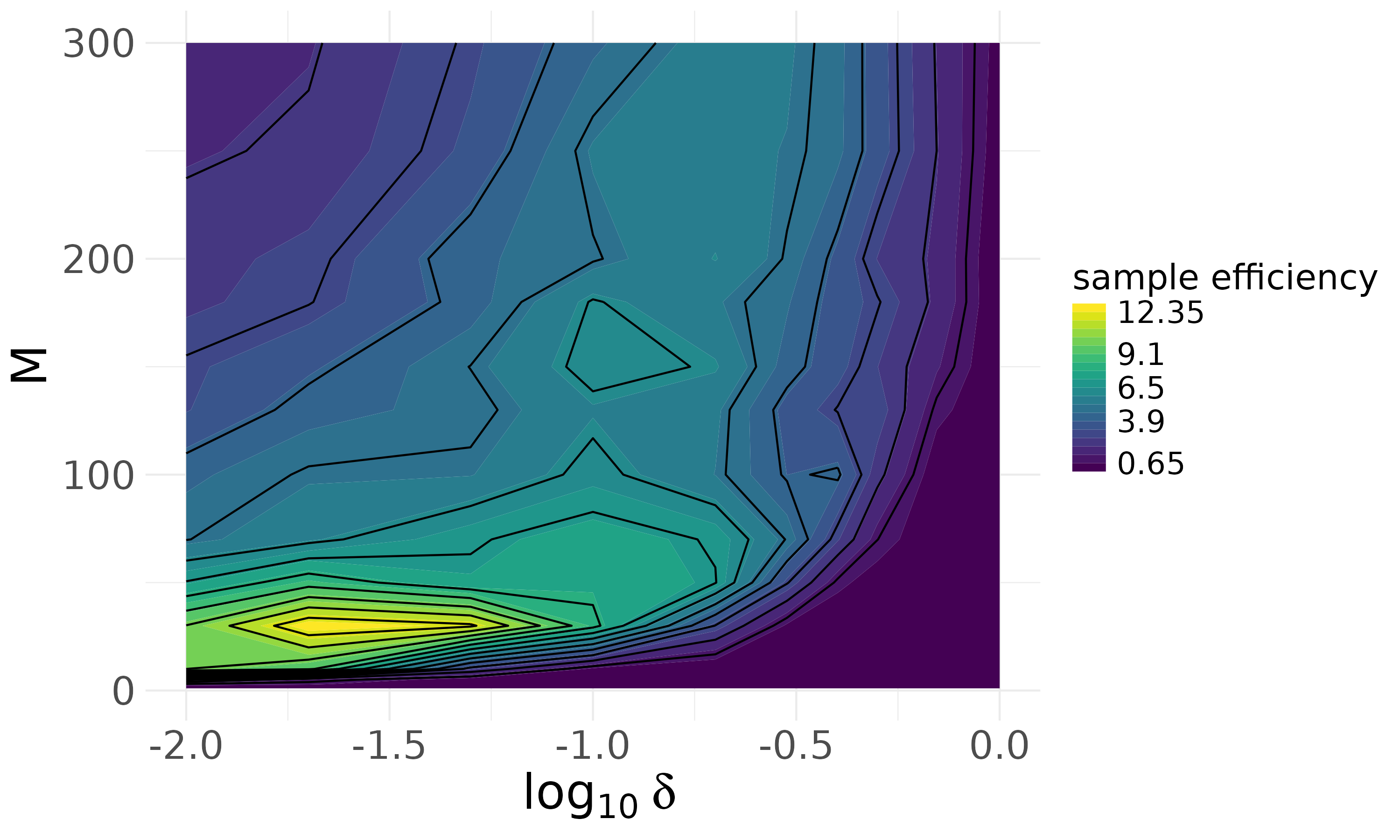}
  \label{fig:SEIR_high4_min}
\end{subfigure}
\hfill
\begin{subfigure}{0.49\linewidth}
  \centering
        \captionsetup{
    justification=centering,
    singlelinecheck=false,
    margin={0pt,12mm}
  }
  \caption{ Minimum across all states, xPGBS}
  \includegraphics[width=\linewidth,height=0.31\textheight,keepaspectratio]{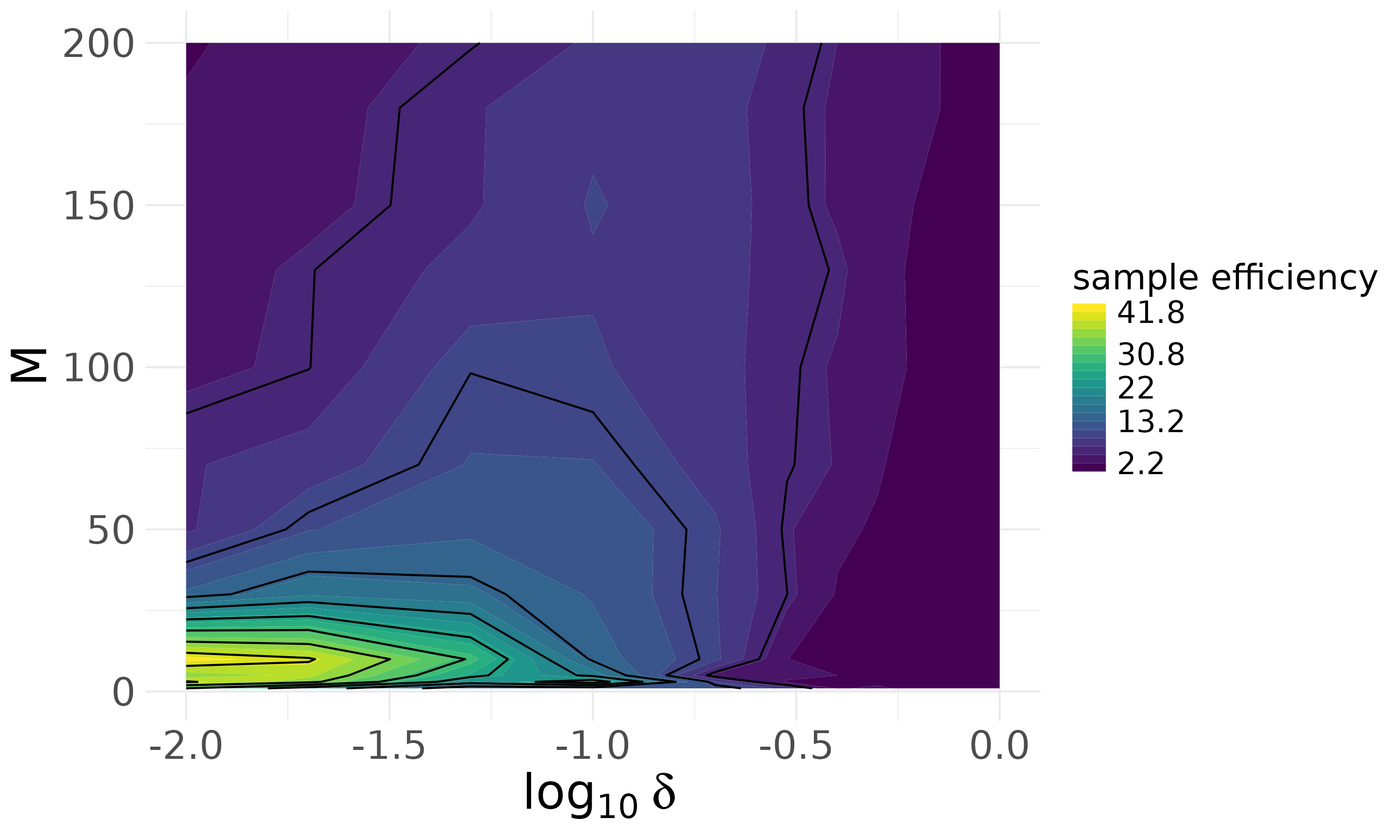}
  \label{fig:SEIR_BS_high4_min}
\end{subfigure}
\caption{
Contour plots of sample efficiency after running xPG and xPGBS for $10^6$ iterations in 4D; latent states are four coupled SEIR models with shared parameters $\beta=0.6$, $\eta=0.6$, and $\gamma=0.2$.
}
\label{fig:SEIR_xPG_xPGBS}
\end{figure}

Figures~\ref{fig:SEIR_high4_min} and \ref{fig:SEIR_BS_high4_min} present the results of applying xPG and xPGBS to the simulated data, respectively. Both plots show the sample efficiency for the state with the lowest ESS across all states. Corresponding plots for the susceptible, exposed and infectious states are provided in Appendix~\ref{sec:additional_results}.

% In Figure~\ref{fig: SEIR_BS_four}, we exclude the region where $M\leq 5$, since the corresponding xPG chains exhibit poor mixing and do not provide sufficiently reliable approximations to the posterior distribution within the simulation length considered.

 % \CS{LANYA: presumably you mean mixing efficiency?}
The mixing efficiency of the susceptible state is much lower than that obtained in Section~\ref{subsec: SIR}. In the SIR model, the number of individuals in the susceptible state is relatively tightly constrained given observations of the number of infected individuals. In contrast, in the SEIR model, the inclusion of the exposed state introduces additional flexibility, so there are many more possible combinations of the numbers of susceptible and exposed individuals that are consistent with the observed infectious counts. As a result, the high-probability region of the posterior distribution of susceptible states is wider than that in the SIR model. At a given observation time point, the tight conditional distribution of the observed number infectious \eqref{eq.BinObs} requires a small value of $\delta$ to explore the high-probability region of the posterior distribution of the infectious state effectively. However, a small value of $\delta$ also leads to small movement when exploring the posterior distributions of the susceptible and exposed states, which results in poor mixing of their Markov chains. 

As in the previous section, we define the pair of tuning parameters that give the highest minimum sample efficiency across all states as the optimal parameters, $(\hat{\delta}, \hat{M})$. Table~\ref{tab:alg_SEIR_comparison} displays the optimal values of the tuning parameters together with six sample-efficiency metrics. Across all six metrics, the algorithms exhibit the same ordering of sample efficiency: PG $<$ PGBS $<$ xPG $<$ xPGBS, which matches the results obtained for the SIR model (Section~\ref{subsec: SIR}). In particular, PGBS is only slightly better for the lowest sample efficiency of the model parameters for PG ($\eta$), xPG is $21$ times better, and xPGBS is $76$ times better. For the lowest sample efficiency of the states for PG (exposed state), PGBS again offers only a modest improvement over PG, whereas both xPG and xPGBS achieve improvements of roughly one order of magnitude.

% Can you also give some idea of the improvement in sample eff for xPG, PGBS and xPGBS over PG? E.g. PGBS is 1.5x better for the lowest sample eff for PG (eta), xPG 2.5x better, and xPGBS over 15x better, and for the states PGBS is only slightly better, while xPG and xPGBS are both an order of magnitude better.

In the results for the SIR model (Section~\ref{subsec: SIR}) and the SEIR model, the optimal values of $M$ for xPGBS are substantially smaller than those for xPG. As explained above, both xPG and xPGBS attain their optima at relatively small values of $\delta$. With a small value of $\delta$, the set of possible states generated by xPG or xPGBS is small, which means it is not necessary to produce many particles to effectively cover all of these states when simulating forward using the conditional particle filter. Moreover, compared to xPG, xPGBS makes better use of each particle because of the backward simulation step. Consequently, xPGBS with a relatively small value of $M$ is able to explore the latent state space efficiently.

For the SIR and SEIR models, the minimum ESS over all latent states and observation time points does not necessarily occur at the first observation time. As a result, the tuning strategy may overestimate the number of particles required. Suppose the minimum ESS occurs at time $t_{\hat{\ell}}$, with $\hat\ell>1$. To maximise the sample efficiency, we would ideally tune the acceptance rate at this time point, $\alpha(t_{\hat\ell})$, to match the optimal acceptance rate $\hat{\alpha}$ derived in Section \ref{subsec: optimal acceptance rate}. However, since $t_{\hat\ell}$ cannot be identified in advance, we instead tune the acceptance rate at the first observation time, $t_1$, to be $\hat{\alpha}$. Because the acceptance rates are non-decreasing over time, we have $\alpha(t_\ell)\ge \alpha(t_1)\approx \hat{\alpha}$. Therefore, for a fixed value of $\delta$, tuning $\alpha(t_1)$ to $\hat{\alpha}$ may lead to an overestimation of the required number of particles $M$. This behaviour is illustrated in Appendix~\ref{appendix: optimal acceptance rate}. 

\begin{table}[ht]
\centering
\caption{Efficiency of different particle Gibbs algorithms run on simulated data from four coupled SEIR models. All runs were for $10^6$ iterations with thinning interval of $10$.}
\renewcommand{\arraystretch}{1.0} % increase row height
\begin{tabular}{lcccc}
\hline
Alg & PG & xPG & PGBS & xPGBS \\
\hline
$\hat{M}$ & $350^{*}$ & $30$ & $350^{\dagger}$ & $10$ \\
$\hat{\delta}$ & $1$ & $0.02$ & $1$ & $0.01$ \\
$\min_{t, d} \text{ESS}(S_t^d)/\hat{M}$ & $0.90$ & $17.77$ &  $1.02$ & $42.86$ \\
$\min_{t, d} \text{ESS}(E_t^d)/\hat{M}$ & $0.62$ & $12.78$ &  $1.02$& $45.88$ \\
$\min_{t, d} \text{ESS}(I_t^d)/\hat{M}$ & $1.27$ & $216.35$ & $1.57$ & $1307.32$ \\
$\text{ESS}(\beta)/\hat{M}$ & $0.57$ & $10.56$ & $0.64$ & $21.12$ \\
$\text{ESS}(\eta)/\hat{M}$ & $0.23$ & $4.81$ & $0.37$ & $17.59$ \\
$\text{ESS}(\gamma)/\hat{M}$ & $18.10$ & $305.72$ & $21.77$ & $641.75$ \\
\hline
\end{tabular}
\label{tab:alg_SEIR_comparison}
\vspace{1mm}
\begin{minipage}{0.9\linewidth}

\footnotesize
$^{*}(\dagger)$ $\hat{M}$ may be $>350$ because PGBS and xPGBS were only run up to $M=350$ as it took around $7$ days to run PG (or PGBS) with $M=350$.
\end{minipage}
\end{table}

%Importantly, this overestimation is less problematic than underestimation, since c
Choosing an $M$ that is too large is generally less problematic than choosing one that is too small. If the tuning rule errs by overestimating $M$, the main consequence is an increase in computational cost, with a possible modest improvement in mixing. By contrast, underestimating $M$ may reduce computational cost, but can have a substantially detrimental effect on the mixing of the Markov chain. Indeed, Figures~\ref{fig: SIS_four}, \ref{fig:SIR_high4_S}, \ref{fig:SIR_high4_I} and \ref{fig:SEIR_high4_min} all show that, for any fixed value of $\delta\in (0, 1]$, the contours are much denser in the region where $M$ is smaller than its optimal value than in the region where $M$ is larger than its optimal value. Similar findings for another particle MCMC method, the pseudo-marginal Metropolis--Hastings (PMMH) algorithm, are reported in \cite{AOS2013} and \cite{biomet2015}, where the mixing efficiency is shown to decrease rapidly as $\sigma$ increases, with $\sigma \propto 1/M$.

\subsection{Autoregulatory model}
\begin{figure*}[t]
    \centering
    \includegraphics[width=0.7\linewidth]{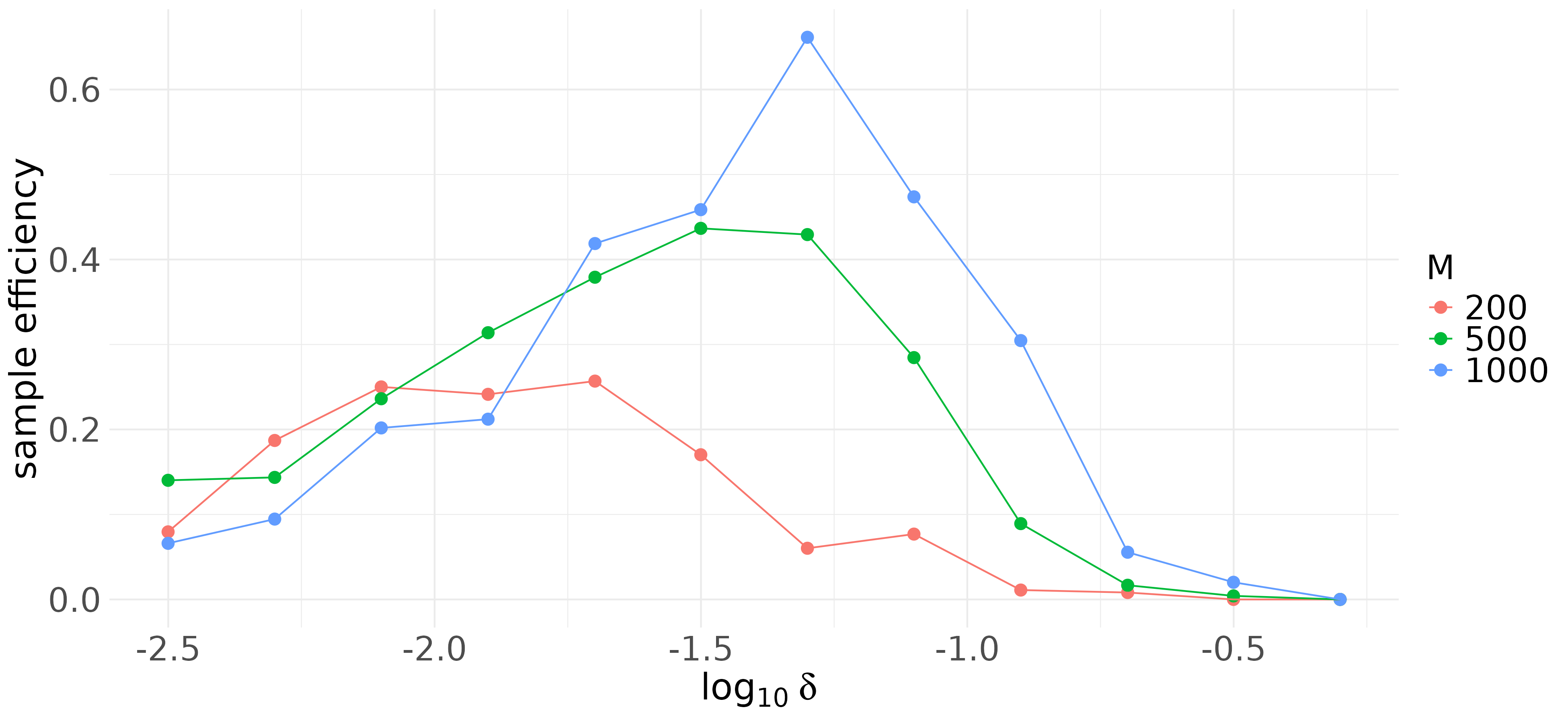}
    \caption{Sample efficiency of xPG run for $10^5$ iterations against $\log_{10}\delta$ for the autoregulatory model. Each curve corresponds to a different value of $M$; the observations are the exact molecule counts of $\mathsf{RNA}$, $\mathsf{P}$ and $\mathsf{P_2}$ simulated using model parameters $\theta_{1:8}
    =(0.2, 1.4, 0.7, 0.4, 0.2, 1.8, 0.6, 0.2)$, and the initial molecule counts, $X_{1:5}=(5, 8, 8, 8, 2)$.  }
    \label{fig: Autoreg_efficiency}
\end{figure*}

In this section, xPG is applied to the autoregulatory model with five species and eight types of reaction introduced in Section~\ref{subsec: autoregulatory model intro}. We simulate the data using $(\theta_1,\dots,\theta_8)=(0.2, 1.4, 0.7, 0.4, 0.2, 1.8, 0.6, 0.2)$. %the following parameter values
%\begin{align*}
%    (\theta_1, \theta_2, \theta_3, \theta_4, \theta_5, \theta_6, \theta_7, \theta_8)
%    =(0.2, 1.4, 0.7, 0.4, 0.2, 1.8, 0.6, 0.2)
%\end{align*}
 The initial values of the first three states are fixed at $(X_1, X_2, X_3)=(5, 8, 8)$, with $G=10$. The prior distribution of $X_4$ (or $X_5$ because $X_5=G-X_4$) is assumed to be uniform over $\{0,1, \dots, G\}$.
The marginal priors are independent with $\theta_1, \theta_5 \sim\text{Gamma}(100, 500)$
%\[
%\theta_1, \theta_5 \sim\text{Gamma}(100, 500),
%\]
for identifiability, and $\theta_i\sim\text{Gamma}(1, 1)$ for $i\in\{2, 3, 4, 6, 7, 8\}$. 
%\[
%\theta_i\sim\text{Gamma}(1, 1),\qquad i\in\{2, 3, 4, 6, 7, 8\}.
%\]
Given the exact observations of $X_1$, $X_2$ and $X_3$ at time $t=0, 1, \dots, 20$, we run xPG with $M\in \{200, 500, 1000\}$ and $\delta\in \{10^{-0.2k}, k=0, 1, \dots, 12\}$. Figure~\ref{fig: Autoreg_efficiency} shows the sample efficiency for $X_4$ (or $X_5$) against $\log_{10}\delta$. This indicates that the optimal value of $\delta$ decreases as $M$ decreases, as for xPG applied to epidemic models.

% Add a comment here about how it doesn't work for PG. Do you have values for the sample efficiency for delta=1 you can quote, or are they essentially 0? If they're 0, you can just give values for the smallest delta shown on the plot and comment on how getting a non-negligible sample eff for delta=1 would require a huge M and therefore take prohibitively long to run.

At each observation time, only proposals which hit the three-dimensional observation receive non-zero weight and can therefore survive resampling. The standard PG is unlikely to generate such latent states. Consequently, most particles receive zero weight, leading to severe particle degeneracy in PG. In contrast, xPG mitigates this issue by proposing paths that are highly correlated with the reference path. In the simulation study, when $M=1000$, xPG attains the maximum ESS for $X_4$ at time zero when $\delta=0.05$, whereas xPG with $\delta=0.5$ fails to move $X_4$ from its initial value, suggesting that obtaining a non-negligible sample efficiency for PG (i.e. xPG with $\delta=1$) would require a huge M and therefore take prohibitively long to run.

\section{Discussion}\label{sec: discussion}
% You can then write in the discussion about how this is the first time PG has been run on the autoreg...
We have introduced xPG and xPGBS methods for estimating the model parameters and latent states of hidden Markov models whose latent processes are Markov jump processes. Our xPG algorithm builds on the xPG method for diffusions introduced in \cite{malory2021bayesian} and \cite{Finke2023}, but focuses on Markov jump processes discretised by the popular tau-leap method. On a variety of reaction-networks, we demonstrate that it mitigates the particle degeneracy and mixing issues for which it is designed. The xPGBS algorithm adds an additional backward sampling step to further improve the sample efficiency of both the latent states and the model parameters. However, this approach is only applicable when the number of reactions of each type can be uniquely identified from two states at consecutive time points, as in the SIR and SEIR epidemic models. 

In addition, we have suggested a strategy for tuning the two parameters: $\delta$, which controls the correlation between the proposed paths and the reference path in the conditional particle filter (smaller $\delta\implies$ more correlated), and $M$, the number of alternative proposed paths. The strategy targets an optimal acceptance rate for the worst mixing state, which is not only valid for xPG, but can also be used to find the optimal number of particles $M$ for any particle Gibbs algorithm. The sample efficiency plots for the SIS models (Figure \ref{fig: SIS_four}) show that picking $\delta$ and $M$ following the tuning rule gives close to optimal sample efficiency.

We applied xPG to simulated data from models composed of multiple independent SIS sub-models; since SIS is one-dimensional, the state dimension is the number of models. In each dimension, inference was performed for each point in a grid of ($\delta,M$) values. The optimal $\delta$ decreased (optimal correlation increased) as the model dimension increased and, with more than three SIS models, xPG had far higher sample efficiency than PG. In seven dimensions, the good mixing of xPGibbs contrasted with a sample efficiency of $0$ for PG with $M=650$, showing that xPG can be applied in settings where achieving a reasonable ESS with particle Gibbs is too computationally expensive. 

We then compared the performance of particle Gibbs and xPG, with and without backwards sampling, on data simulated from four-region coupled SIR and SEIR models. These experiments further illustrate the efficiency gains achieved by our methods. For both the SIR and SEIR models, adding backward sampling to particle Gibbs improves sample efficiency, but much less than using xPG. xPGBS achieves the highest sample efficiency, suggesting that this should be the method of choice when backward sampling is possible. 

The benefit of xPG over PG is most pronounced when the observations are highly informative. In such settings, xPG can explore regions of high likelihood more effectively by constraining the proposed particles to remain close to the reference trajectory. In contrast, when the observations are less informative, standard particle Gibbs may already produce a reasonably large proportion of particles with high weights.

To the best of our knowledge, xPG is also the first algorithm to make exact joint inference of the latent states and rate parameters of the five-dimensional autoregulatory model of \cite{GolightlyWilkinson2005} from a sequence of exact state observations feasible. This setting is particularly challenging for many particle MCMC methods because the low observation noise can lead to severe particle degeneracy.  By contrast, alternative Bayesian inference approaches, such as the nearly minimal extended state-space algorithm (nMESA) \citep{Sherlock2023}, are designed primarily for inference on the rate parameters rather than the latent state trajectories. Running xPG with $M=1000$ for $10^5$ iterations takes approximately $14$ hours. At the optimal value $\delta=0.05$, it yields a minimum ESS of $1570$ across the ESS of all eight model parameters. By comparison, \cite{Sherlock2023} report that running nMESA for $2 \times 10^5$ iterations requires around 40 hours and achieves a minimum ESS of $1491$. These results suggest that, for the autoregulatory model, xPG achieves higher sample efficiency than nMESA. 

Our method is based on the tau-leap approximation of MJPs. Smaller $\tau$ gives a more accurate approximation but increases computational cost. Moreover, boundary conditions, such as all the states being non-negative must be enforced when applying xPG and xPGBS. An xPG method based on exact simulation of MJPs would remove the bias caused by discretisation and forced boundaries, and is a promising direction for future research. %In that case, boundary corrections would no longer be required, since each reaction event would be simulated exactly.

% \bibliographystyle{unsrtnat}
% \bibliography{references}

\bibliographystyle{Chicago}

\bibliography{references.bib}

\newpage
\onecolumn

%TC:ignore
\appendix
% \begin{center}
% \begin{Large}
%   \textbf{\textrm{Supplementary material to:} Scalable Bayesian inference for time-discretised Markov jump processes}\\

%   % by

% %  \emph{Chris Sherlock, Andrew Golightly and Anthony Lee}
% \end{Large}
% \end{center}

\section{Further details on reaction networks}\label{appendix:reactionNetworks}

The stoichiometry matrix for the SIS model is

$$
\Gamma =
\begin{bmatrix}
 -1 & 1\\
 1 & -1
\end{bmatrix}.
$$

The stoichiometry matrix for the SIR model is

$$
\Gamma =
\begin{bmatrix}
 -1 & 0\\
 1 & -1\\
 0 & 1
\end{bmatrix}.
$$
When considering four coupled SIR models, we use a connectivity matrix of
\[
C=
\begin{pmatrix}
1   & 0.1  & 0.1  & 0.1 \\
0.2 & 1    & 0.05 & 0.1 \\
0.1 & 0.15 & 1    & 0.1 \\
0.1 & 0.2  & 0.1  & 1
\end{pmatrix}.
\]

The stoichiometry matrix for the SEIR model is

$$
\Gamma =
\begin{bmatrix}
 -1 & 0 & 0\\
 1 & -1 & 0\\
 0 & 1 & -1\\
 0 & 0 & 1
\end{bmatrix}.
$$
When considering four coupled SEIR models, we use a connectivity matrix of 
\[
C=
\begin{pmatrix}
1   & 0.2  & 0.1  & 0.1 \\
0.2 & 1    & 0.15 & 0.2 \\
0.1 & 0.15 & 1    & 0.1 \\
0.1 & 0.2  & 0.1  & 1
\end{pmatrix}.
\]
The reactions in the autoregulatory model are given by

$$
\begin{aligned}
\mathsf{DNA} + \mathsf{P_2}
&\xrightleftharpoons[\theta_2 X_4]{\theta_1 (G - X_4) X_3}
\mathsf{DNA\cdot P_2}, \\[6pt]
\mathsf{DNA}
&\xrightarrow{\theta_3 (G - X_4)}
\mathsf{DNA} + \mathsf{RNA}, \\[6pt]
\mathsf{RNA}
&\xrightarrow{\theta_4 X_1}
\mathsf{RNA} + \mathsf{P}, \\[6pt]
2\mathsf{P}
&\xrightleftharpoons[\theta_6 X_3]{\theta_5 X_2 (X_2 - 1)/2}
\mathsf{P_2}, \\[6pt]
\mathsf{RNA}
&\xrightarrow{\theta_7 X_1}
\varnothing, \\[6pt]
\mathsf{P}
&\xrightarrow{\theta_8 X_2}
\varnothing .
\end{aligned}
$$

Here, $X_{1:5}$ denote the numbers of molecules of $\mathsf{RNA}$, $\mathsf{P}$, $\mathsf{P_2}$, $\mathsf{DNA\cdot P_2}$, and $\mathsf{DNA}$, respectively. The total number of DNA copies is conserved, so $X_4 + X_5 = G$.

\section{Conditional Particle filter}
\begin{algorithm}[H]
\caption{Conditional Particle Filter}
\label{alg: conditionalPF}
\begin{algorithmic}[1]
\Require the parameter, $\theta$; a reference path, $x_{(0:n)\tau}^{(0)}$; the observations $y_{t_{1:L}}$; the proposal distribution $q_0(\cdot|\theta)$ for $X_0$; the proposal distribution $q_{\tau}(\cdot|x_{t-1}, \theta)$ for $X_t$ given $X_{t-1}=x_{t-1}$
\State Sample $x^{(1:M)}_{0}\stackrel{\text{i.i.d.}}{\sim} q_0(x_0|\theta)$ \label{step: 1}
\State For $k=1$ to $K$, sample $x_{k\tau}^{(m)}\sim q_{\tau}(\cdot|x_{(k-1)\tau}^{(m)}, \theta)$, $m=1, \dots, M$ \label{step: 2}
\State Compute the weights of $x^{(0)}_{t_1}, \dots, x^{(M)}_{t_1}$ according to \eqref{eq: weightT} with $\tilde{x}_0^{(m)}=x_0^{(m)}$, $m=0, \dots,M$
\For{$\ell = 2$ to $L$}
    \State Set $\Tilde{x}_{t_{\ell-1}}^{(0)}=x_{t_{\ell-1}}^{(0)}$
    \State Draw $M$ times with replacement from $\{x_{t_{\ell-1}}^{(m)}\}_{m=0}^M$ with probability of obtaining $x_{t_{\ell-1}}^{(m)}$ being proportional to $w_{\ell-1}^m$ to obtain unweighted samples $\{\Tilde{x}_{t_{\ell-1}}^{(m)}\}_{m=1}^{M}$, store $a_{\ell-1}^{1:M}$.
    \For{$k=1$ to $K$}
    \State Sample $x_{t_{\ell-1}+k\tau}^{(m)}\sim q_{\tau}(\cdot|\Tilde{x}_{t_{\ell-1}+(k-1)\tau}^{(m)}, \theta)$, $m=1, \dots, M$ \label{step: 7}
    \State If $k<K$, set $\Tilde{x}_{t_{\ell-1}+k\tau}^{(m)}=x_{t_{\ell-1}+k\tau}^{(m)}$, $m=1, \dots, M$
    \EndFor
    \State Compute the weights of $x_{t_{\ell}}^{(0)},\dots, x_{t_{\ell}}^{(M)}$ by \eqref{eq: weightT}
where $q(\cdot|x_{t_{\ell-1}},\theta) =
\prod_{k=1}^K q_{\tau}(x_{t_{\ell-1}+k\tau}|x_{t_{\ell-1}+(k-1)\tau},\theta).$
\EndFor\\
\Return $\{(x_{(0:n)\tau}^{(m)}, a_{1:(L-1)}^{m}, w^m_{L})\}_{m=0}^M$
\end{algorithmic}
\end{algorithm}

\section{Particle degeneracy}\label{appendix:particleDegeneracy}
\begin{figure}[H]
    \centering
    \includegraphics[width=0.5\linewidth]{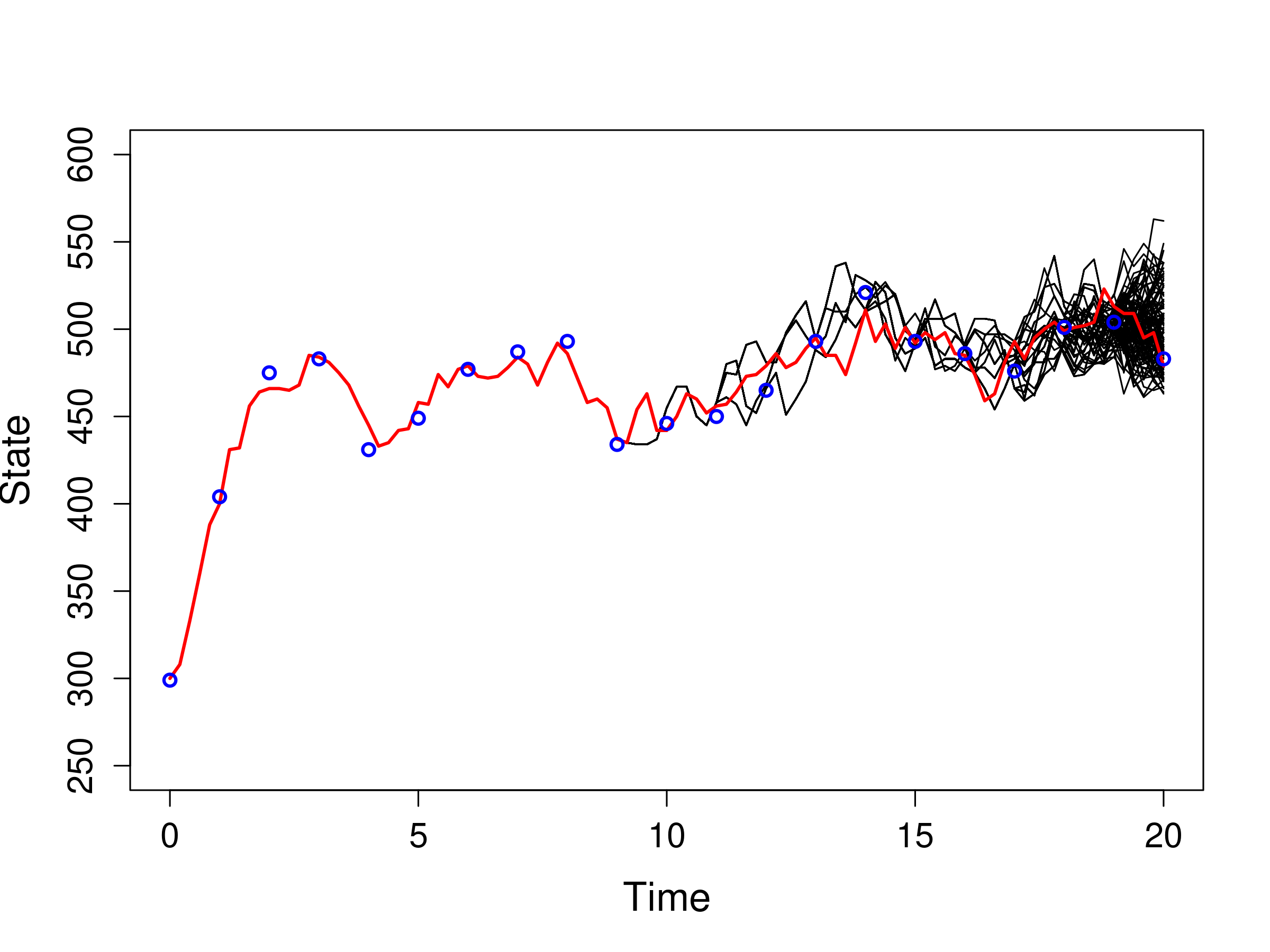}
    \caption{Trajectories traced back from the particles at the terminal time in a single run of the standard conditional particle filter with $M=100$. The red trajectory denotes the reference path, the black trajectories denote the proposed paths, and the blue dots indicate the observations.}
    \label{fig:particle_degeneracy}
\end{figure}

\section{Exchangeable particle Gibbs in continuous state space}\label{appendix: xpg for diffusions}
In practice, Algorithm \ref{alg: conMultiGeneratorXPG_Xt} illustrates how we generate $x_t^{(1:M)}$ given $x_t^{(0)}$ and $x_{t-\tau}^{(0:M)}$. The first step is to recover the standard $d$-dimensional Brownian motion $z^{(0)}$ that drives the reference path from $t-\tau$ to $t$. According to Equation \ref{eq: E-M}, we construct an invertible mapping $h:\mathbb{R}^d\times\mathbb{R}^d \to \mathbb{R}^d$ such that $X=h(Z, x_{t-\tau})$ has marginal distribution $q(\cdot\mid x_{t-\tau}, \theta)$. The mapping is defined as follows:
\begin{align*}
    h(Z,x)=x+\mu(x)\tau+\sqrt{\sigma(x)\tau}Z,
\end{align*}
therefore, 
\[
Z^{(0)}=h^{-1}(X_t;x_{t-\tau})\sim N(0, I_d)
\]
 We then sample $\hat{Z}^{(0)}, \dots, \hat{Z}^{(M)}$ independently from $N_d(0, \epsilon^2I_d)$, and set $Z^*=\sqrt{1-\epsilon^2}Z^{(0)}+\hat{Z}^{(0)}$ and $Z^{(m)}=\sqrt{1-\epsilon^2}Z^*+\hat{Z}^{(m)}$, $m = 1, \dots, M$. Finally we obtain $x^{(m)}_t$ by mapping back via $x^{(m)}_t=h(z^{(m)}, x_{t-\tau}^{(m)})$. Here we have $\text{Cov}(Z^{(i)}, Z^{(j)})=1-\epsilon^2$ for any $i, j\in\{0, \dots, M\}$ and $i\neq j$. We stress that all the proposed samples are independent when $\epsilon$ is $1$, whereas they are identical when $\epsilon$ is $0$. 
 
 In addition, given $x_0^{(0)}$ and the distribution $q_0( x_0|\theta)$, we can also apply Algorithm \ref{alg: conMultiGeneratorXPG_Xt} to generate $M$ samples from the joint distribution $\tilde{q}_0(x^{(1:M)}_0 \mid x^{(0)}_0, \theta)$ by constructing a one-to-one mapping $h:\mathbb{R}^d \to \mathbb{R}^d$ such that, for $Z \sim N_d(0, I_d)$, the random variable $X = h(Z)$ has marginal distribution $q_0(\cdot \mid \theta)$. Next, we obtain $z^{(0)}=h^{-1}(x_0^{(0)})$ in Step~\ref{step: find_z0} of Algorithm~\ref{alg: conMultiGeneratorXPG_Xt}. The remaining steps are applied directly to obtain correlated and exchangeable samples $x^{(0)}_0, \dots, x^{(M)}_0$.

 \begin{algorithm*}
    \caption{Exchangeable Proposal $\Tilde{q}(\cdot|x^{(0)}_t, x^{(0:M)}_{t-\tau}, \theta)$ with marginal $q$ for general d}\label{alg: conMultiGeneratorXPG_Xt}
    \begin{algorithmic}[1]
    \Require An invertible mapping $h: \mathbb{R}^d\times\mathbb{R}^d\to \mathbb{R}^d$ such that $h(Z_{1:d};x_{t-\tau})\sim q(\cdot|x_{t-\tau}, \theta)$ where $Z_{1:d}\sim N(0, I_d)$; the correlation parameter $\epsilon$; the current state $x^{(0)}_t$
    \State Calculate $z^{(0)}=h^{-1}(x^{(0)}_t; x_{t-\tau}^{(m)})$ \label{step: find_z0}
    \State Sample $\hat{z}^{(0)}$ from a $N(0, \epsilon^2I_d)$ distribution
    \State $z^*=\sqrt{1-\epsilon^2}z^{(0)}+\hat{z}^{(0)}$
    \State \textbf{for} $m=1,\dots, M$ \textbf{do}\\
      \quad Sample $\hat{z}^{(m)}$ from a $N(0, \epsilon^2I_d)$ distribution\\
      \quad Set $z^{(m)}=z^*\sqrt{1-\epsilon^2}+\hat{z}^{(m)}$\\
      \quad Set $x^{(m)}_t=h(z^{(m)}, x_{t-\tau}^{(m)})$\\
      \textbf{end for}
    \end{algorithmic}
  \end{algorithm*}
 
 Therefore, in Algorithm \ref{alg: conMultiGeneratorXPG_Xt}, we can prove that, if $Z^{(0)}\sim N(0, I_d)$, then $\mathbf{Z}=(Z^{(0)}, \dots, Z^{(M)})^\top\sim N(\mu_{M+1}, \Sigma_{M+1})$ and $\rho=1-\epsilon^2$. Hence, the joint distribution of $z^{(0)}, \dots, z^{(M)}$ is exchangeable under any permutation of $(z^{(0)}, \dots, z^{(M)})$. Moreover, since $x^{(m)}_0 = h(z^{(m)}_0)$ for $m = 0, \dots, M$, it follows directly that the joint distribution of $x^{(0)}_0, \dots, x^{(M)}_0$ is also exchangeable.

The conditional particle filter within xPG for continuous state spaces follows the standard conditional particle filter, with modifications in steps \ref{step: 1}, \ref{step: 2} and \ref{step: 7}. In step \ref{step: 1}, rather than sampling $x_0^{(1:M)}$ independently, we generate them jointly from the distribution in Equation \eqref{eq: q_tilde_0} using Algorithm \ref{alg: conMultiGeneratorXPG_Xt}. Likewise, in step \ref{step: 2} and \ref{step: 7}, the states $x_{k\tau}^{(1:M)}$ are drawn jointly from the distribution in \eqref{eq: q_tilde} via Algorithm \ref{alg: conMultiGeneratorXPG_Xt}.

Building on particle Gibbs, the exchangeable particle Gibbs method replaces the standard conditional particle filter with the \textit{exchangeable conditional particle filter}. The exchangeability ensures that the detailed balance still holds for xPG, meaning this new method remains valid; see \cite{malory2021bayesian}. The proof of detailed balance for the full algorithm is similar to that presented in Section \ref{sec: proof_thm1}, so it will not be repeated here. 

\section{Exchangeable particle Gibbs}\label{appendix:exPG}
\begin{algorithm}[H]
    \caption{Update proposals in each time step}\label{alg: generateProposedN}
    \begin{algorithmic}[1]
\Require $N_k^{(0)}, \mu_k^{(0)}, \dots, \mu_k^{(M)}$
\State $\Tilde{N}_k^{(0)}\sim \text{Bin}(N_k^{(0)}, 1-\delta)$.
\State Sort $0:M$ according to the value of $\mu_k^{(i)}$, $i=0,\dots,M$ in ascending order to obtain the sorted index set $\mathcal{S}=\{s_i\}_{i=-\mathrm{lo}}^{\mathrm{hi}}$ where $s_0=0$.

\For{$i=0, 1, \dots, \mathrm{hi}-1$}
\State generate $\Tilde{N}_k^{(s_{i+1})}$ based on $\Tilde{N}_k^{(s_{i})}$ using \textbf{Eq} \ref{eq: PoisUpdate}
\EndFor

\For{$i=0, -1, \dots, -\mathrm{lo}+1$}
\State generate $\Tilde{N}_k^{(s_{i-1})}$ based on $\Tilde{N}_k^{(s_{i})}$ using \textbf{Eq} \ref{eq: BinThinning}
\EndFor
\For{$j = 1, \dots, M$}
\State Generate $N_k^{(j)}$ using \textbf{Eq} \ref{eq: finalPoisUpdate}.
\EndFor
\end{algorithmic}
\end{algorithm}
The correlation between $N_k^{(s_{i})}$ and $N_k^{(s_{j})}$, $i<j$, is given by
$$\text{Corr}\left(N_k^{(s_{i})}, N_k^{(s_{j})}\right)=\frac{(1-\delta)\mu^{(s_i)}}{\sqrt{\mu^{(s_i)}_k\mu^{(s_{j})}_k}},$$
where $i, j=-\mathrm{lo}, \dots, \mathrm{hi}-1$.

\section{Justification for xPG}
\subsection{Proof of Theorem \ref{thm1}}\label{sec: proof_thm1}
In this section, we prove the exchangeable particle Gibbs for MJPs satisfies detailed balance with respect to the target posterior distribution of the latent states, $p(x_{(0:n)\tau}|y_{1:L}, \theta)$. For simplicity, we assume observations occur at integer times, $t_\ell=\ell$, $\ell=1, \dots, L$.

We begin by introducing Lemma \ref{lemma1}, which plays a central role in proving Theorem \ref{thm1}. In a conditional particle filter, at each observation time $t_\ell$, we propagate the $M$ unweighted particles $\tilde{x}_{t_\ell}^{(1:M)}$ forward to the next observation time $t_{\ell+1}$. We refer to each such propagated portion of the path as a \textit{segment}.

The lemma states that, within the propagation of a single time step of length $\tau$, the probability of first simulating the number of events for each reaction type for the reference path (i.e., $n^{(0), \mathcal{R}_{1:v}}$) and subsequently proposing $M$ additional event counts for each reaction type, conditional on all information available at the start of the time step, is equal to the probability of first simulating the number of events for each reaction type for the $j$-th segment (i.e., $n^{(j), \mathcal{R}_{1:v}}$), $j=1, \dots, M$, and then proposing the remaining event counts, again conditional on all information available at the start of the time step.

Recall that we defined that $x_{t_\ell}^{(a_\ell^j)}=\tilde{x}_{t_\ell}^{(j)}$. Thus, $x_{t_\ell}^{(a_\ell^j)}$ serves as the initial state at time $t_\ell$ for the $j$-th proposed segment within the interval $[\ell K\tau, (\ell+1)K\tau]$. Let $\mu_{\ell K +k}$ denote the expected number of reactions in the $k$th tau-leap segment after the $\ell$th observation time, with the convention that the $0$th observation time is the initial time, and that $n_{\ell K +k}$ is the corresponding realisation of the number of reactions in that interval.

\begin{lemma}
    For any $\ell\in \{0, 1, \dots, L-1\}$, $k\in\{1, \dots, K\}$ and $j \in \{1, \dots, M\}$, we have
\begin{equation}
    \begin{aligned}
        &p\!\left(
  n_{\ell K+k}^{(0), \mathcal{R}_{1:v}}
  \mid
  x_{t_\ell}^{(0)},
  n^{(0), \mathcal{R}_{1:v}}_{\ell K+1:\ell K+k-1}
\right)  \times
 \tilde{q}\!\left(
  n_{\ell K+k}^{(1:M), \mathcal{R}_{1:v}}
  \mid
  x^{(a_\ell^{0:M})}_{t_\ell},
  n_{\ell K+1:\ell K+k-1}^{(0:M), \mathcal{R}_{1:v}},
  n_{\ell K+k}^{(0), \mathcal{R}_{1:v}}
\right) \\
&=
p\!\left(
  n_{\ell K+k}^{(j), \mathcal{R}_{1:v}}
  \mid
  x_{t_\ell}^{(a_\ell^j)},
  n^{(j), \mathcal{R}_{1:v}}_{\ell K+1:\ell K+k-1}
\right) \times
 \tilde{q}\!\left(
  n_{\ell K+k}^{(-j), \mathcal{R}_{1:v}}
  \mid
  x^{(a_\ell^{0:M})}_{t_\ell},
  n_{\ell K+1:\ell K+k-1}^{(0:M), \mathcal{R}_{1:v}},
  n_{\ell K+k}^{(j), \mathcal{R}_{1:v}}
\right).
    \end{aligned}
    \label{eq: subExch}
\end{equation}
\label{lemma1}
\end{lemma}

Appendix \ref{appendix:proofLemma1} provides the proof of Lemma~\ref{lemma1}.

The remainder of the proof of Theorem~\ref{thm1} follows similarly to the proof of the validity of the standard particle Gibbs, as presented in \cite{Chapter1ChrisPaul}. 

\begin{proof}[Proof of Theorem~\ref{thm1}]
The target posterior probability mass function of the full path $X_{(0:n)\tau}$ is given by 
\begin{equation}
\begin{aligned}
    &p(x_{(0:n)\tau}|y_{1:L}, \theta)\\
    \propto&\; p_0(x_0|\theta)\prod_{k=1}^n p\left(n_k^{\mathcal{R}_{1:v}}|x_0, n^{\mathcal{R}_{1:v}}_{1:(k-1)}\right)\prod_{\ell=1}^Lg(y_\ell|x_{t_\ell})\\
    =&\;p_0(x_0|\theta)\prod_{k=1}^n p\left(n_k^{\mathcal{R}_{1:v}}|\mu_k^{\mathcal{R}_{1:v}}\right)\prod_{\ell=1}^Lg(y_\ell|x_{t_\ell})
\end{aligned}
\label{eq: posterior}
\end{equation}
where $n_{1:0}=\emptyset$. The probability of proposing all of the other random variables in the conditional particle filter, conditional on the reference path, and selecting the path back from $x_T^{(i)}$ is given by
\begingroup
\small
\begin{equation}
    \begin{aligned}
    &\tilde{q}_0\left(x_0^{(1:M)}|x_0^{(0)}\right)
    \times\prod_{\ell=0}^{L-1} \prod_{k=1}^K  \tilde{q}\left(n_{\ell K+k}^{(1:M), \mathcal{R}_{1:v}}\mid  \mu_{\ell K+k}^{(0:M), \mathcal{R}_{1:v}}, n_{\ell K+k}^{(0), \mathcal{R}_{1:v}} \right) \\
    &\qquad\times \left\{\prod_{\ell=1}^{L-1}\prod_{m=1}^{M}\frac{g\left(y_\ell|x_{t_\ell}^{(a_\ell^m)}\right)}{\sum_{j=0}^{M}g\left(y_\ell|x_{t_\ell}^{(j)}\right)} \right\}\frac{g\left(y_L|x_{t_L}^{(i)}\right)}{\sum_{j=0}^{M}g\left(y_L|x_{t_L}^{(j)}\right)}
    \end{aligned}
    \label{eq: proposal}
\end{equation}
\endgroup
where $a_\ell^0 = 0$ for $\ell = 0, \dots, L$. 

Based on \eqref{eq: posterior} and \eqref{eq: proposal}, the product of the posterior density of the current reference path and the probability of generating all the random variables within a single run of the conditional particle filter, followed by selecting the new path back from $x_T^{(i)}$, can be written as the product of three terms. 

The first term corresponds to the prior density of the initial state and the
proposal used to generate the remaining initial particles,
\begin{equation}
    p_0(x_0^{(0)}|\theta)\tilde{q}_0\left(x_0^{(1:M)}|x_0^{(0)}\right).
    \label{eq: item1}
\end{equation}
The second term accounts for the generation of reaction counts within each
interval,
\begin{equation}
\begin{aligned}
    &\prod_{\ell=0}^{L-1} \prod_{k=1}^K  p\left(n_{\ell K+k}^{(0), \mathcal{R}_{1:v}}|\mu^{(0), \mathcal{R}_{1:v}}_{\ell K+k}\right)
    \times \tilde{q}\left(n_{\ell K+k}^{(1:M), \mathcal{R}_{1:v}}\mid \mu^{(0:M), \mathcal{R}_{1:v}}_{\ell K+k}, n_{\ell K+k}^{(0), \mathcal{R}_{1:v}} \right).
\end{aligned}
    \label{eq: item2}
\end{equation}
Finally, the third term corresponds to the contribution of the observation
likelihoods and the path resampling probabilities,
\begin{equation}
\begin{aligned}
     &\prod_{\ell=1}^Lg(y_\ell|x_{t_\ell}^{(0)})\Bigg[ \prod_{\ell=1}^{L-1}\prod_{m=1}^{M}\frac{g\left(y_\ell|x_{t_\ell}^{(a_\ell^m)}\right)}{\sum_{j=0}^{M}g\left(y_\ell|x_{t_\ell}^{(j)}\right)}\Bigg]
     \times \frac{g\left(y_L|x_{t_L}^{(i)}\right)}{\sum_{j=0}^{M}g\left(y_L|x_{t_L}^{(j)}\right)}.
\end{aligned}
    \label{eq: item3}
\end{equation}
From \eqref{eq: exPrior} we know that  \eqref{eq: item1} is invariant to $x_0^{(0)}\leftrightarrow x_0^{(b_0^i)}$. In addition, using Lemma \ref{lemma1}, \eqref{eq: item2} is equal to 
\begin{equation}
\begin{aligned}
    &\prod_{\ell=0}^{L-1} \prod_{k=1}^K  p\left(n_{\ell K+k}^{(b^i_{\ell+1}), \mathcal{R}_{1:v}}|\mu^{(b^i_{\ell+1}), \mathcal{R}_{1:v}}_{\ell K+k}\right)\times \tilde{q}\left(n_{\ell K+k}^{(-b^i_{\ell+1}), \mathcal{R}_{1:v}}\mid \mu^{(0:M), \mathcal{R}_{1:v}}_{\ell K+k}, n_{\ell K+k}^{(b^i_{\ell+1}), \mathcal{R}_{1:v}} \right).
\end{aligned}
    \label{eq: item2_ex}
\end{equation}
In applying Lemma \ref{lemma1}, note that $a_{\ell}^{\,b^i_{\ell+1}} = b_{\ell}^i$, so that $a_\ell^j = b_\ell^i$ whenever $j = b^i_{\ell+1}$, $\ell=1, \dots, L-1$. Since  $a_\ell^0=0$ for $l=1,\dots,L-1$, \eqref{eq: item3} can be rewritten as 
\begin{align*}
    \frac{g\left(y_L|x_{t_L}^{(0)}\right)g\left(y_L|x_{t_L}^{(i)}\right) \prod_{\ell=1}^{L-1}\prod_{m=0}^{M}g\left(y_\ell|x_{t_\ell}^{(a_\ell^m)}\right) }{\left(\prod_{\ell=1}^{L-1}\sum_{j=0}^{M}g\left(y_\ell|x_{t_\ell}^{(j)}\right)\right)^{M}\sum_{j=0}^{M}g\left(y_L|x_{t_L}^{(j)}\right)}
\end{align*}
which is invariant to $x_{t_L}^{(0)}\leftrightarrow x_{t_L}^{(i)}$. Consequently, for any $i \in \{1, \dots, M\}$, the product of the three terms is invariant to switching the reference path $\{x_{k\tau}^{(0)}\}_{k=0}^n$ and the path traced back from $x_T^{(i)}$, \emph{i.e.},  $x_{(0:n)\tau}^{(b_{0:L}^i)}$, 
with $b_L^i = i$. Hence, the exchangeable particle Gibbs algorithm satisfies detailed balance with respect to $p(x_{(0:n)\tau}|y_{0:L}, \theta)$. 
\end{proof}

\subsection{Proof of Lemma \ref{lemma1}}\label{appendix:proofLemma1}
 Firstly, for each $k\in\{1,\dots,K\}$, knowing $x_{t_l}^{(a_\ell^{0:M})}$ and $n_{\ell K +1:\ell K+k-1}^{(0:M), \mathcal{R}_{1:v}}$, we also know $\mu_{\ell K+k}^{(0:M),\mathcal{R}_{1:v}}$. Therefore, for any $j\in\{0, \dots, M\}$, the quantity $(\mu_{\ell K+k}^{(0:M),\mathcal{R}_{1:v}}, n_{\ell K+k}^{(j), R_{1:\nu}})$ are sufficient statistics for $n_{\ell K+k}^{(-j),\mathcal{R}_{1:v}}$, conditional on $(x_{t_l}^{(a_\ell^{0:M})},n_{\ell K +1:\ell K+k-1}^{(0:M), \mathcal{R}_{1:v}}, n_{\ell K+k}^{(j), R_{1:\nu}})$. Secondly, since the system is Markovian, without loss of generality, it suffices to prove the result for $\ell=0$. 
\[
p(n_k^{(0),\mathcal{R}_{1:v}}|\mu_k^{(0),\mathcal{R}_{1:v}})q(n_k^{(1:M),\mathcal{R}_{1:v}}|\mu_k^{(0:M),\mathcal{R}_{1:v}}, n_k^{(0),\mathcal{R}_{1:v}})
=
p(n_k^{(j),\mathcal{R}_{1:v}}|\mu_k^{(j),\mathcal{R}_{1:v}})q(n_k^{(-j),\mathcal{R}_{1:v}}|\mu_k^{(0:M),\mathcal{R}_{1:v}}, n_k^{(j),\mathcal{R}_{1:v}}).
\]
Conditional on $\mu_k^{(0:M)}$, the simulations for each different reaction type are independent of each other. So it suffices to prove the above for a single reaction type. Thus, thirdly, for notational simplicity, we omit the superscript for the reaction; we also omit the subscript, $k$.

For a given reaction type, let $\qthin(\ntil^{(0)}|n^{(0})$, $\qR(\ntil^{(s_{i+1})}|\ntil^{(s_i)})$, $\qL(\ntil^{(s_{i-1})}|\ntil^{(s_i)})$ and $\qfill(n^{(j)}|\ntil^{(j)})$ denote the conditional mass functions from \eqref{eq: firstBinthin}, \eqref{eq: PoisUpdate}, \eqref{eq: BinThinning} and \eqref{eq: finalPoisUpdate}, respectively. Without loss of generality, we suppose that $\mu^{(j)}\ge \mu^{(0)}$ and $j=s_r$ for some $r \ge 0$; otherwise, we relabel $0\leftrightarrow j$. We will show that
\begin{align*}
p(n^{(0)}|\mu^{(0)})&
\qthin(\ntil^{(0)}|n^{(0)})
\prod_{i=0}^{\hi-1}\qR(\ntil^{(s_{i+1})}|\ntil^{(s_i)})
\prod_{i=-\lo+1}^{0}\qL(\ntil^{(s_{i-1})}|\ntil^{(s_i)})
\prod_{i=1}^{M}\qfill(n^{(i)}|\ntil^{(i)})\\
&=
p(n^{(j)}|\mu^{(j)})
\qthin(\ntil^{(j)}|n^{(j)})
\prod_{i=r}^{\hi-1}\qR(\ntil^{(s_{i+1})}|\ntil^{(s_i)})
\prod_{i=-\lo+1}^{r}\qL(\ntil^{(s_{i-1})}|\ntil^{(s_i)})
\prod_{i=0,i\ne j}^{M}\qfill(n^{(i)}|\ntil^{(i)}).
\end{align*}
Integrating out $\ntil^{(0:M)}$ then provides the required result.

Both sides of the equation have the common terms $\prod_{i=r}^{\hi-1}\qR(\ntil^{(s_{i+1})}|\ntil^{(s_i)})$, $\prod_{i=-\lo+1}^{0}\qL(\ntil^{(s_{i-1})}|\ntil^{(s_i)})$ and $\prod_{i=1,i\ne j}^M\qfill(n^{(i)}|\ntil^{(i)})$. Also, $p(n|\mu)\equiv \Pois(n;\mu)$. Thus, we must show
\begin{align}
\Pois(n^{(0)};\mu^{(0)})&
\qthin(\ntil^{(0)}|n^{(0)})
\left\{\prod_{i=0}^{r-1}\qR(\ntil^{(s_{i+1})}|\ntil^{(s_i)})\right\}
~\qfill(n^{(j)}|\ntil^{(j)})\nonumber\\
&=
\Pois(n^{(j)};\mu^{(j)})
\qthin(\ntil^{(j)}|n^{(j)})
\left\{\prod_{i=1}^{r}\qL(\ntil^{(s_{i-1})}|\ntil^{(s_i)})\right\}
~\qfill(n^{(0)}|\ntil^{(0)}).
\label{eqn.bigprod}
\end{align}
Now, defining $\mutil^{(j)}:=(1-\delta)\mu^{(j)}$,
\[
\qR(\ntil^{(s_{i})}|\ntil^{(s_{i-1})})
=
\frac{\left(\mutil^{(s_{i})}-\mutil^{(s_{i-1})}\right)^{\ntil^{(s_i)}-\ntil^{(s_{i-1})}}}
     {\left(\ntil^{(s_i)}-\ntil^{(s_{i-1})}\right)!}
     \exp\left\{-\left(\mutil^{(s_{i})}-\mutil^{(s_{i-1})}\right)\right\}
     \]
     and
     \begin{align*}
     \qL(\ntil^{(s_{i-1})}|\ntil^{(s_{i})})
     &=
     \frac{\ntil^{(s_i)}!}{\ntil^{(s_{i-1})}!(\ntil^{(s_i)}-\ntil^{(s_{i-1})})!}\left(\frac{\mutil^{(s_{i-1})}}{\mutil^{(s_{i})}}\right)^{\ntil^{(s_{i-1})}}
     \left(\frac{\mutil^{(s_{i})}-\mutil^{(s_{i-1})}}{\mutil^{(s_i)}}\right)^{\ntil^{(s_i)}-\ntil^{(s_{i-1})}}\\
     &=
     \qR(\ntil^{(s_{i})}|\ntil^{(s_{i-1})})
     \frac{\ntil^{(s_i)}!}{\ntil^{(s_{i-1})}!}\frac{\left(\mutil^{(s_{i-1})}\right)^{\ntil^{(s_{i-1})}}}{\left(\mutil^{(s_{i})}\right)^{\ntil^{(s_{i})}}}
          \exp\left(\mutil^{(s_{i})}-\mutil^{(s_{i-1})}\right).
     \end{align*}
     So,
     \begin{equation}
      \begin{aligned}
\prod_{i=1}^r\qL(\ntil^{(s_{i-1})}|\ntil^{(s_{i})})
&=
\left\{\prod_{i=1}^r\qR(\ntil^{(s_{i})}|\ntil^{(s_{i-1})})\right\}
\times
     \frac{\ntil^{(s_r)}!}{\ntil^{(s_{0})}!}\frac{\left(\mutil^{(s_{0})}\right)^{\ntil^{(s_{0})}}}{\left(\mutil^{(s_{r})}\right)^{\ntil^{(s_{r})}}}
          \exp\left(\mutil^{(s_{r})}-\mutil^{(s_{0})}\right)\nonumber\\
          &=
          \left\{\prod_{i=1}^r\qR(\ntil^{(s_{i})}|\ntil^{(s_{i-1})})\right\}
          \times
     \frac{\ntil^{(j)}!}{\ntil^{(0)}!}\frac{\left(\mutil^{(0)}\right)^{\ntil^{(0)}}}{\left(\mutil^{(j)}\right)^{\ntil^{(j)}}}
     \exp\left(\mutil^{(j)}-\mutil^{(0)}\right)\nonumber\\
     &=
          \left\{\prod_{i=1}^r\qR(\ntil^{(s_{i})}|\ntil^{(s_{i-1})})\right\}
          \times
     \frac{\ntil^{(j)}!}{\ntil^{(0)}!}\frac{\left(\mu^{(0)}\right)^{\ntil^{(0)}}}{\left(\mu^{(j)}\right)^{\ntil^{(j)}}}\left(1-\delta\right)^{\ntil^{(0)}-\ntil^{(j)}}\nonumber\\
&\times     \exp\left\{(1-\delta)\mu^{(j)}-(1-\delta)\mu^{(0)}\right\}.\label{eqn.multiplier}
     \end{aligned}
      \end{equation}
Let
     \[
     b:=(\mu^{(0)})^{n^{(0)}}(\mu^{(j)})^{n^{(j)}}\delta^{n^{(0)}-\ntil^{(0)}+n^{(j)}-\ntil^{(j)}}\exp\{-\mu^{(0)}-\mu^{(j)}\}/[(n^{(0)}-\ntil^{(0)})!(n^{(j)}-\ntil^{(j)})!].
     \]
     The multiplier of $\prod_{i=1}^r q_L(\ntil^{(s_{i-1})}|\ntil^{(s_{i})})$ in \eqref{eqn.bigprod} is
\begin{align*}
\frac{\left(\mu^{(j)}\right)^{n^{(j)}}}{n^{(j)}!}\exp\{-\mu^{(j)}\}&
\frac{n^{(j)}!}{\ntil^{(j)}!(n^{(j)}-\ntil^{(j)})!}(1-\delta)^{\ntil^{(j)}}\delta^{n^{(j)}-\ntil^{(j)}}
\frac{(\mu^{(0)}\delta)^{n^{(0)}-\ntil^{(0)}}}
     {(n^{(0)}-\ntil^{(0)})!}
     \exp\left\{-\delta \mu^{(0)}\right\}\\
   &  =
b\times     \frac{1}{\ntil^{(j)}!}
     (1-\delta)^{\ntil^{(j)}}
     (\mu^{(0)})^{-\ntil^{(0)}}\exp\{(1-\delta)\mu^{(0)}\}
\end{align*}
That is, the right hand side of \eqref{eqn.bigprod} is
\[
b\times \prod_{i=1}^r q_L(\ntil^{(s_{i-1})}|\ntil^{(s_{i})})\times
 \frac{1}{\ntil^{(j)}!}
     (1-\delta)^{\ntil^{(j)}}
     (\mu^{(0)})^{-\ntil^{(0)}}\exp\{(1-\delta)\mu^{(0)}\}.
 \]
 However, by \eqref{eqn.multiplier}, this is
 \[
b\times \prod_{i=1}^r q_R(\ntil^{(s_{i})}|\ntil^{(s_{i-1})})\times
 \frac{1}{\ntil^{(0)}!}
     (1-\delta)^{\ntil^{(0)}}
     (\mu^{(j)})^{-\ntil^{(j)}}\exp\{(1-\delta)\mu^{(j)}\},
 \]
which is the left hand side of \eqref{eqn.bigprod}, completing the proof.

\section{Justification for Backward Sampling}\label{Appendix: Backward sampling}
We prove the validity of xPGBS in this section. The proof is similar to the theoretical justification of Ancestor Sampling in \cite{Lindsten2014}. 
Recall from Section \ref{subsec: xpgbs} that for Backwards sampling, the state at observation time $t_1$ is $\mathbf{x}_{1}=(x_0, x_{\tau}, \dots, x_{t_\ell})^\top$ and the state at observation time $t_\ell$ is $\mathbf{x}_{\ell}=(x_{t_{\ell-1}+\tau}, x_{t_{\ell-1}+2\tau}, \dots, x_{t_\ell})^\top$, $\ell=2, \dots, L$. For simplicity, suppose there is only one type of reaction, the proof for multiple reaction channels follows naturally. Following \cite{Lindsten2014}, let us define a sequence of unnormalized densities, for $\ell=1, \dots, L$,
\begin{equation}
    \begin{aligned}
        \gamma_{\theta, \ell}\left( \mathbf{x}_{1:\ell}\right)&=p( \mathbf{x}_{1:\ell}, y_{1:\ell}\mid \theta)\\
        &=p_0(x_0\mid \theta)\prod_{s=0}^{\ell-1}\prod_{k=1}^K p(n_{sK+k}\mid \mu_{sK+k})\prod_{s=1}^{\ell}g(y_s\mid x_{t_s}, \theta),
    \end{aligned}
    \label{eq: marginal_gamma}
\end{equation}
where $\mu_{s K +k}$ and $n_{s K +k}$ denote the expected and realised numbers of reactions, respectively, as defined in Sec~\ref{sec: proof_thm1}. When $\ell=0$, it follows that 
\[
 \bar{\gamma}_{\theta, 0}\left(x_0\right):= \gamma_{\theta, 0}\left(x_0\right)=p_0(x_0\mid\theta).
\]
For $\ell \ge 1$, the corresponding normalized density is defined as 
\[
\bar{\gamma}_{\theta, \ell}\left( \mathbf{x}_{1:\ell}\right)=\frac{\gamma_{\theta, \ell}\left( \mathbf{x}_{1:\ell}\right)}{Z_{\theta, \ell}}=p(\mathbf{x}_{1:\ell} \mid y_{1:\ell}, \theta),
\]
where $Z_{\theta,\ell}:=\int \gamma_{\theta,\ell}(x_0,\mathbf{x}_{1:\ell}) \md \mathbf{x}_{1:\ell}$.
Following \cite{Lindsten2014}, we define a probability function $\phi_\theta$:
\[
\phi_\theta\left( \mathbf{x}_{1:L}^{(0:M)}, a_{1:L-1}^{(0:M)}, b_L \right)=\phi_\theta\left(  \mathbf{x}_{1:L}^{(b_{1:L})}, b_{1:L}\right)\phi_\theta\left(\mathbf{x}_{1:L}^{(-b_{1:L})}, a_{1:L-1}^{(-b_{2:L})}\mid \mathbf{x}_{1:L}^{(b_{1:L})}, b_{1:L}\right)
\]
where the first term on the right hand side is given by
\[
\phi_\theta\left(  \mathbf{x}_{1:L}^{(b_{1:L})}, b_{1:L}\right)=\frac{\bar{\gamma}_{\theta, \ell}\left(\mathbf{x}_{1:L}^{(b_{1:L})}\right)}{(M+1)^{L}};
\]
that is, we first sample the path $\left( \mathbf{x}_{1:L}^{(b_{1:L})}\right)$ from its marginal distribution $\bar{\gamma}_{\theta, \ell}\left(\mathbf{x}_{1:L}^{(b_{1:L})}\right)$, and sample the corresponding indices $b_{1:L}$ independently from the uniform distribution on set $\{0, 1, \dots, M\}$. The second term is defined to be the conditional distribution of all non-reference path particles and ancestors that is obtained when the conditional particle filter is run forwards.

Similar to the proof in \cite{Lindsten2014}, we first show that the the xPGBS algorithm is equivalent to the following steps for sampling from the full joint probability function $\phi_\theta$. 

\begin{procedure}[Instrumental reformulation of xPGBS]
Given $\mathbf{x}_{1:L}^{\prime(b^\prime_{1:L})}$ and $b^\prime_{1:L} \in \{0,\ldots,M\}^{L}.$ 
\begin{enumerate}
    \item Run the conditional particle filter forwards by drawing:\label{procedure1 step1}
   \[ 
   \left \{ \mathbf{x}_1^{(-b_1^\prime)}\right\}
    \sim
    \phi_\theta\left(
    \cdot
    \mid
   \mathbf{x}_{1:L}^{\prime(b^\prime_{1:L})},
    b_{1:L}^\prime
    \right),\]
    
    and for $\ell=2, \dots, L$, 
    \[
    \left\{\mathbf{x}_\ell^{(-b_\ell^\prime)}, a_{\ell-1}^{-b_{\ell}^\prime}\right\}
    \sim
    \phi_\theta\left(
    \cdot
    \mid 
    \mathbf{x}_{1:\ell-1}^{(-b_{1:\ell-1}^\prime)},
    a_{1:\ell-2}^{(0:M)},
    \mathbf{x}_{1:L}^{\prime( b_{1:L}^\prime)},
    b_{\ell-1:L}^\prime
    \right),
    \]
    \item Draw a new reference path via backward sampling: \label{procedure1 step2}
    \[
    b_L
    \sim
    \phi_\theta\left(
    \cdot
    \mid
    \mathbf{x}_{1:L}^{(-b_{1:L}^\prime)},
    a_{1:L-1}^{(0:M)},
    \mathbf{x}_{1:L}^{\prime(b_{1:L}^\prime)}
    \right),
    \]
    and for $\ell=L-1, \dots, 1$, 
    \begin{equation}
        b_\ell \equiv a_{\ell}^{b_{\ell+1}}\sim \phi_\theta\left(\cdot\mid 
    \mathbf{x}_{1:\ell}^{(-b_{1:\ell}^\prime)},
    a_{1:\ell-1}^{(0:M)}, 
    \mathbf{x}_{1:\ell}^{\prime(b_{1:\ell}^\prime)},
    \mathbf{x}_{\ell+1:L}^{(b_{\ell+1:L})}, b_{\ell+1:L}
    \right).
    \label{eq: procedure1 step2}
    \end{equation}
\end{enumerate}
\label{procedure: 1}
\end{procedure}

\begin{lemma}
        Procedure~\ref{procedure: 1} leaves the joint distribution
    \[
    \phi_\theta\left(
    \mathbf{x}_{1:L}^{(0:M)},
    a_{1:L-1}^{(0:M)},
    b_L
    \right)
    \]
    invariant.
\end{lemma}
\begin{proof}
    By the chain rule for conditional distributions, it the straightforward to verify that step~\ref{procedure1 step1} samples $\{\mathbf{x}_{1:L}^{(-b_{1:L})}, a_{1:L-1}^{(-b_{2:L})}\}$ from its full conditional distribution with respect to $\phi_\theta$:
    \[
    \left\{\mathbf{x}_{1:L}^{(-b_{1:L})}, a_{1:L-1}^{(-b_{2:L})}\right\}\sim \phi_\theta\left(\cdot\mid  \mathbf{x}_{1:L}^{\prime(b^\prime_{1:L})},
    b_{1:L}^\prime\right).
    \]
    In addition, \eqref{eq: procedure1 step2} in step~\ref{procedure1 step2} is equivalent to sampling $b_\ell$ from its full conditional distribution with respect to $\phi_\theta$:
    \[
     b_\ell \equiv a_{\ell}^{b_{\ell+1}}\sim \phi_\theta\left(\cdot\mid 
    \mathbf{x}_{1:\ell}^{(-b_{1:\ell}^\prime)},
    a_{1:\ell-1}^{(0:M)}, 
    \mathbf{x}_{1:\ell}^{\prime(b_{1:\ell}^\prime)},
    \mathbf{x}_{\ell+1:L}^{(b_{\ell+1:L})}, b_{\ell+1:L}, \underbrace{{\mathbf{x}}_{\ell+1:L}^{(-b_{\ell+1:L})}, {a}_{\ell:L-1}^{-b_{\ell+1:L}}}
    \right).
    \]
    because $b_\ell$ is conditionally independent of the variables marked with the underbrace, given $ \{\mathbf{x}_{\ell+1:L}^{(b_{\ell+1:L})}, b_{\ell+1:L} \}$ and these, therefore, do not affect the full conditional distribution of $b_\ell$. Hence, step~\ref{procedure1 step2} can be interpreted as component-wise draws that together constitute a full vector Gibbs update of $b_{1:L}$ with respect to $\phi_\theta$.
\end{proof}

\begin{lemma}
The algorithm of xPGBS is equivalent to Procedure~\ref{procedure: 1}, conditionally on $x_0^{\prime (b_0)}=x_0^{(0)}$
$\mathbf{x}_{1:L}^{\prime(b_{1:T}^\prime)}=\mathbf{x}_{1:L}^{(0)}$ and
$b_{0:L}^\prime=(0,\ldots,0)$.
\end{lemma}
\begin{proof}
    According to the algorithm of xPG, we have
    \begin{align*}
        \phi_\theta\left(\mathbf{x}_{1:L}^{(-b_{1:L})}, a_{1:L-1}^{-b_{2:L}}\mid \mathbf{x}_{1:L}^{(b_{1:L})}, b_{1:L}\right)&=\tilde{q}_0\left(x_0^{(-b_1)}\mid x_0^{(b_1)}\right)\prod_{\ell=0}^{L-1}\prod_{k=1}^Kq\left(n_{\ell K+k}^{-b_{\ell+1}}\mid \mu_{\ell K+k}^{(0:M)}, n_{\ell K+k}^{b_{\ell+1}}\right)\\
        &\qquad\times \left\{\prod_{\ell=1}^{L-1}\prod_{\substack{m=0\\m\neq b_{\ell+1}}}^{M}\frac{g\left(y_\ell\mid x_{t_\ell}^{(a_\ell^m)}\right)}{\sum_{j=0}^Mg\left(y_\ell\mid x_{t_\ell}^{(j)}\right)}\right\}.
    \end{align*}
    Therefore, we have
    \[
    \phi_\theta\left( \mathbf{x}_1^{(-b_1)} \mid  \mathbf{x}_{1:L}^{(b_{1:L})}, b_{1:L}\right)=\tilde{q}_0(x_0^{(-b_1)}\mid x_0^{(b_1)})\prod_{k=1}^Kq\left(n_{k}^{-b_{1}}\mid \mu_{k}^{(0:M)}, n_{k}^{b_{1}}\right).
    \]
    and 
    \begin{align*}
        \phi_\theta\left(\mathbf{x}_\ell^{(-b_\ell)}, a_{\ell-1}^{-b_\ell}\mid \mathbf{x}_{1:\ell-1}^{(-b_{1:\ell-1})}, a_{1:\ell-2}^{-b_{2:\ell-1}}, \mathbf{x}_{1:L}^{(b_{1:L})}, b_{1:L} \right)&=\prod_{k=1}^Kq\left(n_{(\ell-1) K+k}^{-b_{\ell}}\mid \mu_{(\ell-1) K+k}^{(0:M)}, n_{(\ell-1) K+k}^{b_{\ell}}\right)\\
        &\qquad\times\prod_{\substack{m=0\\m\neq b_{\ell}}}^{M}\frac{g\left(y_{\ell-1}\mid x_{t_{\ell-1}}^{(a_{\ell-1}^m)}\right)}{\sum_{j=0}^Mg\left(y_{\ell-1}\mid x_{t_{\ell-1}}^{(j)}\right)}.
    \end{align*}
With $b_\ell=0$ for $\ell=1, \dots, L$, the above conditional distributions coincide with the forward simulation steps of xPGBS. 

For the backward sampling step, for $\ell=L-1, \dots, 1$, we use the relation $a_\ell^{b_{\ell+1}}=b_\ell$. Hence, for $\ell=2, \dots, L$, we have $b_{1:\ell-1}=a_{1:\ell-1}^{b_{2:\ell}}$. Moreover, noting that $a_{1:\ell-1}^{-b_{2:\ell}}$ is the collection of $a_1^{-b_2}, a_2^{-b_3}, \dots, a_{\ell-1}^{-b_{\ell}}$, we obtain
\begin{equation}
\begin{aligned}
    \phi_\theta&\left(a_\ell^{b_{\ell+1}}\mid \mathbf{x}_{1:\ell}^{(0:M)}, a_{1:\ell-1}^{(0:M)}, \mathbf{x}_{\ell+1:L}^{(b_{\ell+1:L})}, b_{\ell+1:L}\right)\propto \phi_\theta\left( \mathbf{x}_{1:\ell}^{(0:M)}, a_{1:\ell-1}^{(0:M)}, \mathbf{x}_{\ell+1:L}^{(b_{\ell+1:L})}, b_{\ell:L}\right)\\
    & = \phi_\theta\left(\mathbf{x}_{1:L}^{(b_{1:L})}, b_{1:L}\right)\phi_\theta\left( \mathbf{x}_{1:\ell}^{(-b_{1:\ell})}, a_{1:\ell-1}^{-b_{2:\ell}}\mid \mathbf{x}_{1:L}^{(b_{1:L})}, b_{1:L}\right)\\
    &= \frac{\bar{\gamma}_{\theta, L}\left(\mathbf{x}_{1:L}^{(b_{1:L})}\right)}{(M+1)^{L+1}} \tilde{q}_0\left(x_0^{(-b_1)}\mid x_0^{(b_1)}\right)\prod_{s=0}^{\ell-1}\prod_{k=1}^Kq\left(n_{s K+k}^{-b_{s+1}}\mid \mu_{s K+k}^{(0:M)}, n_{s K+k}^{b_{s+1}}\right)
         \left\{\prod_{s=1}^{\ell-1}\prod_{\substack{m=0\\m\neq b_{s+1}}}^{M}\frac{g\left(y_s\mid x_{t_s}^{(a_s^m)}\right)}{\sum_{j=0}^Mg\left(y_s\mid x_{t_s}^{(j)}\right)}\right\}\\
         &=\frac{\bar{\gamma}_{\theta, L}\left( \mathbf{x}_{1:L}^{(b_{1:L})}\right) \gamma_{\theta, \ell}\left( \mathbf{x}_{1:\ell}^{(b_{1:\ell})}\right)}{(M+1)^{L+1} \gamma_{\theta, \ell}\left( \mathbf{x}_{1:\ell}^{(b_{1:\ell})}\right)} \tilde{q}_0\left(x_0^{(-b_1)}\mid x_0^{(b_1)}\right)\prod_{s=0}^{\ell-1}\prod_{k=1}^Kq\left(n_{s K+k}^{-b_{s+1}}\mid \mu_{s K+k}^{(0:M)}, n_{s K+k}^{b_{s+1}}\right)\\
         &\qquad\times\left\{\prod_{s=1}^{\ell-1}\prod_{\substack{m=0\\m\neq b_{s+1}}}^{M}\frac{g\left(y_s\mid x_{t_s}^{(a_s^m)}\right)}{\sum_{j=0}^Mg\left(y_s\mid x_{t_s}^{(j)}\right)}\right\}.
\end{aligned}
\label{eq: joint_phi}
\end{equation}

Using Equation~\eqref{eq: marginal_gamma} to expand $\gamma_{\theta,\ell}\left( \mathbf{x}_{1:\ell}^{(b_{1:\ell})}
\right)$ in the numerator in Equation~\eqref{eq: joint_phi}, we find that the expression above is proportional to the ratio 
\[
\frac{\bar{\gamma}_{\theta, L}\left( \mathbf{x}_{1:L}^{(b_{1:L})}\right)}{ \gamma_{\theta, \ell}\left( \mathbf{x}_{1:\ell}^{(b_{1:\ell})}\right)}
\]
multiplied by the product of three terms. 

The first term accounts for both the prior density of the initial state and the proposal distribution for the remaining initial states,
\begin{equation}
    q_0(x_0^{(0:M)})\equiv p_0(x_0^{(b_1)})\tilde{q}_0(x_0^{(-b_1)}).
\label{eq:BS_term1}
\end{equation}
The second term combines the marginal distribution of the reaction counts with the proposal distribution for the remaining reaction counts at each time step up to $t=\ell K$, 
\begin{equation}
    \prod_{s=0}^{\ell-1}\prod_{k=1}^Kp\left(n_{sK+k}^{b_{s+1}}\mid \mu_{sK+k}^{b_{s+1}}\right)\tilde{q}\left(n_{s K+k}^{-b_{s+1}}\mid \mu_{s K+k}^{(0:M)}, n_{s K+k}^{b_{s+1}}\right).
    \label{eq:BS_term2}
\end{equation}
Finally, the third term captures the contributions from the observation likelihoods and path resampling probabilities up to $t=\ell K$,
\begin{equation}
\begin{aligned}
     g\left(y_\ell\mid x_{t_\ell}^{(b_\ell)}\right)\prod_{s=1}^{\ell-1}\left\{g\left(y_s\mid x_{t_s}^{(b_s)}\right) \prod_{\substack{m=0\\m\neq b_{s+1}}}^{M}\frac{g\left(y_s\mid x_{t_s}^{(a_s^m)}\right)}{\sum_{j=0}^Mg\left(y_s\mid x_{t_s}^{(j)}\right)}\right\}\\
    =g\left(y_\ell\mid x_{t_\ell}^{(b_\ell)}\right)\prod_{s=1}^{\ell-1}\left\{\left(\sum_{j=0}^Mg\left(y_s\mid x_{t_s}^{(j)}\right)\right)^{-M}\prod_{m=0}^{M}g\left(y_s\mid x_{t_s}^{(a_s^m)}\right)\right\} 
\end{aligned}
\label{eq:BS_term3}
\end{equation}
since $a_s^{b_{s+1}}=b_s$. As shown in Section~\ref{sec: proof_thm1}, both \eqref{eq:BS_term1} and \eqref{eq:BS_term2} are invariant under any permutation $\sigma(0:M)$ of the indices. The first term in \eqref{eq:BS_term3} also satisfy this invariance. For the fixed particle system and ancestor indices, these factors are therefore common to all choices of the distinguished ancestral path traced back from $b_\ell$ and cancel when normalising the sampling probabilities.

Sampling from \eqref{eq: procedure1 step2} is equivalent to selecting $b_\ell$ from the candidates $b_\ell^{(0:M)}$, with the probability of choosing $b_\ell^{(i)}$ proportional to the expression in \eqref{eq: procedure1 step2} evaluated at $b_\ell=b_\ell^{(i)}$. Consequently, the probability of selecting $b_\ell^{(i)}$ is proportional to 
\begin{equation}
    g\left(y_\ell\mid x_{t_\ell}^{(b_\ell^{(i)})}\right)\frac{\bar{\gamma}_{\theta, L}\left(( \mathbf{x}_{1:\ell}^{(b_{1:\ell}^{(i)})}, \mathbf{x}_{\ell+1:L}^{(b_{\ell+1:L})})\right)}{ \gamma_{\theta, \ell}\left( \mathbf{x}_{1:\ell}^{(b_{1:\ell}^{(i)})}\right)}  
    \label{eq:bs_weight_final}
\end{equation}
where $b_{1:\ell}^{(i)}$ denotes the sequence of ancestral indices obtained by tracing back from $b_\ell=i$.

Fixing $b_s=\tilde{b}_s$ for $s=\ell+1,\dots,L$ and applying \eqref{eq: marginal_gamma}, since $b_\ell^{(i)}=i$ and $w_{\ell}^i=g(y_\ell\mid x_{t_\ell}^{(i)})$, we find that \eqref{eq:bs_weight_final} is proportional to
\[
g\left(y_\ell\mid x_{t_\ell}^{(i)}\right)p\left(n_{\ell K+1}^{\tilde{b}_{\ell +1}}\mid x_{t_\ell}^{(i)}, \theta\right)=w_{\ell}^i \cdot p\left(x_{t_{\ell}+\tau}^{(\tilde{b}_{\ell+1})} \mid x^{(i)}_{t_\ell}, \theta\right),
\]
which corresponds to the backward sampling step described in Secton~\ref{subsec: xpgbs}. 
\end{proof}

% \CS{LANYA: what dark magic is this? I can see the $p_0 \tilde{q}_0$ turns into an exchangeable $q_0$; similarly each $p \tilde{q}$ (should it be this, rather than $pq$?) turns into an exchangeable $q$. But where does it all disappear? Please can you provide more explanation? In fact, FIRST, please explain each change from (29) to the first line separately. For example}
% $$
% \prod_{\substack{m=0\\m\neq b_{s+1}}}^{M}\frac{g\left(y_s\mid x_{t_s}^{(a_s^m)}\right)}{\sum_{j=0}^Mg\left(y_s\mid x_{t_s}^{(j)}\right)}
% =
% \prod_{m=0}^{M}\frac{g\left(y_s\mid x_{t_s}^{(a_s^m)}\right)}{\sum_{j=0}^Mg\left(y_s\mid x_{t_s}^{(j)}\right)}
% \times \frac{\sum_{j=0}^Mg\left(y_s\mid x_{t_s}^{(j)}\right)}{g\left(y_s|x_{t_s}^{(b_s)}\right)},
% $$
% since $a_s^{b_{s+1}}=b_s$.
% \CS{So, we first need an explanation after (29) of where the next main expression comes from. Then we need an explanation of how it all magically disappears to give (30). Or at least, I do.}

\section{Justification of the tuning recommendations}
\subsection{Rationale for the assumptions}\label{appndix_subsec:assumptions}

Assumption \ref{assumption: average weight} states that, whether or not a particle has the reference path in its ancestry does not affect its average weight just prior to the current resampling step. The mixing of the Markov chain $\{X_t\}_{t\in [0, T]}$ suggests that if the most recent reference-path ancestor was at time $s\ll t$ then particles at time $t$ with the reference path in their ancestry should have similar properties to those without reference-path ancestry. When $s$ is closer to $t$, for particle Gibbs, the assumption might not be valid as the reference path is likely to fit a given observation better than one of the proposed paths. However, as $\delta$ gets smaller, there is more and more similarity between the proposed reference paths, making Assumption \ref{assumption: average weight} reasonable for exchangeable particle Gibbs. 

The intuition behind Assumption~\ref{assumption: independence} is as follows. After a large number of MCMC iterations, the full reference path would achieve its stationary distribution. Since the observations are fixed, it is therefore natural to assume that the average of the particle weights also has its own stationary distribution which does not change substantially with the number of proposed particles \(M\). There is no reason why the weight of the reference path should be strictly independent of $M$, but, equally, we do not expect any strong dependence. 

% In addition, by the exchangeability property of xPG, at each observation time, every particle has the same marginal distribution as the reference-path particle, and their correlation is not largely affected by $M$. Therefore, we do not expect any strong dependence between $\overline{W}_\ell$ and $M$. 

Assumption~\ref{assumption: independence} also states that the number of particles from/not from the reference path at the \textit{previous} observation time (after resampling) is independent of the current weight of the reference path and the average weight of all other particles at the current observation time. Again, we do not expect the assumption to hold precisely, but rather that any dependence is weak enough to be ignored.
Suppose the latent process $\{X_{t_\ell}\}_{\ell= 1}^L$ has a stationary distribution. When it is stationary, under the HMM, it implies a stationary distribution for the observations $\{Y_\ell\}_{\ell= 1}^L$. Hence, $\{(X_{t_\ell}, Y_\ell)^\top\}_{\ell=1}^L$ is a Markov chain with a stationary distribution. Similarly, $\{(X_{t_\ell}^{(1:M)}, Y_\ell)^\top\}_{\ell=1}^L$ is a Markov chain as well. Since the reference-path weight and the average proposed-particle weight are given by $W_\ell^{(0)} = g(Y_\ell \mid X_{t_\ell}^{(0)},\theta)$ and $\overline{W}_\ell = \frac{1}{M}\sum_{i=1}^M g(Y_\ell \mid X_{t_\ell}^{(i)},\theta)$, 
both $\{W_\ell^0\}_{\ell=1}^L$ and $\{\overline{W}_\ell\}_{\ell=1}^L$ are stationary, and so is their quotient. Of course, our analysis conditions on a particular sequence of observations, $y_{1:L}$, so the assumption is not strictly true. Part 2 asks for the stochastic process to be reasonably well behaved.

\subsection{Proof of Lemma \ref{lemma.expectPoverMG}}
\label{app.proveExpectPoverMG}
   At the first observation time $t_1$, we have 
   \begin{align*}
       & M_1^G\sim \text{Bin}(M, p_1(G)), \\ 
   &\text{where } p_1(G)  =\frac{M \overline{W}_1}{M\overline{W}_1 + W_1^{(0)}}=P_1^{(M)}. 
   \end{align*}
    At observation time $t_\ell$, $\ell=2, \dots, L$, conditional on $M^G_{\ell-1}$, let $p_\ell(G\mid M_{\ell-1}^G)$ denote the conditional probability that a particle sampled, with probability proportional to its weight, from the $M+1$ particles at time $t_\ell$ is not the reference path and does not have the reference path in its ancestry. Then conditional on $M_{\ell-1}^G$, 
    \[
     M_\ell^G\sim \text{Bin}\left(M,p_{\ell}(G\mid M^G_{\ell-1})\right),
    \]
    \[
         \text{where } p_\ell(G\mid M^G_{\ell-1})=\frac{M_{\ell-1}^G \overline{W}_{\ell}^G}{M\overline{W}_{\ell}+W_{\ell}^{(0)}}=\frac{M_{\ell-1}^G \overline{W}_{\ell}}{M\overline{W}_\ell+W^{(0)}_\ell},
    \]
where the last equality follows from Assumption \ref{assumption: average weight}.
    We prove the result by induction on $\ell$. According to Assumption \ref{assumption: independence}, the claim is immediate for \(\ell=2\) by
    \[
    \mathbb{E}_{M_1^G}[p_2(G)]=\mathbb{E}_{M_1^G}\left[p_2(G\mid M_1^G)\right]=\frac{\mathbb{E}[M_1^G]\overline{W}_2}{M\overline{W}_2+W_2^{(0)}}=\frac{P_1^{(M)}M\overline{W}_2}{M\overline{W}_2+W_2^{(0)}}=P_1^{(M)}P_2^{(M)}.
    \]
     Suppose now that it holds for some \(2\le\ell \le L-1\). We then show that it also holds for \(\ell+1\). By the total expectation theorem, $\mathbb{E}_{M_{1:\ell}^G }[p_{\ell+1}(G)]$ is equal to 
     \begin{align*}
          &\mathbb{E}_{M_\ell^G}\left[\mathbb{E}_{M_{1:(\ell-1)}^G}\left[p_{\ell+1}(G\mid M_\ell^G)\right]\right]\\
          =&\mathbb{E}_{M_\ell^G} \left[\mathbb{E}_{M_{1:(\ell-1)}^G}\left[M_\ell^G\right]\frac{\overline{W}_{\ell+1}}{M\overline{W}_{\ell+1}+W^{(0)}_{\ell+1}}\right]\\
          =&M\mathbb{E}_{M_{1:(\ell-1)}^G}\left[ p_\ell(G)\right]\frac{\overline{W}_{\ell+1}}{M\overline{W}_{\ell+1}+W^{(0)}_{\ell+1}} \\
          =& \left\{\prod_{k=1}^{\ell} P^{(M)}_{k}\right\} \cdot P_{\ell+1}^{(M)}
     \end{align*}
Therefore, by induction, the claim holds for all \(\ell\in\{2, \dots, L\}\). $\square$

\subsection{Proof of Lemma \ref{lemma.bound}}
\label{Appendix: Tuning suggestion}
To simplify notation, we abbreviate $P^{(L,M)}$ and $E^{(L,M)}$ to $P^{(M)}$ and $E^{(M)}$. 
\begin{proposition}
For \(x \ge 0\) and \(y \ge 0\),
\[
\left| e^{-x} - e^{-y} \right| \le |x-y|.
\]
\label{proposition 1}
\end{proposition}

\begin{proof}
Let \(f(x)=e^{-x}\). Then
\[
|f'(x)| = e^{-x} \le 1, \qquad x \ge 0.
\]
By the mean value theorem,
\[
f(x)-f(y) = (x-y)f'(\xi)
\]
for some \(\xi\) between \(x\) and \(y\). Hence
\[
|f(x)-f(y)| = |x-y|\,|f'(\xi)| \le |x-y|,
\]
since \(\xi \ge 0\).
\end{proof}

\begin{proposition}
For \(x \ge 0\),
\[
1-x \le e^{-x} \le \frac{1}{1+x}.
\]
\label{proposition 2}
\end{proposition}

\begin{proof}
For the first inequality, apply Taylor's theorem to \(f(x)=e^{-x}\) about \(0\):
\[
f(x)=f(0)+xf'(0)+\frac{x^2}{2}f''(\xi)
\]
for some \(\xi\in(0,x)\). Since \(f(0)=1\), \(f'(0)=-1\), and \(f''(\xi)=e^{-\xi}\ge 0\), we get
\[
e^{-x}=1-x+\frac{x^2}{2}e^{-\xi}\ge 1-x.
\]

For the second inequality, apply the same argument to \(g(x)=e^x\). Taylor's theorem gives
\[
e^x = 1+x+\frac{x^2}{2}e^\eta
\]
for some \(\eta\in(0,x)\), so
\[
e^x \ge 1+x.
\]
Taking reciprocals yields
\[
e^{-x}\le \frac{1}{1+x}.
\]
This proves the result.
\end{proof}

\begin{lemma}
\[
\mathbb{E}\!\left[
\left|
\exp\!\left(-\frac{1}{M}\sum_{\ell=1}^L R_\ell\right)
-
\exp\!\left(-\frac{1}{M}\sum_{\ell=1}^L E_\ell\right)
\right|
\right]
\le
\frac{1}{M}
\operatorname{Var}\!\left(\sum_{\ell=1}^L R_\ell\right)^{1/2}.
\]
\label{lemma: approximation1}
\end{lemma}

\begin{proof}
From Proposition~\ref{proposition 1}, 
\[
\left|
\exp\!\left(-\frac{1}{M}\sum_{\ell=1}^L R_\ell\right)
-
\exp\!\left(-\frac{1}{M}\sum_{\ell=1}^L E_\ell\right)
\right|
\le
\frac{1}{M}
\left|
\sum_{\ell=1}^L R_\ell- \sum_{\ell=1}^L E_\ell
\right|.
\]
So, by Jensen's inequality,
\[
\mathbb{E}\!\left[
\left|
\exp\!\left(-\frac{1}{M}\sum_{\ell=1}^L R_\ell\right)
-
\exp\!\left(-\frac{1}{M}\sum_{\ell=1}^L E_\ell\right)
\right|
\right]
\le
\frac{1}{M}
\mathbb{E}\!\left[
\left|
\sum_{\ell=1}^L R_\ell - \sum_{\ell=1}^L E_\ell
\right|
\right]
\]
\[
\le
\frac{1}{M}
\mathbb{E}\!\left[
\left(
\sum_{\ell=1}^L R_\ell - \sum_{\ell=1}^L E_\ell
\right)^2
\right]^{1/2}.
\]
Since \(E_\ell=\mathbb{E}[R_\ell]\), this is
\[
\frac{1}{M}
\operatorname{Var}\!\left(\sum_{\ell=1}^L R_\ell\right)^{1/2},
\]
which proves the result.
\end{proof}

\begin{lemma}
    \[
    \left|P^{(M)}-\exp\left(-\frac{1}{M}\sum_{\ell=1}^LR_\ell\right)\right|\le \frac{1}{M^2}\sum_{\ell=1}^L\{R_\ell\}^2
    \]
    \label{lemma: approximation2}
\end{lemma}
\begin{proof}
Write
\[
P^{(M)} = \prod_{\ell=1}^L a_\ell,
\qquad \text{where } a_\ell := P_\ell^{(M)} \text{ and } 0 \le a_\ell \le 1.
\]
Write 
\[
\exp\left(-\frac{1}{M}\sum_{\ell=1}^LR_\ell\right)=\prod_{\ell=1}^L b_\ell,
\qquad \text{where } b_\ell := \exp\left(-\frac{R_\ell}{M}\right) \text{ and } 0 \le b_\ell \le 1.
\]
Then, using the convention that the product over an empty set is \(1\),
\[
\prod_{\ell=1}^L a_\ell - \prod_{\ell=1}^L b_\ell
=
\sum_{j=1}^L
\left(\prod_{\ell=1}^{j-1} a_\ell\right)
(a_j-b_j)
\left(\prod_{\ell=j+1}^L b_\ell\right).
\]
So
\[
\left|
\prod_{\ell=1}^L a_\ell - \prod_{\ell=1}^L b_\ell
\right|
\le
\sum_{j=1}^L
\left|
\left(\prod_{\ell=1}^{j-1} a_\ell\right)
(a_j-b_j)
\left(\prod_{\ell=j+1}^L b_\ell\right)
\right|
\le
\sum_{j=1}^L |a_j-b_j|,
\]
because all \(a_j\) and \(b_j\) lie between \(0\) and \(1\).

    From Proposition~\ref{proposition 2},
\[
\left|
\frac{1}{1 + R_\ell/M}
-
\exp\!\left(-\frac{1}{M}R_\ell\right)
\right|
\le
\left|
\frac{1}{1 + R_\ell/M}
-
\left(1-\frac{1}{M}R_\ell\right)
\right|
=
\frac{R_\ell^2/M^2}{1 + R_\ell/M}
\le
\frac{1}{M^2}R_\ell^2,
\]
Therefore, we have 
\[
    \left|P^{(M)}-\exp\left(-\frac{1}{M}\sum_{\ell=1}^LR_\ell\right)\right|\le\sum_{\ell=1}^L\left|P_\ell^{(M)}-\exp\left(-\frac{R_\ell}{M}\right)\right| \le \frac{1}{M^2}\sum_{\ell=1}^L\{R_\ell\}^2.
    \]
\end{proof}

We now combine Lemmas \ref{lemma: approximation1} and \ref{lemma: approximation2}  to prove Lemma \ref{lemma.bound}. First define $R^{(M)}:=\exp\!\left(-\frac{1}{M}\sum_{\ell=1}^L R_\ell\right)$.
    Then,
    \begin{align*}
\mathbb{E}\!\left[
\left|
P^{(M)}-E^{(M)}
\right|
\right]
&\le
\mathbb{E}\!\left[
\left|
P^{(M)}-R^{(M)}
\right|
\right] 
 +
\mathbb{E}\!\left[
\left|
R^{(M)}
-
E^{(M)}
\right|
\right] \\
&\le
\frac{1}{M^2}\,
\mathbb{E}\!\left[\sum_{\ell=0}^L R_\ell^2\right]
+
\frac{1}{M}\,
\operatorname{Var}\!\left[\sum_{\ell=0}^L R_\ell\right]^{\frac{1}{2}}.
\end{align*}
% \label{lemma: combination}

\section{Inferred numbers infectious over time}

\begin{landscape}
\begin{figure}
    \centering
    \includegraphics[width=1.0\linewidth]{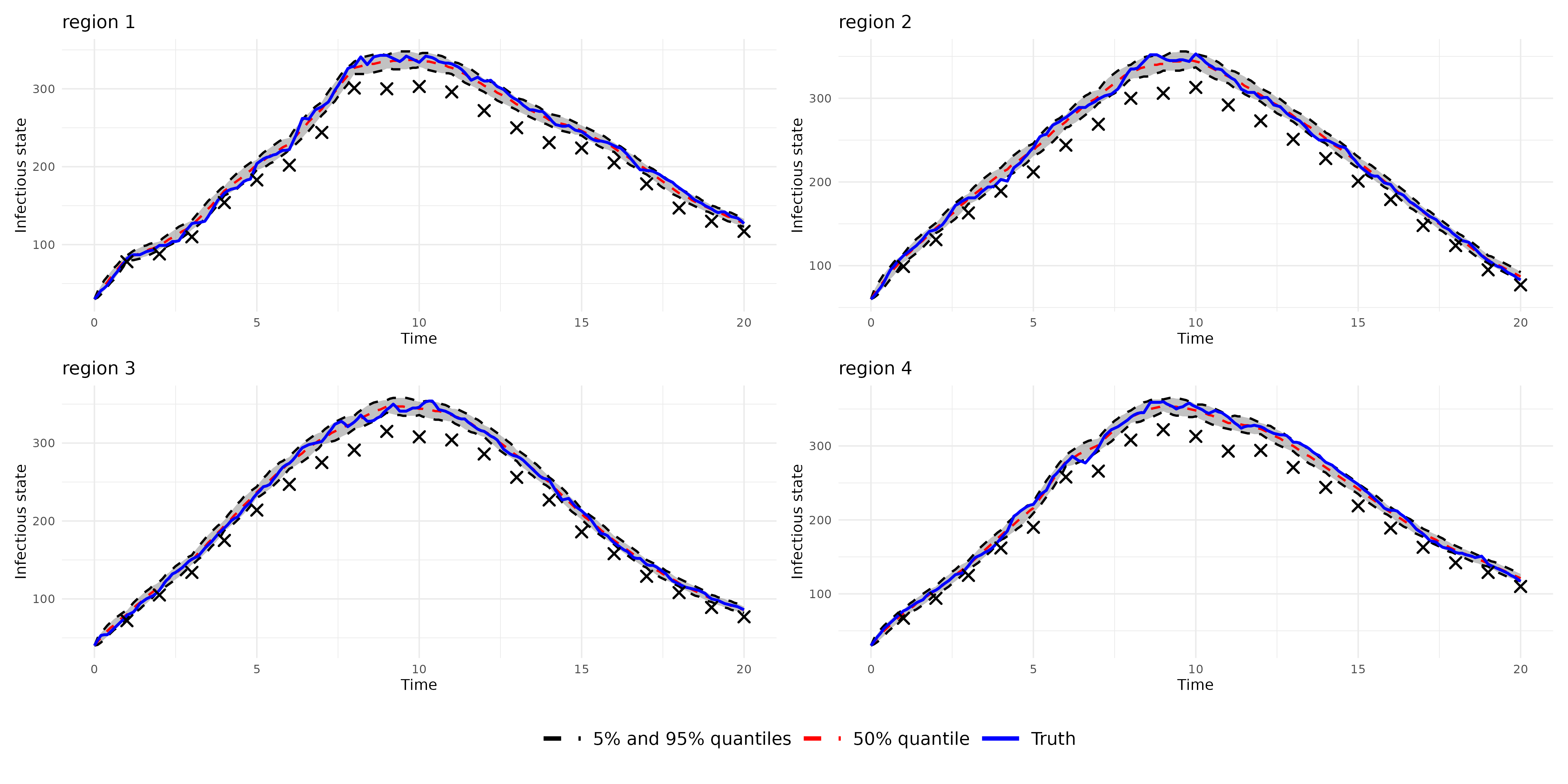}
    \caption{Plots of the numbers infectious over time in four regions in the SEIR model after running xPG for $10^6$ iterations with a thinning interval of $10$. The true parameters are $\beta = 0.6$, $\eta = 0.6$, and $\gamma = 0.2$. Black crosses indicate the observations, black dashed lines show the $5\%$ and $95\%$ quantiles of the latent states inferred by xPG, the red dashed line shows the median, and the blue solid line represents the true latent states from the four coupled SEIR models.
 }
    \label{fig:high4_SEIR_xpg_data}
\end{figure}
\end{landscape}

\begin{landscape}
\begin{figure}
    \centering
    \includegraphics[width=1.0\linewidth]{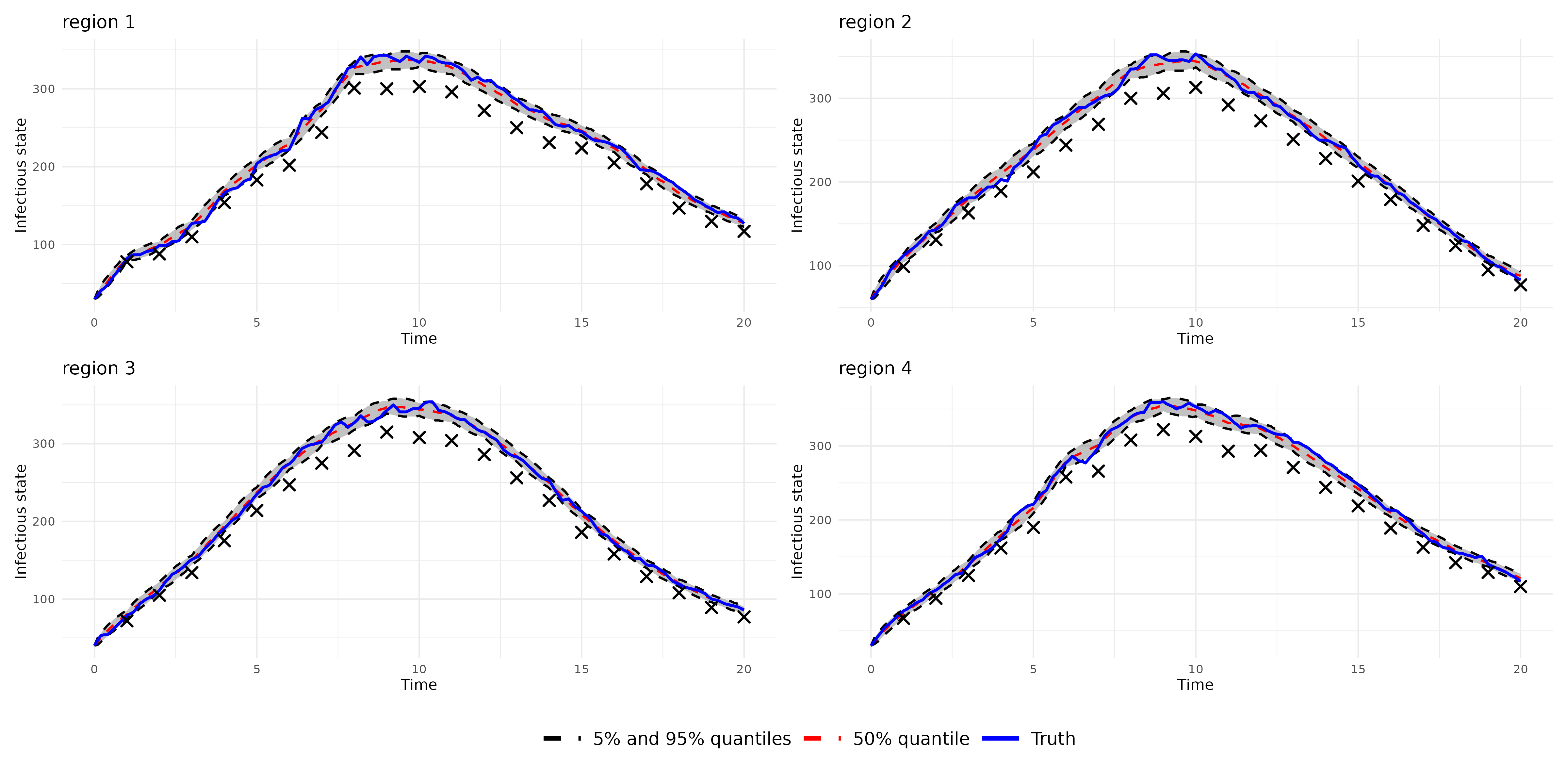}
    \caption{Plots of the numbers infectious over time in four regions in the SEIR model after running xPGBS for $10^6$ iterations with a thinning interval of $10$. The true parameters are $\beta = 0.6$, $\eta = 0.6$, and $\gamma = 0.2$. Black crosses indicate the observations, black dashed lines show the $5\%$ and $95\%$ quantiles of the latent states inferred by xPGBS, the red dashed line shows the median, and the blue solid line represents the true latent states from the four coupled SEIR models.
 }
    \label{fig:high4_SEIR_xpgbs_data}
\end{figure}
\end{landscape}

\section{Posterior distributions of model parameters}
\begin{figure}[H]
    \centering
    \includegraphics[width=1.0\linewidth]{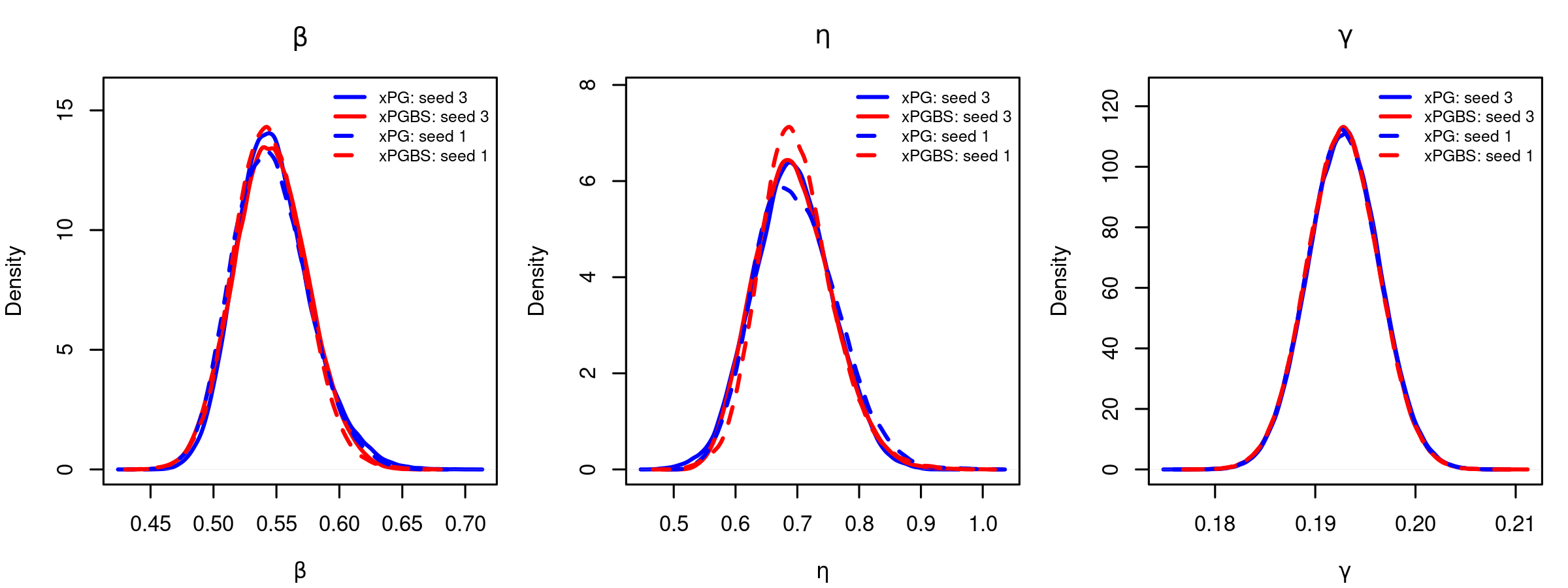}
    \caption{Density estimates of the posterior distributions of $\beta$, $\eta$, and $\gamma$ generated by xPG with $M=30$ and $\delta=0.02$, and by xPGBS with $M=10$ and $\delta=0.02$. Each algorithm was run for $2\times10^6$ iterations, retaining every tenth sample. The blue solid and dashed lines correspond to xPG runs with seeds 3 and 1, respectively, while the red solid and dashed lines correspond to xPGBS runs with seeds 3 and 1, respectively; the differences between xPG and xPGBS are comparable in magnitude to the differences between runs with different seeds for either method.}
    \label{fig: SEIR_high4_xpg_vs_xpgbs_thetas}
\end{figure}

\section{Tuning diagnostics}\label{appendix: optimal acceptance rate}
\begin{figure}[h]
\captionsetup{font=footnotesize} 
\captionsetup[subfigure]{labelformat=empty}
\centering
\begin{subfigure}{0.48\hsize}
  \centering
        \captionsetup{
    justification=centering,
    singlelinecheck=false,
    margin={0pt,12mm}
  }
  \caption{\small a}
  \includegraphics[width=\hsize]{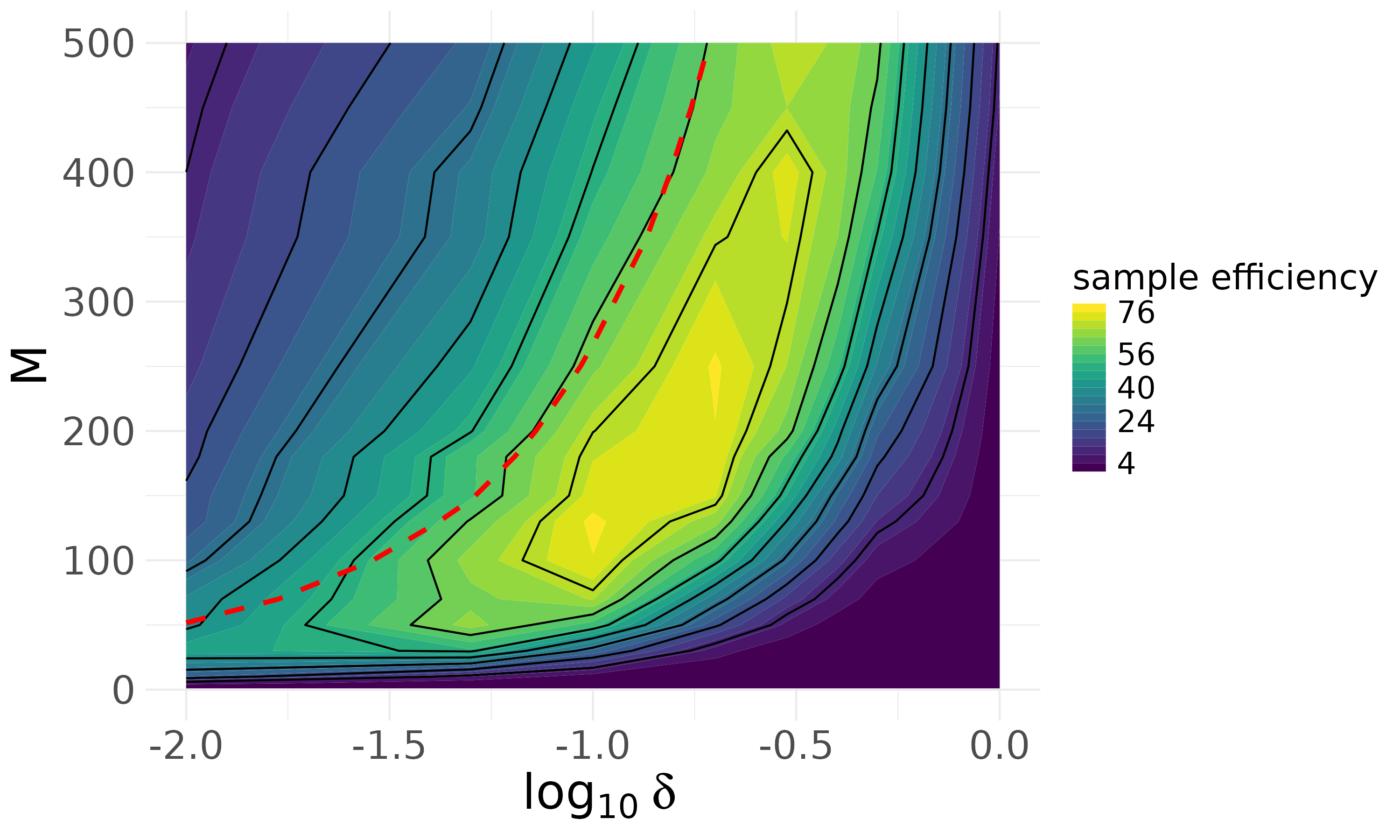}
  \label{fig:SIR_high4_S_tune}
\end{subfigure}%
\hspace{0.02\hsize}
\begin{subfigure}{0.48\hsize}
  \centering
        \captionsetup{
    justification=centering,
    singlelinecheck=false,
    margin={0pt,12mm}
  }
   \caption{\small b}
  \includegraphics[width=\hsize]{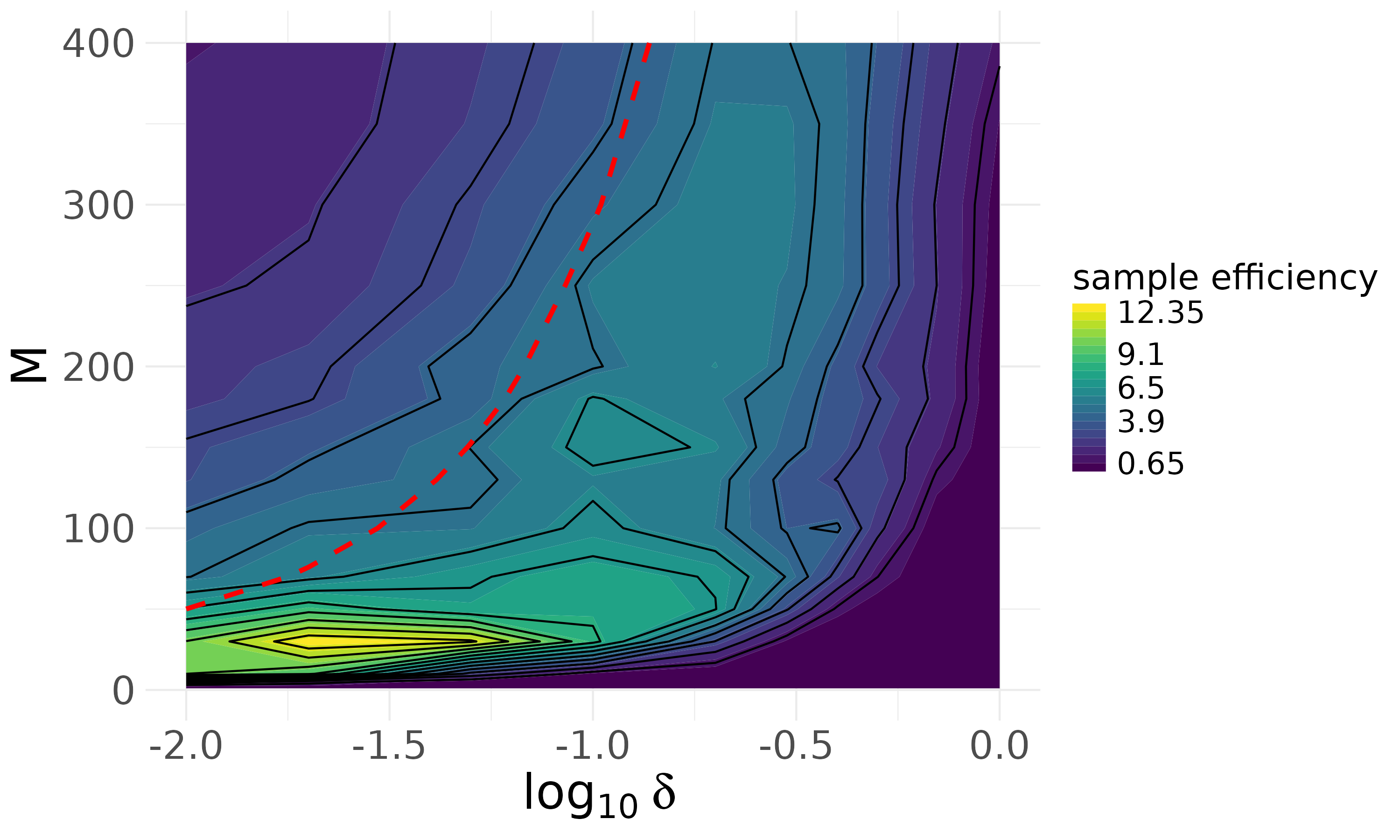}
  \label{fig:SEIR_high4_min_tune}
\end{subfigure}
% \vspace{0.02\hsize}
\vspace{-3mm}

\begin{subfigure}{0.48\hsize}
  \centering
        \captionsetup{
    justification=centering,
    singlelinecheck=false,
    margin={0pt,12mm}
  }
  \caption{\small c}
  \includegraphics[width=\hsize]{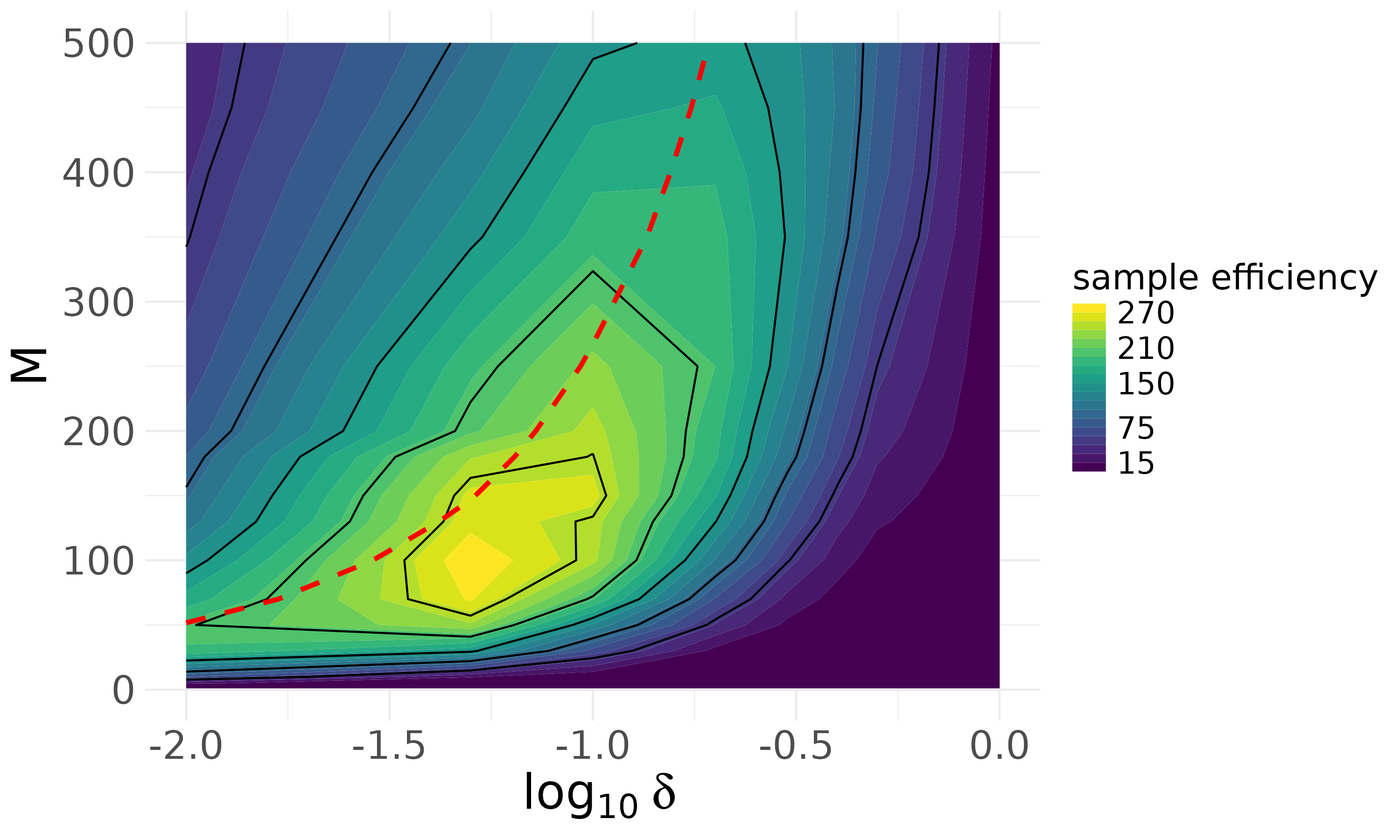}
  \label{fig:SIR_high4_I_tune}
\end{subfigure}%
\hspace{0.02\hsize}
\begin{subfigure}{0.48\hsize}
  \centering
        \captionsetup{
    justification=centering,
    singlelinecheck=false,
    margin={0pt,12mm}
  }
   \caption{\small d}
  \includegraphics[width=\hsize]{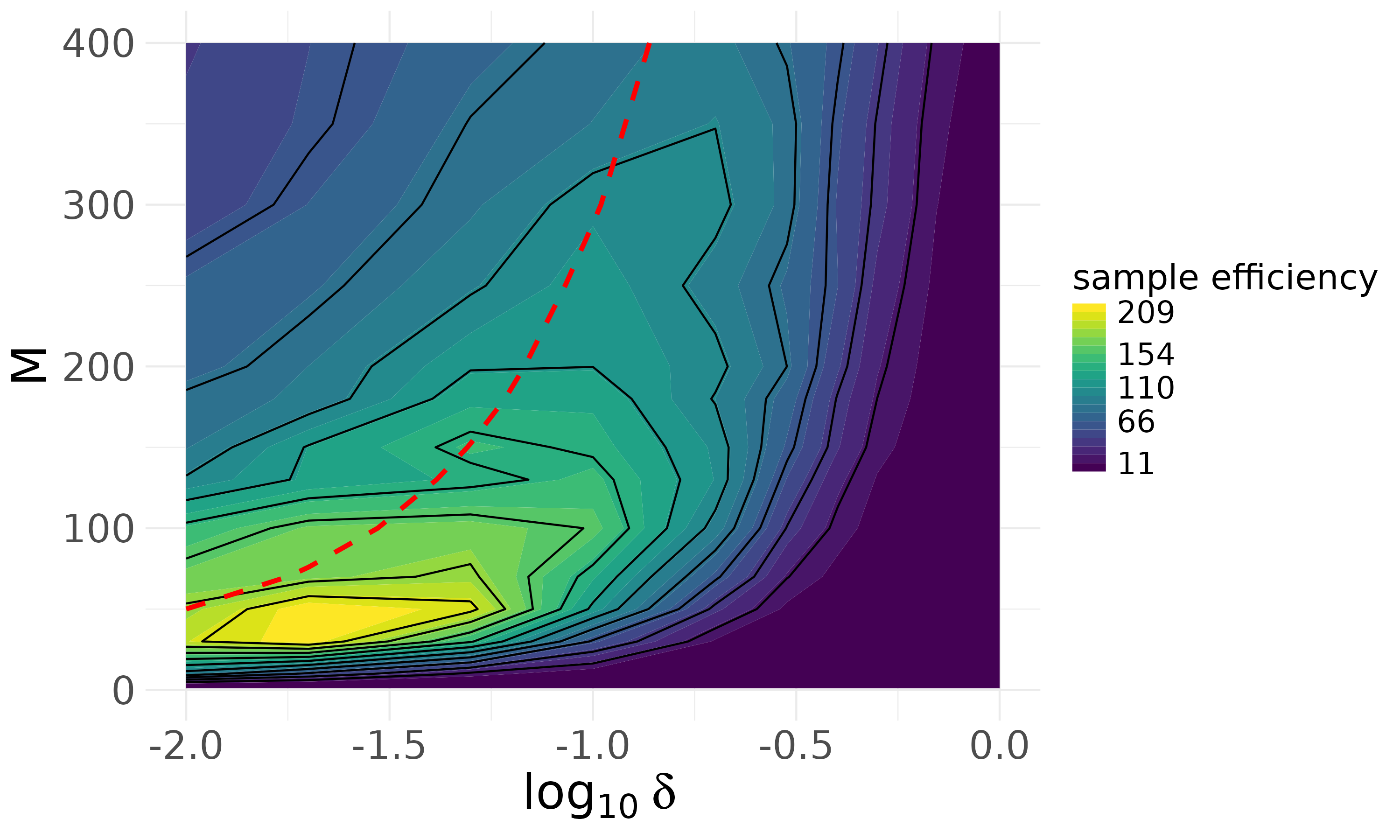}
  \label{fig:SEIR_high4_I_tune}
\end{subfigure}
\vspace{-3mm}

\begin{subfigure}{0.48\hsize}
  \centering
        \captionsetup{
    justification=centering,
    singlelinecheck=false,
    margin={0pt,12mm}
  }
  \caption{\small e}
  \includegraphics[width=\hsize]{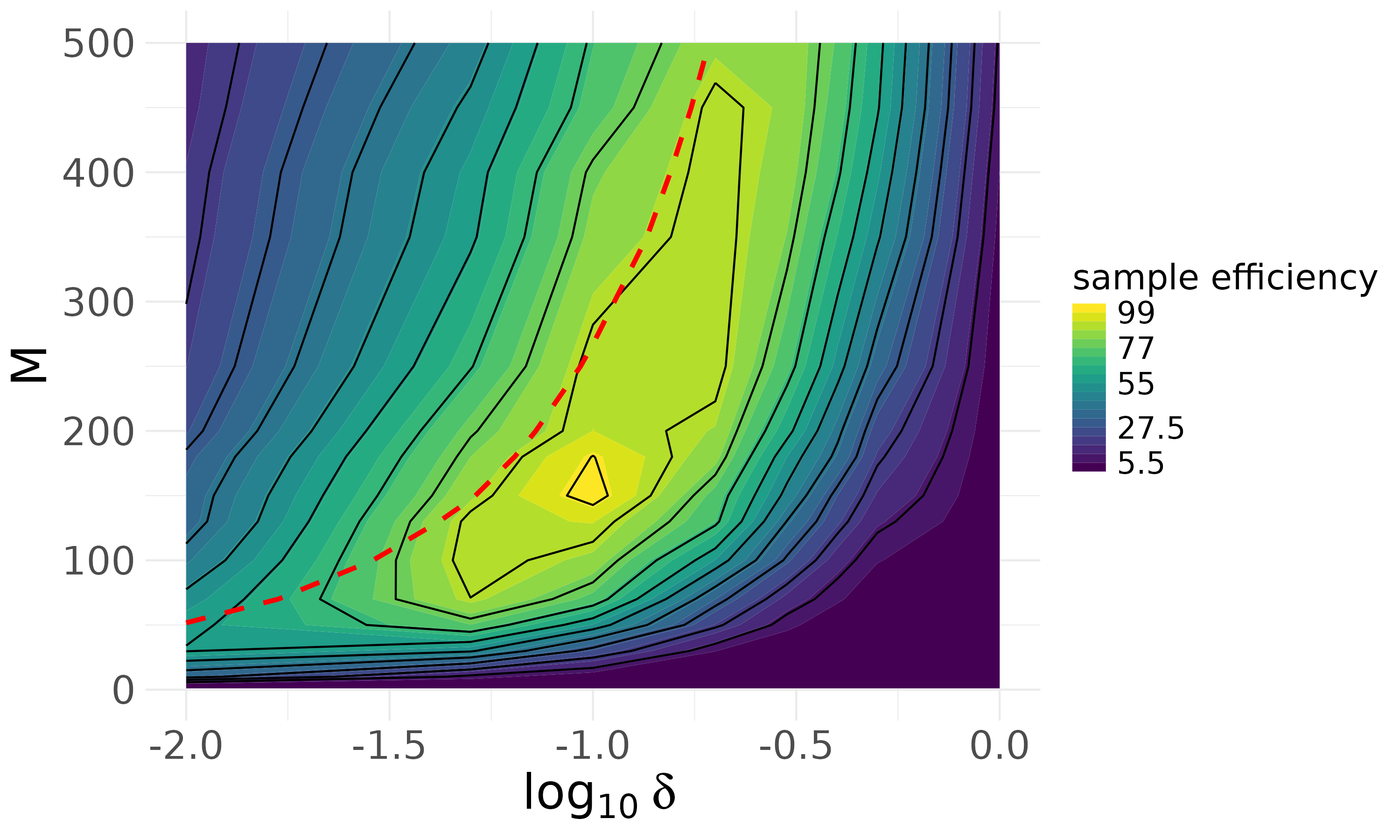}
  \label{fig:SIR_high4_t1_tune}
\end{subfigure}%
\hspace{0.02\hsize}
\begin{subfigure}{0.48\hsize}
  \centering
        \captionsetup{
    justification=centering,
    singlelinecheck=false,
    margin={0pt,12mm}
  }
   \caption{\small f}
  \includegraphics[width=\hsize]{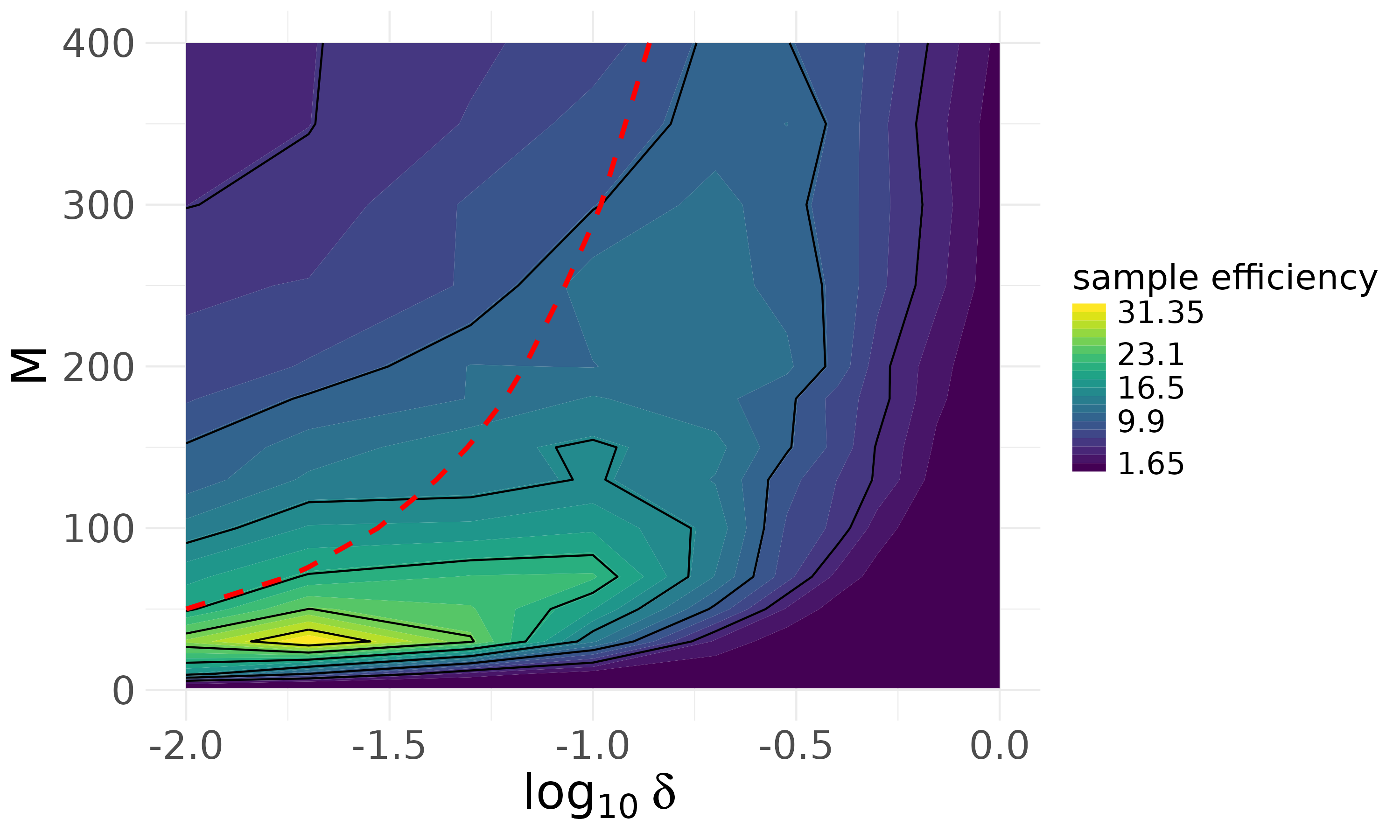}
  \label{fig:SEIR_high4_t1_tune}
\end{subfigure}
% \hspace{0.02\hsize}
\caption{\small
Contour plots of sample efficiency for xPG applied to simulated data from four coupled SIR models (left column) and four coupled SEIR models (right column), using $10^6$ iterations.
\subref{fig:SIR_high4_S_tune} and \subref{fig:SEIR_high4_min_tune}
show the efficiency for the worst mixing state over all regions and observation times;
\subref{fig:SIR_high4_I_tune} and \subref{fig:SEIR_high4_I_tune}
show the efficiency for the number of infectious individuals;
\subref{fig:SIR_high4_t1_tune} and \subref{fig:SEIR_high4_t1_tune}
show the efficiency for the worst mixing state at the first observation time across all regions.
The red dashed line indicates the points at which the acceptance rate at the first observation time $t_1$ is \(\exp(-1)\).}
\end{figure}

As is illustrated in Section~\ref{subsec: SEIR}, Figures~\ref{fig:SIR_high4_S_tune} and ~\ref{fig:SEIR_high4_min_tune} show that the value of $M$ suggested by the tuning rule is larger than the empirically optimal value when maximising sample efficiency of the worst mixing state over all latent states and observation times. In contrast, Figures~\ref{fig:SIR_high4_I_tune} and~\ref{fig:SEIR_high4_I_tune}, together with Figures~\ref{fig:SIR_high4_t1_tune} and~\ref{fig:SEIR_high4_t1_tune}, indicate that the suggested value of $M$ approximately matches the optimal $M$ when maximising either the sample efficiency of the worst mixing infectious state (the number of infectious individuals) across all regions and observation times, or the sample efficiency of the worst mixing state across all species, averaged across all regions, at the first observation time.

\begin{figure}[h]
    \centering
    \includegraphics[width=0.7\linewidth]{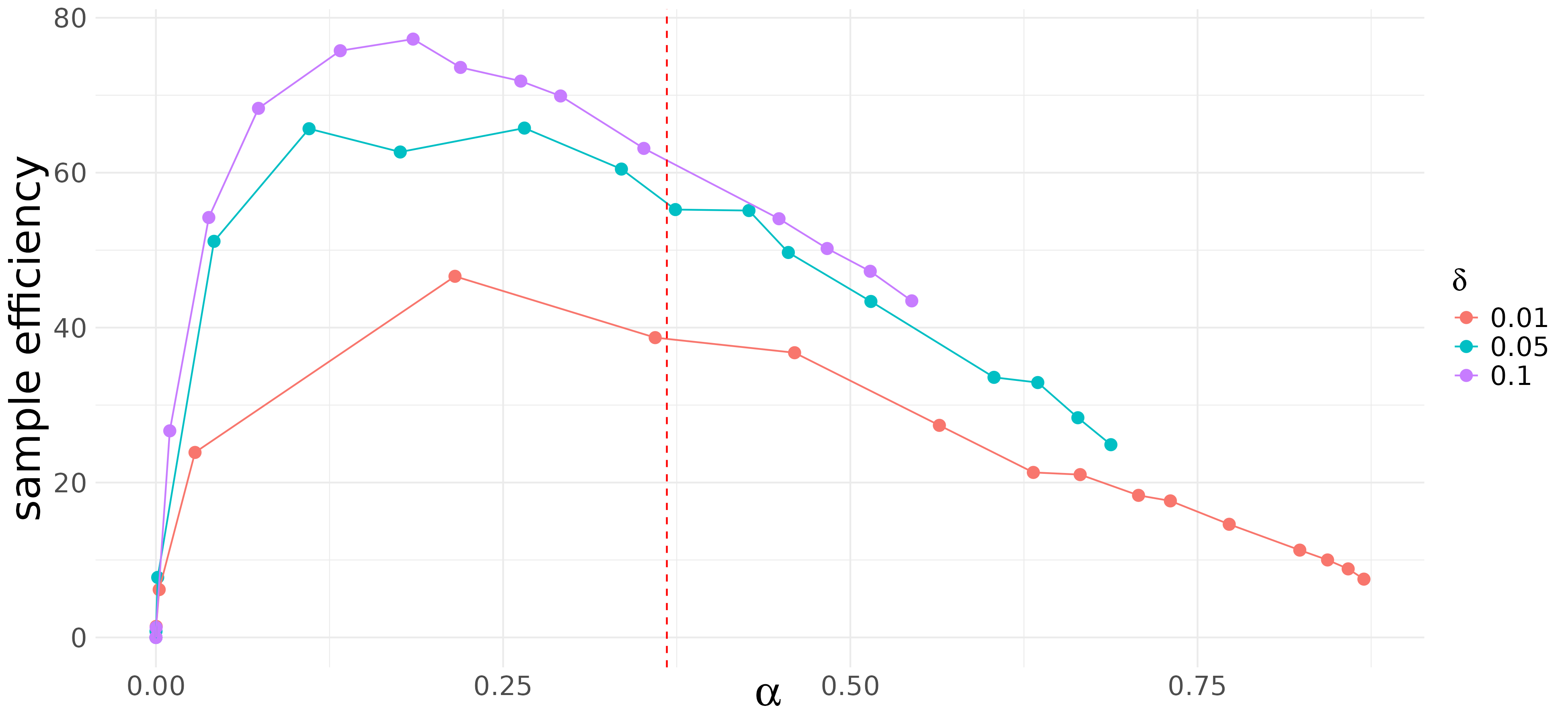}
    \caption{Sample efficiency of the worst mixing state over all regions and time points (the number of susceptible individuals) versus the acceptance rate at the first observation time for xPG applied to the four-coupled SIR model, based on $10^6$ iterations}
    \label{fig:high4_SIR_acc}
\end{figure}

\begin{figure}[h]
    \centering
    \includegraphics[width=0.7\linewidth]{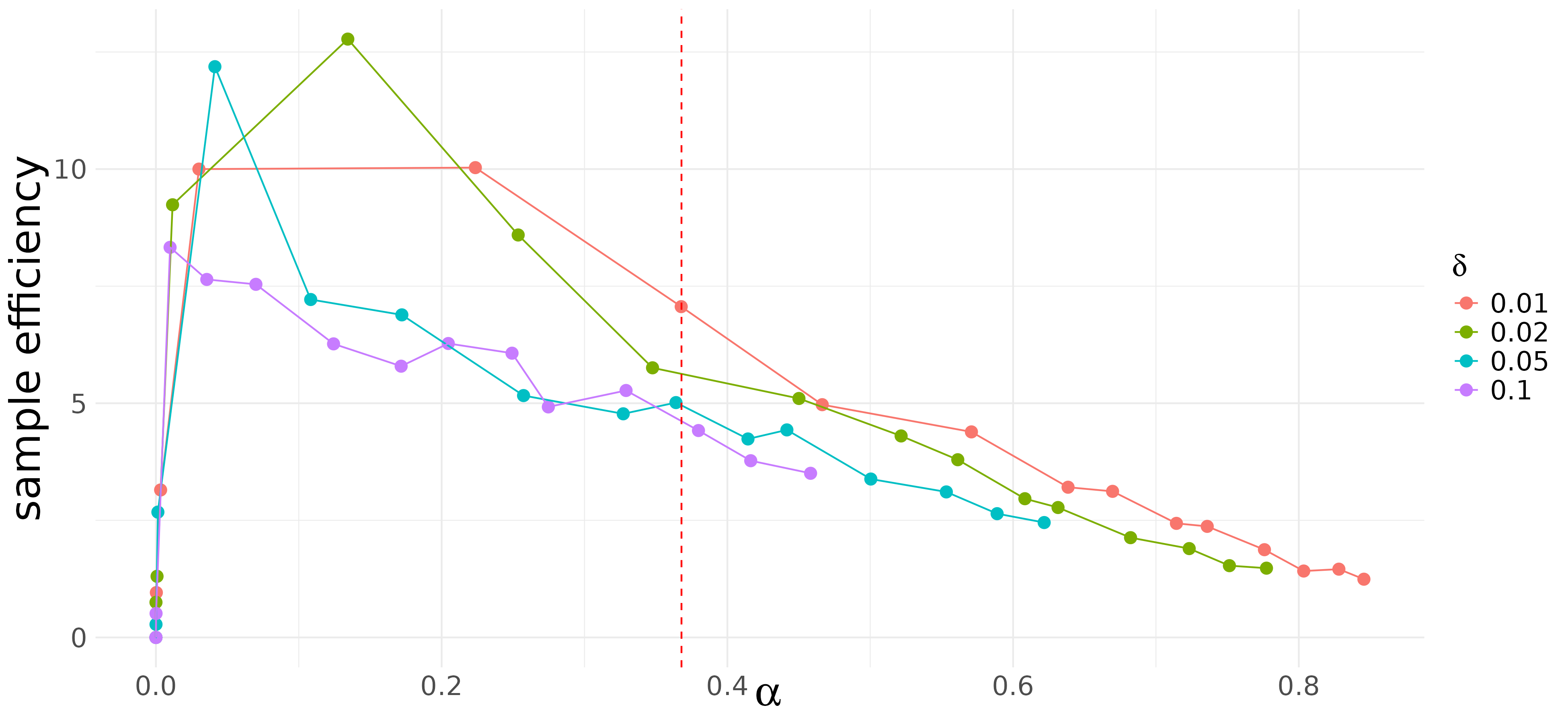}
    \caption{Sample efficiency of the worst mixing state over all regions and time points versus the acceptance rate at the first observation time for xPG applied to the four-coupled SEIR model, based on $10^6$ iterations}
    \label{fig:high4_SEIR_acc}
\end{figure}

Despite the overestimation of the optimal value of $M$, the sample efficiency achieved at the suggested value remains reasonably high, at approximately three quarters of the maximum for the four coupled SIR models and one half of the maximum for the four coupled SEIR models, as shown in Figure~\ref{fig:high4_SIR_acc} and ~\ref{fig:high4_SEIR_acc}.

\newpage
\section{Supplementary inference results}\label{sec:additional_results}

\begin{figure}[h]
\captionsetup{font=footnotesize} 
\captionsetup[subfigure]{labelformat=empty}
\centering
\begin{subfigure}{0.48\hsize}
  \centering
  \captionsetup{
    justification=centering,
    singlelinecheck=false,
    margin={0pt,12mm}
  }
  \caption{\small a}
  \includegraphics[width=\hsize]{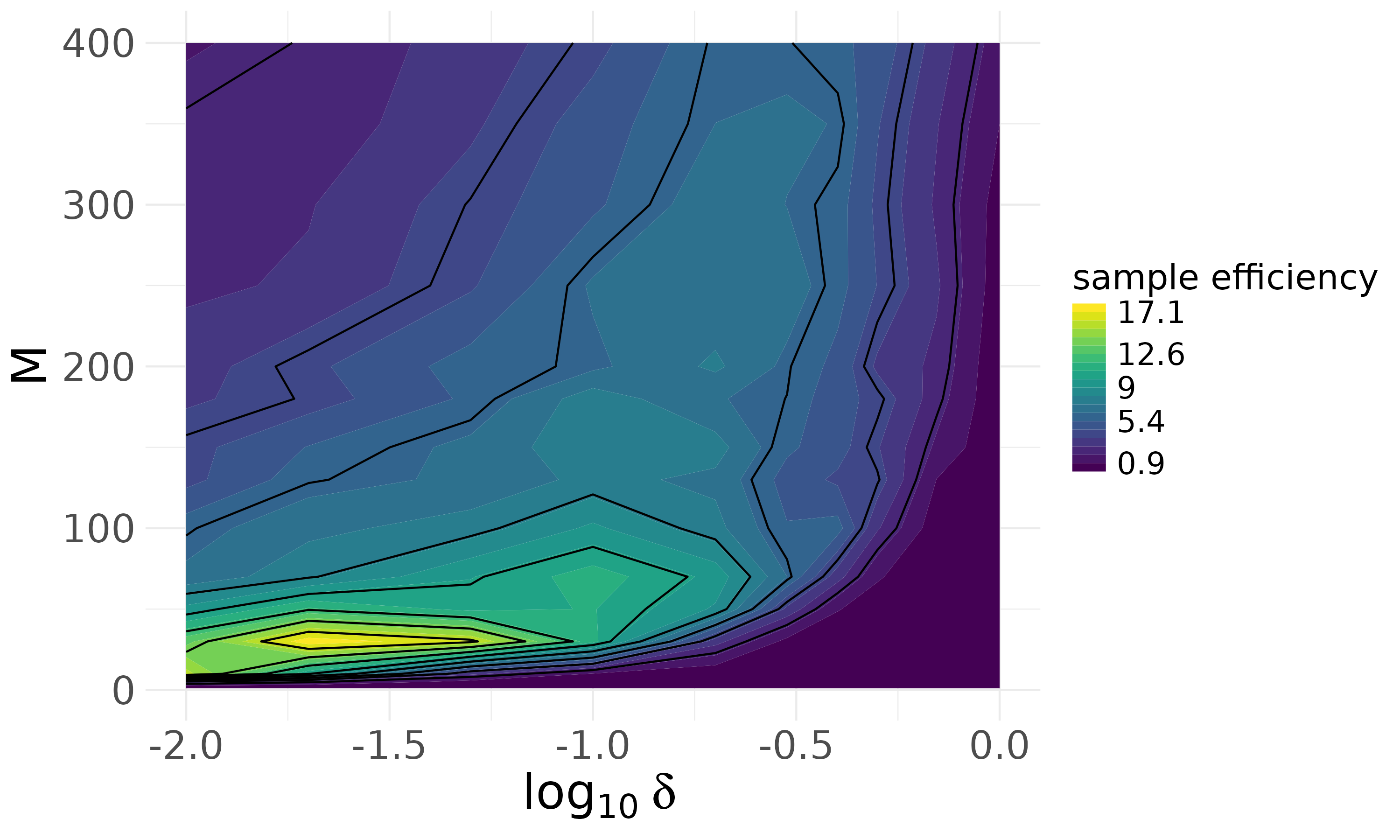}
  \label{fig:SEIR_high4_S}
\end{subfigure}%
\hspace{0.02\hsize}
\begin{subfigure}{0.48\hsize}
  \centering
  \captionsetup{
    justification=centering,
    singlelinecheck=false,
    margin={0pt,12mm}
  }
   \caption{\small b}
  \includegraphics[width=\hsize]{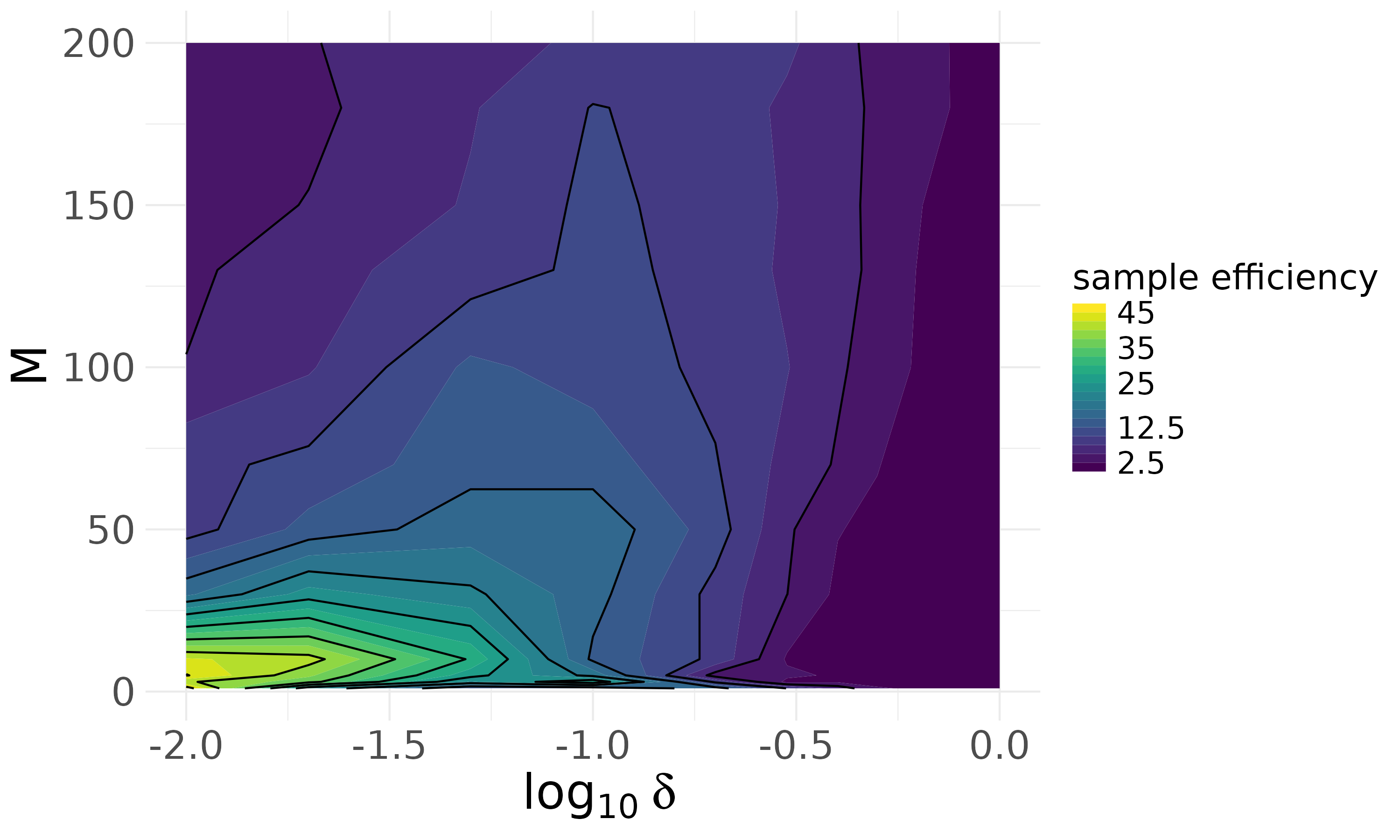}
  \label{fig:SEIR_BS_high4_S}
\end{subfigure}
% \vspace{0.02\hsize}
\vspace{-3mm}

\begin{subfigure}{0.48\hsize}
  \centering
  \captionsetup{
    justification=centering,
    singlelinecheck=false,
    margin={0pt,12mm}
  }
  \caption{\small c}
  \includegraphics[width=\hsize]{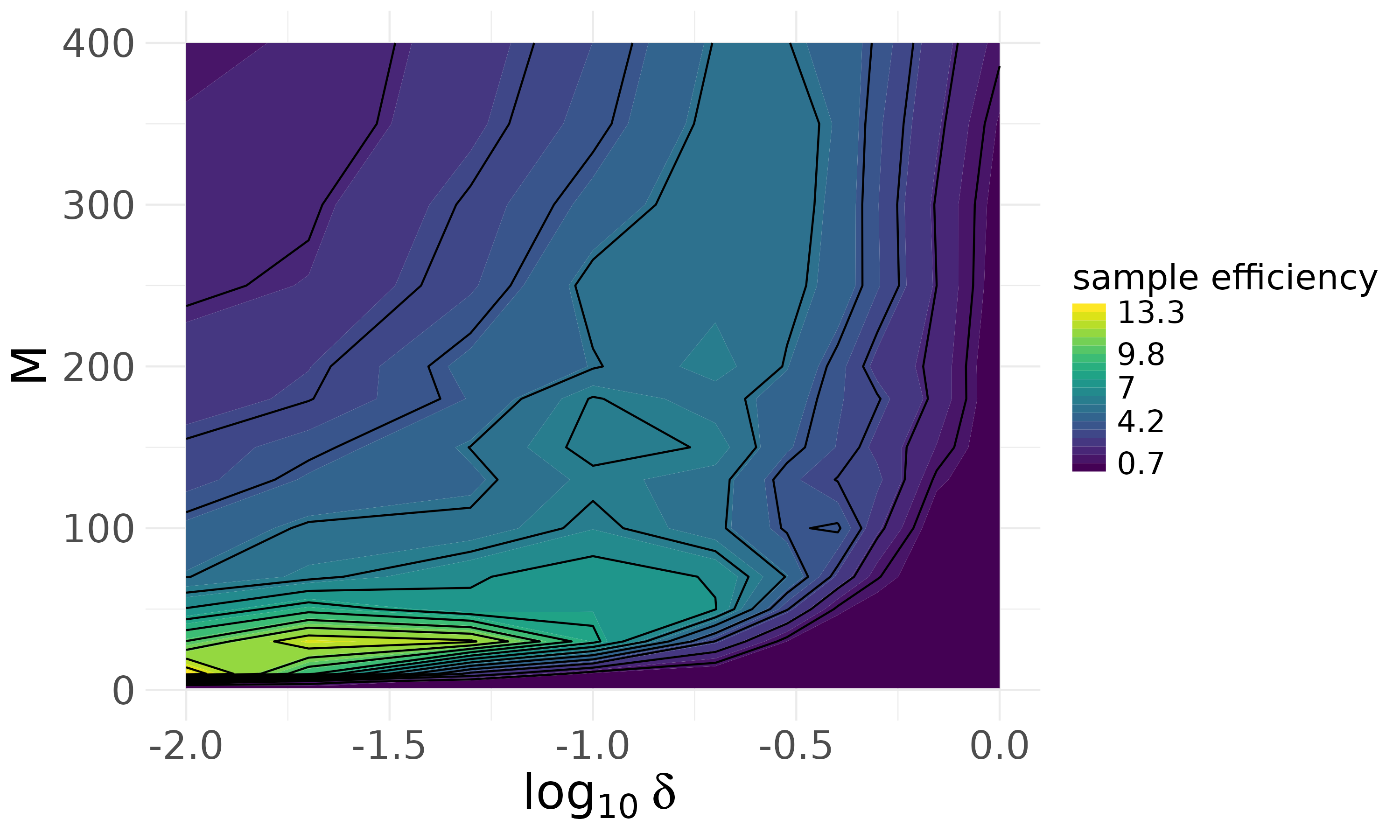}
  \label{fig:SEIR_high4_E}
\end{subfigure}%
\hspace{0.02\hsize}
\begin{subfigure}{0.48\hsize}
  \centering
  \captionsetup{
    justification=centering,
    singlelinecheck=false,
    margin={0pt,12mm}
  }
   \caption{\small d}
  \includegraphics[width=\hsize]{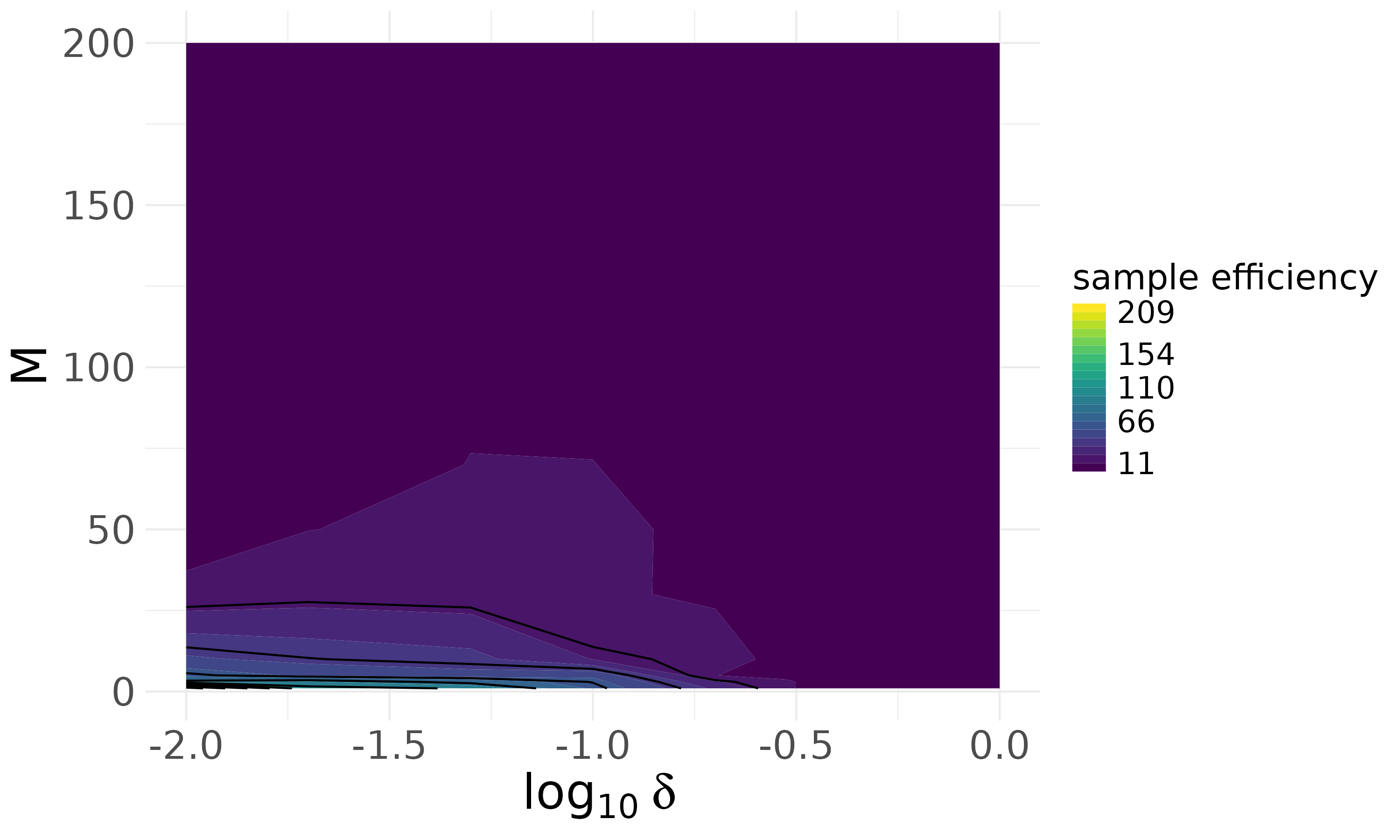}
  \label{fig:SEIR_BS_high4_E}
\end{subfigure}
\vspace{-3mm}

\begin{subfigure}{0.48\hsize}
  \centering
  \captionsetup{
    justification=centering,
    singlelinecheck=false,
    margin={0pt,12mm}
  }
  \caption{\small e}
  \includegraphics[width=\hsize]{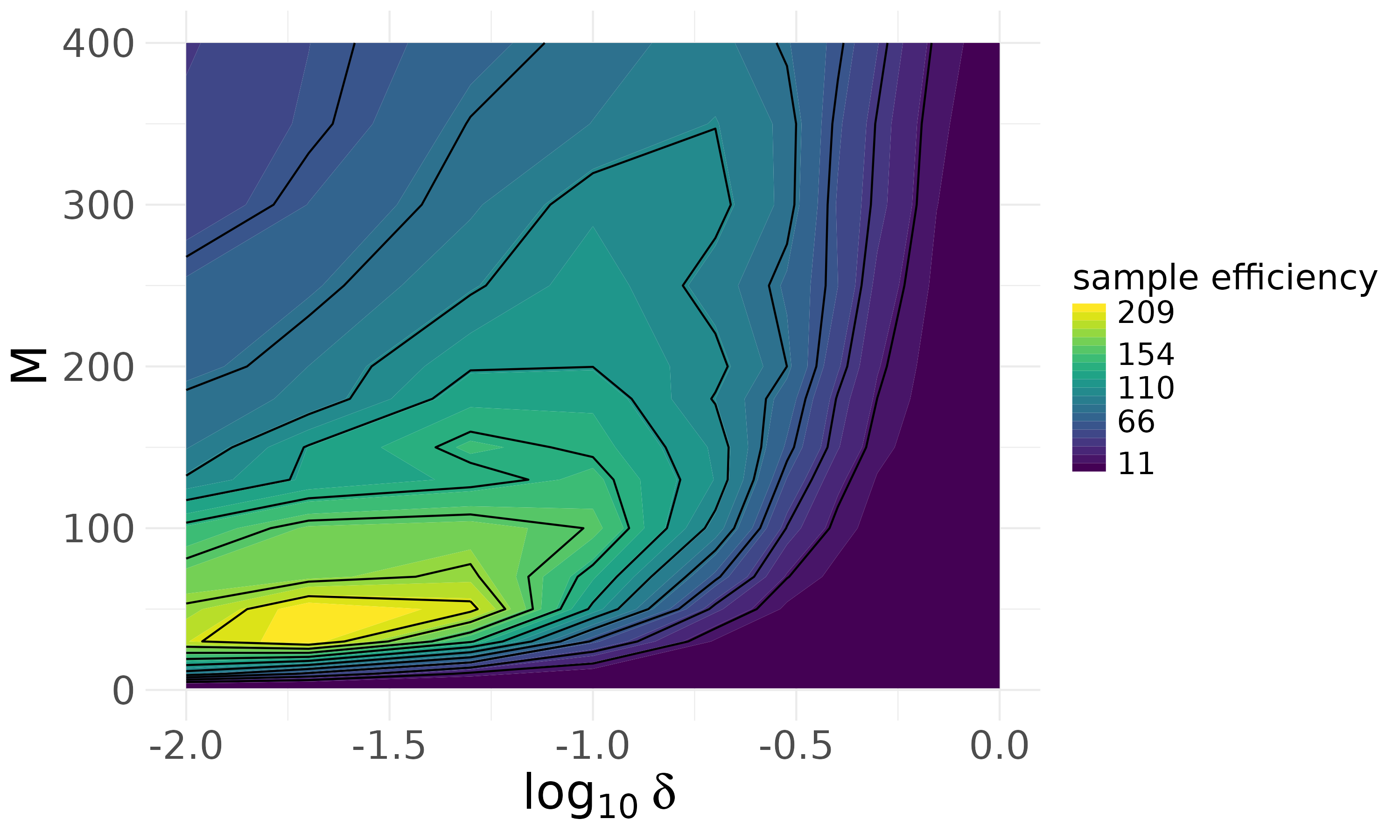}
  \label{fig:SEIR_high4_I}
\end{subfigure}%
\hspace{0.02\hsize}
\begin{subfigure}{0.48\hsize}
  \centering
  \captionsetup{
    justification=centering,
    singlelinecheck=false,
    margin={0pt,12mm}
  }
   \caption{\small f}
  \includegraphics[width=\hsize]{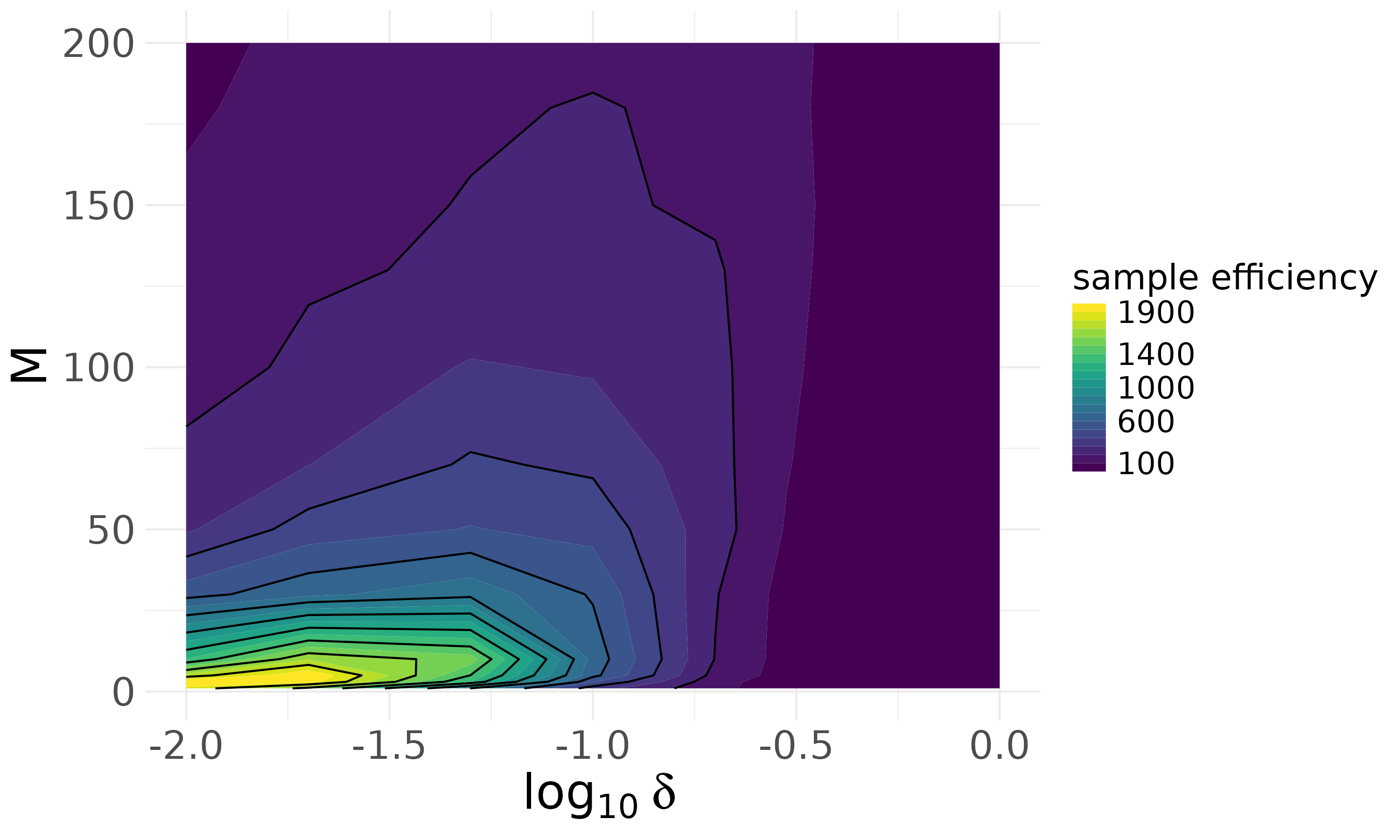}
  \label{fig:SEIR_BS_high4_I}
\end{subfigure}
% \hspace{0.02\hsize}
\caption{\small
Contour plots of sample efficiency for xPG (left column) and xPGBS (right column) applied to simulated data from four coupled SEIR models, using $10^6$ iterations.
\subref{fig:SEIR_high4_S} and \subref{fig:SEIR_BS_high4_S}
show the efficiency for the number of susceptible individuals;
\subref{fig:SEIR_high4_E} and \subref{fig:SEIR_BS_high4_E}
show the efficiency for the number of exposed individuals;
\subref{fig:SEIR_high4_I} and \subref{fig:SEIR_BS_high4_I}
show the efficiency for the number of infectious individuals.}
\label{fig: SIR_three}
\end{figure}

%TC:endignore

\end{document}